\documentclass[draft,smallcondensed]{svjour3}

\usepackage{textcomp}
\usepackage{xspace}

\let\OldTexttrademark\texttrademark
\renewcommand{\texttrademark}{\OldTexttrademark\xspace}%

\usepackage[T1]{fontenc}

\usepackage[utf8]{inputenc}

\usepackage[ruled]{algorithm2e} % For algorithms

\usepackage{todonotes}
\usepackage{paralist}
\usepackage{fixltx2e} % Required on old systems

\newcounter{challenge}
\newenvironment{challenge}[1]{\refstepcounter{challenge}\noindent%\medskip
   \textbf{C\thechallenge. #1.} \rmfamily}{\medskip}

\title{A Survey of Challenges for Runtime Verification from 
  Advanced Application Domains (Beyond Software)
  \thanks{
    Corresponding authors: C\'esar S\'anchez \email{<cesar.sanchez@imdea.org>} 
    and Gerardo Schneider \email{<gersch@chalmers.se>}.  
}
}

\newcommand{\hi}[1]{}

\hi{
\author{
   C\'esar S\'anchez\inst{1} \and
   Gerardo Schneider\inst{2} \and 
   Wolfgang Ahrendt\inst{3} \and
   Ezio Bartocci\inst{4} \and \\
   Domenico Bianculli\inst{5} \and
   Christian Colombo\inst{6} \and
   Yli\'es Falcone\inst{7} \and
   Adrian Francalanza\inst{6} \and \\ 
   Sr\dj{}an Krsti\'{c}\inst{8} \and
   Dejan Nickovic\inst{9} \and 
   Gordon J. Pace\inst{6} \and 
   Jose Rufino\inst{10} \and 
   Julien Signoles\inst{11} \and \\
   Dmitriy Traytel\inst{8} \and
   Alexander Weiss\inst{12}}
}
\hi{
\institute{
  IMDEA Software Institute, Spain \\
  \and 
  University of Gothenburg, Sweden \\
  \and
  Chalmers University of Technology, Sweden \\
  \and
  TU Wien, Austria \\
  \and
  University of Luxembourg, Luxembourg\\
  \and
  University of Malta, Malta\\
  \and
  Univ. Grenoble Alpes, CNRS, Inria, LIG, France\\
    \and
  ETH Z\"urich, Switzerland \\
  \and
  Austrian Institute of Technology, Austria \\
  \and
  Universidade de Lisboa, Portugal \\
  \and
  CEA LIST, Software Reliability \& Security Lab, France \\
  \and 
  Accemic Technologies GmbH, Germany \\
}
}

\author{
   C\'esar S\'anchez \and
   Gerardo Schneider  \and 
   Wolfgang Ahrendt \and
   Ezio Bartocci \and
   Domenico Bianculli \and
   Christian Colombo \and 
   Yli\'es Falcone \and
   Adrian Francalanza \and \\
   Sr\dj{}an Krsti\'{c} \and 
   Jo\H{a}o M. Louren\c{c}o \and \\
   Dejan Nickovic \and 
   Gordon J. Pace \and \\
   Jose Rufino \and 
   Julien Signoles \and \\
   Dmitriy Traytel \and
   Alexander Weiss}
\institute{
  C. S\'anchez \at IMDEA Software Institute, Spain
  \and 
  G. Schneider \at University of Gothenburg, Sweden 
  \and
  W. Ahrendt \at   Chalmers University of Technology, Sweden  \and
  E. Bartocci \at TU Wien, Austria \and
  D. Bianculli \at  University of Luxembourg, Luxembourg  \and
  C. Colombo \and A. Francalanza \and G. Pace \at   University of Malta, Malta \and
  Y. Falcone \at   Univ. Grenoble Alpes, CNRS, Inria, LIG, France \and
  S. Krsti\'{c} \and D. Traytel \at    ETH Z\"urich, Switzerland \and
  J. Louren\c{c}o \at Universidade Nova de Lisboa, Portugal  \and
  D. Nickovic \at   Austrian Institute of Technology, Austria \and 
  J. Rufino \at  Universidade de Lisboa, Portugal \and 
  J. Signoles \at  CEA LIST, Software Reliability \& Security Lab, France \and 
  A. Weiss \at  Accemic Technologies GmbH, Germany 
}
 
\titlerunning{A Survey of Challengers for RV}

\usepackage{hyperref}

\begin{document}

\maketitle

\begin{abstract}
  Runtime verification is an area of formal methods that studies the
  dynamic analysis of execution traces against formal specifications.
  Typically, the two main activities in runtime verification efforts
  are the process of creating monitors from specifications, and the
  algorithms for the evaluation of traces against the generated
  monitors.
  Other activities involve the instrumentation of the system to
  generate the trace and the communication between the system under
  analysis and the monitor.
  
  Most of the applications in runtime verification have been focused
  on the dynamic analysis of software, even though there are many more
  potential applications to other computational devices and target
  systems.
  In this paper we present a collection of challenges for runtime
  verification extracted from concrete application domains, focusing
  on the difficulties that must be overcome to tackle these specific
  challenges.
  The computational models that characterize these domains require to
  devise new techniques beyond the current state of the art in runtime
  verification.
\end{abstract}

% !TEX root = main.tex

\section{Introduction}
\label{sec:introduction}

{\em Runtime verification} (RV) is a computing analysis paradigm based
on observing executions of a system to check its expected behaviour. 
The typical aspects of an RV application are the generation of a monitor
from a specification and then the use of the monitor to analyse the
dynamics of the system under study.
RV has been used as a practical application of formal verification, and as
a less {\em ad-hoc} approach complementing conventional testing and debugging.
Compared to static formal verification, RV gains applicability by
sacrificing completeness as not all traces are observed and typically
only a prefix of a potentially infinite computation is processed.
See~\cite{havelund05verify,leucker09brief} for surveys on RV, and the
recent book~\cite{DBLP:series/lncs/10457}.

Most of the practical motivations and applications of RV have been
related to the analysis of software.
However, there is a great potential for applicability of RV beyond
software reliability if one generalises to new domains beyond computer
programs (like hardware, devices, cloud computing and even human
centric systems).
Novel applications of RV to these areas can have an enormous impact in
terms of the new class of designs enabled and they potential increase
in reliability in a cost effective manner.
Many system failures through history have exposed the limitations of
existing engineering methodologies and encouraged the study and
development of novel formal methods.
Ideally, one would like to validate a computational system prior to
its execution.
However, current static validation methods, such as model checking,
suffer from practical limitations preventing their wide use in real
large-scale applications.
For instance, those techniques are often bound to the design stage of
a system and suffer from the state-explosion problem (the
unfeasibility to exhaustively explore all system states statically),
or cannot handle many interesting behavioural properties. 
Thus, as of today many verification tasks can only realistically be
undertaken by complementary dynamic analysis methods. 
RV is the discipline of formal dynamic analysis that studies how to
detect and ensure, at execution time, that a system meets a desirable
behavior.

Even though research on runtime verification has flourished in the
last decade\footnote{See the the conference series at
\url{http://runtime-verification.org}.}, a big part of the (European) community in the area has recently been gathered via a EU COST action initiative\footnote{{\em Runtime Verification beyond Monitoring (ARVI)}: ICT COST Action IC1402 (\url{http://www.cost.eu/COST\_Actions/ict/IC1402})} in order to explore, among other things, potential areas of application of RV, including finances, medical devices, legaltech, security and privacy, and embedded, cloud and distributed systems.

In this survey paper we concentrate in the description of different challenging and exciting  application domains for RV, others than programming languages.  
In particular we consider 
monitoring in the following application domains:
\begin{description}
\item[Distributed systems:] where the timing of observations may vary
  widely in a non-synchronised manner. (Section \ref{sec:distributed})
\item[Hybrid and embedded systems:] where continuous and discrete behaviour coexist and  the resources of the monitor are constrained. (Section \ref{sec:hybrid})
\item[Hardware:] where the timing must be precise and the monitor must operate
  non disruptively. (Section \ref{sec:hardware})
\item[Security \& Privacy:] where a suitable combination between static and dynamic analysis is needed (Section \ref{sec:security})  
\item[Transactional Information Systems:] where the behaviour of modern
  information systems is monitored, and the monitors must compromise
  between expressivity and non-intrusiveness (Section~\ref{sec:transactions}).
\item[Contracts \& Policies:] where the connection between the legal world and the technical is paramount (Section \ref{sec:contracts})  
\item[Huge, unreliable or approximated domains:] where we consider
  systems that are not reliable, or aggregation or sampling is
  necessary due to large amounts of data. (Section \ref{sec:huge})
\end{description}

In all these cases, we first provide an overview of the domain,
present sufficient background to present the context and scope 
Then we introduce the subareas of interest addressed in the section,
and identify challenges and opportunities from the RV point of view.
We do not aim for completeness in the identification of the challenges
and admittedly only identify a subset of the potential challenges to
be addressed by the RV research community in the next years.
We identify the challenges listed as some of the most important.

%%% Local Variables:
%%% mode: latex
%%% TeX-master: t
%%% End:

%\input{samplebody-journals}

% !TEX root = ../main.tex

\section{Distributed and Decentralized Runtime Verification}
\label{sec:distributed}

Distributed systems are generally defined as computational artifacts that
run into execution units placed at different physical locations, and
that exchange information to achieve a common goal.
A localized unit of computation in such a setup is generally assigned
its own process of control (possibly composed of multiple threads),
but does not execute in isolation.
Instead, the process interacts and exchanges information with other such remote
units using the communication infrastructure imposed by the
distributed architecture, such as a computer
network~\cite{attiya04distributed,coulouris11distributed,garg02elements}.
%
% Due to their concurrent nature and to the other aspects of
% distribution, it is well-known that

Distributed systems are notoriously difficult to design, implement and
reason about.
Below, we list some of these difficulties.
\begin{itemize}
\item Multiple stakeholders impose their own requirements on the
  system and the components, which results in disparate specifications
  expressed in widely different formats and logics that often concern
  themselves with different layers of abstraction.
\item Implementing distributed systems often involves collaboration
  across multiple development teams and the use of various
  technologies.
\item The components of the distributed system may be more or less
  accessible to analysis, as they often evolve independently, may
  involve legacy systems, binaries, or even use remote proprietary
  services.
\item The sheer size of distributed systems, their numerous possible execution interleaving, and
  unpredictability due to the inherent dynamic nature of the
  underlying architecture makes them hard to test and
  verify using traditional pre-deployment methods.
  Moreover, distributed computation is often characterized by a high degree of dynamicity where all of the components that comprise the
  system are not known at deployment (for example, the dynamic
  discovery of web services) --- this dynamicity further complicates
  system analysis.
\end{itemize}
Runtime Verification (RV) is very promising to address these
difficulties because it offers mechanisms for correctness analysis
\emph{after} a system is deployed, and can thus be used in a
multi-pronged approach towards assessing system correctness.
It is well-known that even after extensive analysis at static time,
latent bugs often reveal themselves once the system is
deployed.
A better detection of such errors at runtime using dynamic techniques,
particularly if the monitor can provide the runtime data that leads to
the error, can aid system engineers to take remedial action when
necessary.
Dynamic analysis can also provide invaluable information for diagnosing
and correcting the source of the error.
Finally, runtime monitors can use runtime and diagnosis information to
trigger reaction mechanisms correcting or mitigating the errors.

We discuss here challenges from the domain of \emph{Distributed and
  Decentralized Runtime Verification} (DDRV), a broad area of research
that studies runtime verification in connection with distributed or
decentralized systems, or when the runtime verification process is
decentralized.
That it, this body of work includes the monitoring of distributed
systems as well as the use of distributed systems for monitoring.

Solutions to these research efforts exist (see for instance ~\cite{basin15failure,bundala_complexity_2014,polyLarva-CFMP12,falcone15runtime,DBLP:conf/dlt/FengLQ15,kuhtz_ltl_2009,kuhtz_efficient_2012}).
We refer to~\cite{FrancalanzaPS18} for a recent survey on this topic.

\subsection{Context and Areas of Interest}

\subsubsection{Characteristics}

There are a number of characteristics that set DDRV apart from
non-distributed RV.
These characteristics also justify the claim that traditional RV
solutions and approaches commonly do not necessarily (or readily)
apply to DDRV.
This, in turn, motivates the need for new mechanisms, theories, and
techniques.
Some characteristics were identified
in~\cite{DBLP:conf/isola/BonakdarpourFRT16,francalanza13distributed,DBLP:journals/tocs/JoyceLSU87}, and recently revisited in~\cite{FrancalanzaPS18}.

\begin{description}
\item[Heterogeneity and Dynamicity.] 
  One of the reasons that makes distributed systems hard to design,
  implement and understand is that there are typically many
  participants involved.
  Each participant imposes its own requirements ending in a variety of
  specifications expressed in different formats.
  In turn, the implementation often involves the collaboration of
  multiple development teams using a variety of technologies.
  Additionally, the size and dynamic characteristics of the execution
  platform of distributed systems allow many possible execution
  interleavings, which leads to an inherent unpredictability of the
  executions.
  Consequently, testing and verification with traditional
  pre-deployment methods are typically ineffective.
\item[Distributed Clocks and Latency.]
  Distributed systems can be classified according to the nature of the
  clocks: from (1) synchronous systems, where the computation proceeds
  in rounds, (2) timed asynchronous systems, where messages can take
  arbitrary long but there is a synchronized global clock, (3)
  asynchronous distributed systems.
  In an asynchronous distributed system, nodes are loosely coupled,
  each having its own computational clock, due to the impracticality
  of keeping individual clock synchronized with one another.
  As a result of this asynchrony, the order of computational events
  occurring at distinct execution units may not be easy (or even
  possible) to discern.
\item[Partial Failure.] A requirement of any long-running distributed system is that, when execution units (occasionally) fail, the overall
  computation is able to withstand the failure.
  However, the independence of failure between the different
  components of a distributed system and the unavailability of
  accurately detecting remote failures, makes designing fail tolerant
  systems challenging.
\item[Non-Determinism.]
  Asynchrony implies fluctuations in latency, which creates
  unpredictability in the global execution of a distributed system.
  In addition, resource availability (\emph{e.g.,} free memory) at
  individual execution units is hard to anticipate and guarantee.
  These sources of unpredictable asynchrony often induce
  non-deterministic behavior in distributed computations.
\item[Multiple Administrative Domains and Multiple Accessibility.]
    In a distributed system, computation often crosses administrative
    boundaries that restrict unfettered computation due to security
    and trust issues (\emph{e.g.,} mistrusted code spawned or
    downloaded from a different administrative domain may be executed
    in a sandbox).
    Administrative boundaries also limit the migration and sharing of
    data across these boundaries for reasons of confidentiality.
    Also, different components may feature different accessibility
    when it comes to analysis, maintenance, monitorability,
    instrumentation, and enforcement.
    The connected technologies may range from proprietary to the public
domain, from available source code to binaries only, from well-documented developments to sparsely documented (legacy) systems.
  \item[Mixed Criticality.]
    The components and features of a distributed system may
    not be equally critical for the overall goal of the system.
    The consequences of malfunctioning of certain components are more
    severe than the malfunctioning of others.
    For instance, the failure of one client is less critical than the failure of a server which many clients connect
    to.
    Also, some components could be critical for preventing data or
    financial loss, or alike, whereas others may only affect
    performance or customer satisfaction.
  \item[Evolving Requirements.]
    The execution of a distributed system is typically characterized
    by a series of long-running reactive computational entities
    (\emph{e.g.,} a web server that should ideally never stop handling
    client requests).
    Such components are often recomposed into different configurations
    (for example, service-oriented architectures) where their intended
    users change.
    In such settings, it is reasonable to expect the correctness
    specifications and demands to change over the execution of the
    system, and to be composed of smaller specifications obtained from
    different users and views.
\end{description}

%We now describe some applications of DDRV studied in the literature.

%\input{1.distributed/dist-applic-areas}
\subsubsection{Applications}

We briefly mention some of the existing or envisioned application
areas of DDRV, namely
concurrent software, new programming paradigms such as reversible computing~\cite{FraMezTuo18:DAIS}, the verification of distributed algorithms or
distributed data bases, privacy and security (intrusion detection systems,
auditing of policies on system logs~\cite{basin_monitoring_2012,GargJD11},
decentralized access control~\cite{TsankovMDB14}),
blockchain technology~\cite{Nakamoto2008}, monitoring software-defined
networks with software defined monitoring, robotics (e.g., distributed
swarms of autonomous robots), and home automation.

\subsubsection*{Enforcing Interleavings}
%
% Testing
%
Sometimes the system that one analyzes dynamically ---using runtime
verification--- is distributed in nature.
For example, multithreaded programs can suffer from concurrency
errors, particularly when executing in modern hardware platforms, as
multicore and multiprocessor architectures are very close to
distributed systems.
This makes the testing of concurrent programs notoriously difficult
because it is very hard to explore the interleavings that lead to
errors.
The work in~\cite{luo13enforceMOP} proposes to use enforcement
exploiting user-specified properties to generate local monitors that
can influence the executions. The goal is to improve testing by
forcing promising schedules that can lead to violations, even though
violations of the specified property can also be prevented by blocking
individual threads whose execution may lead to a violation.  The
process for generating monitors described in~\cite{luo13enforceMOP}
involves the decomposition of the property into local decentralized
monitors for each of the threads.

\subsubsection*{Observing Distributed Computations}
Checking general predicates in a distributed system is hard, since one
has to consider all possible interleavings.
Techniques like \emph{computation
  slices}~\cite{alagar01techniques,AttardF17:SEFM,chauhan13distributed,mittal07solving}
have been invented as a datatype for the efficient distributed
detection of predicates.
Slices are a concise approximation of the computation, which are
precise enough to detect the predicate because slices guarantee that
if a predicate is present in a slice of a computation then the
predicate occurred in some state of the computation.

Predicate detection can involve a long runtime and large memory
overhead~\cite{chauhan13distributed} except for properties with
specific structure (that is, for some fragments of the language of
predicates).
Current efficient solutions only deal with sub-classes of safety properties
like linear, relational, regular and co-regular, and stable properties.
Even though most techniques for predicate
detection~(\cite{alagar01techniques,cooper91consistent,mittal07solving})
send all local events to a central process for inspection of its
interleavings, some approaches (like~\cite{chauhan13distributed}) consider
purely distributed detection.

\subsubsection*{Monitor Decomposition and Coordination}

Most approaches to monitoring distributed systems consider that the
system is a black-box that emits events of interest, while others use
manual instrumentation and monitor placement.
Some exceptions, for example~\cite{falcone15runtime,francalanza13distributed,FraSey15}, investigate how
to exploit the hierarchical description of the system to generate
monitors that are then composed back with the original system.
The modified system shares the original decomposition (of course
implementing its functionality) and includes the monitors embedded,
but this approach requires to have access to the system description
and is specific to a given development language.
Although the work in~\cite{falcone15runtime} does not specifically
target distributed systems, the compiler can generate a distributed
system in which case the monitor will be distributed as well.
A similar approach is presented
in~\cite{AttardF16,detecterRV15,cassar16implementing,FraSey15}, where a framework for
monitoring asynchronous component-based systems is presented based on
actors.

\subsubsection*{Monitoring Efficiency}

Most RV works assume a single monitor that receives all events and
calculates the verdicts.
Even though a single monitor can be implemented for decentralized and
distributed systems by sending all information to a central monitor,
distribution itself can be exploited to coordinate the monitoring task
more efficiently.
Many research efforts study how to gain more efficient solutions by exploiting
the locality in the observations to also perform partially the
monitoring task locally as much as possible.
For example, the approaches in~\cite{AttardF16,detecterRV15,cassar16implementing,falcone15runtime}
 exploit the hierarchical structure of the system to generate local
 monitors, and \cite{CasFS:15,Fra:Sey:13,FraSey15} exploit the
 structure and semantics of the specification.
Lowering overheads is also pursued in~\cite{polyLarva-CFMP12} by
offloading part of the monitoring computation to the computing
resources of another machine.

When atomic observations of the monitored system occur locally,
 monitors can be organized hierarchically according to the structure
 of the original
 specification~\cite{bauer12decentralised,bauer16decentralised,francalanza13distributed}.
Substantial savings in communication overheads are obtained because
often a verdict is already reached in a sub-formula.
All these results are limited to LTL, and later extended to all
regular languages in~\cite{falcone14efficient}.
Decentralized monitoring assumes that the computation proceeds in
rounds, so distributed observations are synchronized and messages
eventually arrive.
The assumption of bounded message delivery is relaxed
in~\cite{colombo16organising}.

\subsubsection*{Fault Tolerance}
One of the main and most difficult characteristics of distributed
systems is that failures can happen independently~(see
\cite{FH08:InfComp}).
Most of the RV efforts that consider distributed systems assume that
there are no errors, that is, nodes do not crash and messages are not
corrupted, lost, duplicated or reordered.
Even worse, failure dependencies between components can be intricate
and the resulting patterns of behaviors can be difficult to predict
and explain.
At the same time, one of the common techniques for fault tolerance
is the replication of components so this is a promising approach
for monitoring too~\cite{FrancalanzaH07:JLP}.
For example,~\cite{fraigniaud14onthenumber} studies the problem of distributed
monitoring with crash failures, where events can be observed from more
than one monitor, and where the distributed monitoring algorithm tries
to reach a verdict among the surviving monitors.

Another source of failure is network errors, studied
in~\cite{Basin-inconsistencies-2013,basin15failure}, which targets the
incomplete knowledge caused by network failures and message
corruptions and attempts to handle the resulting disagreements.
Node crashes are handled because message losses can simulate node
crashes by ignoring all messages from the crashed node.

\subsection{Challenges}
\label{sec:distributed-challenges}
The characteristics outlined above bring additional challenges to obtain  effective DDRV setups.

\begin{challenge}{Distributed Specifications}
  It is a well-established fact that certain specifications cannot be
  adequately verified at runtime
 ~\cite{AcetoAFI18,Chang:92:ALP,FalconeFM12:STTT,FraAI17:FMSD,pnueli06psl}.
  The partial ordering on certain distributed events, due to
  \emph{distributed clocks} hinders the monitoring of temporal
  specifications requiring a specific relative ordering of these
  events~\cite{AttardF17:SEFM}.
  As such, the lack of a global system view means that even
  fewer specifications can be monitored at runtime.
  Even though some work exists proposing specific languages tailored
  to distributed systems~\cite{sen04efficient}, the quest for expressive and tractable
  languages is an important and challenging goal.
\end{challenge}

\begin{challenge}{Monitor Decomposition, Placement, and Control}
  The runtime analysis carried out by monitors needs to be distributed
  and managed across multiple execution nodes.
  As argued originally in~\cite{francalanza13distributed}, and later
  investigated empirically in works such as~\cite{AttardF17:SEFM,bauer16decentralised}, the decomposition and placement of monitoring analysis is an important engineering
  decision that affects substantially the overheads incurred such as
  the number and size of messages, the communication delay, the spread
  of computation across monitors~\cite{El-HokayemF17a}.
  Such placement also affects the administrative domains under which
  event data is analyzed and may compromise confidentiality
  restrictions and lead to security violations that may be due to the
  communication needed by monitors to reach a verdict (for instance if
  monitors communicate partial observations or partial evaluations of
  the monitored properties).
\end{challenge}

\begin{challenge}{Restricted Observability}
  The flip side of security and confidentiality constraints in
  distributed systems translates into additional observability
  constraints that further limit what specifications can be monitored
  in practice.
  Distributed monitors may need to contend with traces whose event
  data may be obfuscated or removed in order to preserve
  confidentiality which, in turn, affects the nature of the verdicts
  that may be given~\cite{AcetoAFI17:FSTTCS,GrigoreK:18:Concur}.
\end{challenge}

\begin{challenge}{Fault Tolerance}
  DDRV has to contend with the eventuality of failure in a distributed
  system~\cite{basin15failure}.
  Using techniques involving replication and dynamic reconfiguration
  of monitors, DDRV can be made tolerant to \emph{partial failure}.
  More interestingly, fault-tolerant monitoring algorithms could
  provide reliability to the monitors.
  A theory allowing to determine which specifications combined with
  which monitoring algorithms could determine the guarantees that should be
  investigated.
\end{challenge}

\begin{challenge}{Deterministic Analysis}
  Since monitoring needs to be carried out over a distributed
  architecture, this will inherently induce non-deterministic
  computation.
  In spite of this, the monitoring analysis and the verdicts reported
  need to feature aspects such as strong eventual consistency~\cite{El-HokayemF17a} or observational verdict determinism~\cite{Fra17:Concur}, and conceal any internal non-determinism.
  In practice, this may be hard to attain (\emph{e.g.,} standard
  determinization techniques on monitors incur triple exponential
  blowup~\cite{AcetoAFIK17:CIAA});
  non-deterministic monitor behavior could also compromise the
  correctness of RV setup and the validity of the verdicts reported
 ~\cite{Fra16:Fossacs}.
  % Overlaps too much with Monitor Decomposition and Control:
  % \item[Monitoring Overhead]
  % %
  % The overhead of monitoring is common to all areas of runtime verification. However, in the distributed setting, it can be particularly prohibitive. With centralised monitors, the additional communication overhead can be significant, whereas decentalised monitors can slow down each and every (monitired) component.
\end{challenge}

\begin{challenge}{Limits of Monitorability}
  Distributed systems impose further limitations on the class of
  properties that can be detected
(see~\cite{AcetoAFI18,Bauer:ltl,dangelo05lola,FalconeFM12:STTT,FraAI17:FMSD,pnueli06psl,viswanathan00foundations}
  for notions of monitorability for non-distributed systems and~\cite{AttardF17:SEFM,El-HokayemF17a} for decentralized systems~\cite{corr/abs-1808-02692}).
  Associated with the challenge of exploring new specification
  languages for monitoring distributed systems, there is the need to
  discern the limitations of what can be detected dynamically.
\end{challenge}

% \begin{enumerate}
% \item Multiple traces (one at each location) which leads to partial
%   ordering of trace events.
% \item Decentralization of monitoring introducing issues related to
%   concurrency, monitor correctness, monitor locality.
% \item Restricted or conflicting views of events, stemming from event
%   reordering, trace event dropping, node failure (partial failure),
%   security issues. This affects monitor expressivity and introduces
%   aspects such as fault-tolerance
% \end{enumerate}

%\subsection{Existing Solutions}
%\label{sec:distributed-existing}
%\input{1.distributed/dist-existing}

%!TEX root = ../main.tex

\section{Hybrid Systems}\label{sec:hybrid}
   
{\em Hybrid systems} (HS)~\cite{Henzinger95} are a powerful formal framework to model and to 
reason about systems exhibiting a sequence of piecewise continuous behaviors interleaved 
with discrete jumps.  In particular, {\em hybrid automata} (HA) extend finite state-based 
machines with continuous dynamics (generally represented as ordinary differential equations) 
in each state (also called {\em mode}).  HS are suitable modelling techniques to analyze safety 
requirements of {\em Cyber-Physical Systems} (CPS).  CPS consist of computational and physical 
components that are tightly integrated.  Examples include engineered (i.e., self-driving 
cars), physical and biological systems~\cite{BartocciL16,Bartocci2009} that are monitored 
and/or controlled through sensors and actuators by a computational embedded core. 
The behavior of CPS is characterised by the real-time progressions of physical 
quantities interleaved by the transition of discrete software and hardware states. 
HA are typically employed to model the behavior of CPS and to evaluate at design-time 
the correctness of the system, and its efficiency and robustness with respect to the desired safety 
requirements.   

HA are called \emph{safe} whenever given an initial set of states, the possible trajectories 
originated from these initial conditions are not able to reach a bad set of states.
Proving a safety requirement requires indeed to solve a reachability analysis problem 
that is generally undecidable~\cite{AMP+07,Henzinger95} for hybrid systems.  However, this did not 
stop researchers to develop, in the last two decades, semi-decidable efficient reachability 
analysis techniques for particular classes of hybrid systems~\cite{Althoff2013,ASY07tcs1,chen2013flow,ddm04,DangGM09,FranzleHTRS07,Fre08,helicopter,gkr04,kong2015dreach,RayGDBBG15}.

Despite all this progress, the complexity to perform a precise 
reachability analysis of HS is still limited in practice to small problem instances (e.g., \cite{AMP95rad,AMP+07,ASY01drp,Henzinger95}).  
Furthermore, the models of the physical systems may be inaccurate or partially available.  
The same may happen when a CPS implementation employs third-party software components for which neither the 
source code or the model is available.  

A more practical solution, close to testing, is to monitor and to predict 
CPS  behaviors at simulation-time or at runtime~\cite{BartocciDDFMNS18}.
The monitoring technology include the techniques to specify what we
want to detect and to measure and how to instrument the system.
Monitoring can be applied to:
\begin{itemize}
\item Real systems during their execution, where the behavioral observations are 
constructed from sensor readings.
\item System models during their design, where behaviors correspond to simulation traces.
\end{itemize} 

In the following, we provide an overview of the main specification-based monitoring techniques 
available for CPS and HS.  We also show the main applications of the monitoring techniques 
in system design and finally we discuss the main open challenges in this research field.

\subsection{Context and Areas of Interest}

\subsubsection{Specification Languages}

One of the main specification language that has been used in the research community for 
the formal specification of continuous and hybrid systems is 
{\em Signal Temporal Logic} (STL)~\cite{maler2004monitoring,maler13monitoring}.
STL extends {\em Metric Interval Temporal Logic}
(MITL)~\cite{alur96thebenefits}, a dense-time specification formalism,
with predicates over real-valued variables.
This mild addition to MITL has an important consequence, despite its
simplicity---the alphabet in the logic has an order and admits a
natural notion of a distance metric. 
Given a numerical predicate over a real-valued variable and a variable
valuation, we can henceforth answer the question on how far the
valuation is from satisfying or violating the predicate. 
This rich feedback is in contrast to the classical yes/no answer that
we typically get from reasoning about Boolean formulas.
The quantitative property of numerical predicates can be extended to
the temporal case, giving rise to the quantitative semantics for
STL~\cite{fainekos09robustness,donze2010robust}.

We can use with ease STL to specify real-time constraints and complex
temporal relations between events occurring in continuous
signals.
These events can be trivial threshold crossings, but also more
intricate patterns, identified by specific shapes and durations.
We are typically struggling to provide elegant and precise description
of such patterns in STL.
We can also observe that these same patterns can be naturally
specified with regular expressions, as time-constrained sequences
(concatenations) of simple behavior descriptions.

Timed Regular Expressions (TRE)~\cite{asarin02timed}, a dense-time
extension of regular expressions, seem to fit well our need of talking
about continuous signal patterns.
While admitting natural specification of patterns, regular expressions
are terribly inadequate for specification of properties that need universal
quantification over time.
For instance, it is very difficult to express the classical
requirement ``every request is eventually followed by a grant'' with
conventional regular expressions (without negation and intersection
operators).
It follows that TRE complements STL, rather than replacing it.

CPS consist of software and physical components that are generally
spatially distributed (e.g., smart grids, robotics teams) and
networked at every scale.  In such scenario, temporal logics may not be
 sufficient to capture not only time but also topological and
spatial requirements.  In the past five years, there has been a great
effort to extend STL for expressing spatio-temporal requirements.
Examples include \emph{Spatial-Temporal Logic}
(SpaTeL)~\cite{Bartocci2016TNCS,spatel}, the \emph{Signal
  Spatio-Temporal Logic} (SSTL)~\cite{sstl2,sstl} and the
\emph{Spatio-Temporal Reach and Escape Logic}
(STREL)~\cite{Bartocci17memocode}.

\subsubsection{Monitoring Continuous and Hybrid Systems}

We first discuss some issues that are specific to the analysis of
continuous and hybrid behaviors.
We also provide an overview of different methods for monitoring STL
with qualitative and quantitative semantics and matching TRE patterns.

\paragraph{Handling Numerical Predicates}

In order to implement monitoring and measuring procedures for STL and
TRE, we need to address the problem of the computer representation of
continuous and hybrid behaviors.
Both STL and TRE have a dense-time interpretation of continuous
behaviors which are assumed to be ideal mathematical objects.
This is in contrast with the actual behaviors obtained from simulators
or measurement devices and which are represented as a finite
collection of value-timestamp pairs  $(w(t), t)$, where $w(t)$ is the
observed behavior. 
The values of $w$ at two consecutive sample points $t$ and $t'$ do not
precisely determine the values of $w$ inside the interval $(t,t')$. 
To handle this issue pragmatically, interpolation can be used to
``fill in'' the missing values between consecutive samples.
Some commonly used interpolations to interpreted sampled data are step
and linear interpolation.
Monitoring procedures are sensitive to the interpolation used.

\paragraph{Monitoring STL with Qualitative and Quantitative Semantics}

An offline monitoring procedure for STL properties with qualitative
semantics is proposed in~\cite{maler13monitoring}.  
The procedure is recursive on the structure (parse-tree) of the
formula, propagating the truth values upwards from input behaviors via
super-formulas up to the main formula.
In the same paper, the procedure is extended to an
incremental version that computes the truth value of the sub-formulas
along the observation of new sampling points.  

There are several algorithms available in the literature for computing
robustness degree of STL
formulas~\cite{dokhanchi14online,donze13efficient,donze2010robust,fainekos09robustness,wed,abs-1802-03775,filtering}.
The algorithm for computing the space robustness of a continuous
behavior with respect to a STL specification was originally proposed
in~\cite{fainekos09robustness}. In~\cite{donze13efficient}, the
authors develop a more efficient algorithm for measuring space
robustness by using an optimal streaming algorithm to compute the min
and the max of a numeric sequence over a sliding window and by
rewriting the \emph{timed until operator} as a conjunction of simpler
\emph{timed and untimed operators}.  
The procedure that combines monitoring of both space and time
robustness is presented in~\cite{donze2010robust}.  

Finally, the following two approaches have been proposed to monitor
the space robustness of a signal with respect to an STL specification.
The first approach proposed in~\cite{dokhanchi14online} considers STL
formulas with bounded future and unbounded past operators.  The
unbounded past operators are efficiently evaluated exploiting the fact
that the unbounded history can be stored as a \emph{summary} in a
variable that is updated each time a new value of the signal becomes
available.  For the bounded future operators, the algorithm computes
the number of look-ahead steps necessary to evaluate these operators
and then uses a model to predict the future behavior of the system and
to estimate its robustness.  
The second approach~\cite{online_journal} computes instead an interval
of robustness for STL formulas with bounded future operators.

\paragraph{Matching TRE Patterns}

An offline procedure for computing the set of all matches of a timed
regular expression in a continuous or hybrid behavior was proposed
in~\cite{ulus14timed}. The procedure relies on the observation that any match
set can always be represented as a finite union of two-dimensional
zones, a special class of convex polytopes definable as the
intersection of inequalities of the form
$(x < a)$, $(x > a)$ and $(x - y <a)$. 
This algorithm has been recently extended to enable online matching of
TRE patterns~\cite{ulus16online}.

\subsubsection{Tools}
\label{sec:hybrid-tools}

The following tools are publicly available and they support
both the qualitative and the quantitative semantics for monitoring CPSs.

\begin{enumerate}
    \item AMT 2.0~\cite{Nickovic2018}: available at 
    \url{http://www-verimag.imag.fr/DIST-TOOLS/TEMPO/AMT/content.html}
	\item Breach~\cite{breach}: available at 
	\url{https://github.com/decyphir/breach}
	\item S-Taliro~\cite{annpureddy2011s}: available at 
	\url{https://sites.google.com/a/asu.edu/s-taliro/}
	\item U-Check~\cite{Bortolussi2015}: available at 
	\url{https://github.com/dmilios/U-check}
\end{enumerate}

The AMT 2.0 tool~\cite{Nickovic2018} provides a framework for the
qualitative and quantitative analysis of xSTL, which is an extended
Signal Temporal Logic that integrates TRE with STL requirements over
analog system output signals.
The software tool AMT is a standalone executable with a graphical
interface enabling the user to specify xSTL properties, the signals
and whether the analysis is going to be offline or incremental.  The
new version of the tool provides also the possibility to compute
quantitative measurements over segments of the signals that match the
properties specified using TRE~\cite{measures}. AMT 2.0 offers also a
\emph{trace diagnostics}~\cite{FerrereMN15} mechanism that can be used
to explain property violations.

Breach~\cite{breach} and S-Taliro~\cite{annpureddy2011s} are add-on
Matlab toolboxes developed for black-box testing based
verification~\cite{FainekosGP06formats} of Simulink/Stateflow
models. 
These tools have also been used for other applications including
parameter mining~\cite{Yang2012,mining},
falsification~\cite{AbbasFSIG13tecs} to synthesis~\cite{stlce}.

Finally, U-Check~\cite{Bortolussi2015} is a stand-alone program
written in Java, which deals with statistical model checking of STL
formulas and parameter synthesis for stochastic models
described as Continuous-Time Markov Chains.

\subsubsection{Applications}
\label{sec:hybrid-applications}

Specification-based monitoring of cyber-physical systems (CPS)~\cite{harmonia} has been 
a particularly fertile field for research on runtime verification leading to several
theoretical and  practical applications such as quantitative semantics, 
simulation-guided falsification, real-time online monitoring, system 
design and control.  Here is an overview of the most relevant 
applications in the CPS scenario:

\begin{description}

\item[Real-time Monitoring of CPS.] 
  The complexity of the new generation of digital system-on-chip (SoC)
  and analog/mixed-signal systems (AMS) requires new efficient
  techniques to verify and to validate their behavior both at physical
  and software level.
  The simulation of such systems is now too time-consuming to be
  economically feasible.
  An alternative approach is to monitor the system under test (SUT)
  online by processing the signals and software traces that are
  observable after instrumentation~\cite{harmonia}.  This approach
  leverages the use of dedicated hardware accelerators such as {\em Field
  Programmable Gate Arrays} (FPGA) and of proper synthesis
  tools~\cite{fpga,wed,SelyuninJNRHBNG17,SelyuninNBNG16} that can
  translate temporal logic specifications into hardware monitors.
  This will be discussed in more detail in the next section dedicated
  to hardware supported runtime verification.

\item[Falsification Analysis and Parameter Synthesis.] 
The continuous dynamics of simulated CPS design should be tolerant to
 model approximation errors, noisy inputs, model
uncertain parameters and the initial conditions.
The Boolean semantics of STL, that decides whether a signal is correct 
or not with respect to a given requirement is not often informative enough 
to reason about CPS behavior, because its result may be very sensitive to 
small changes of the parameter values or to small perturbations in 
the received inputs. 
In the last decade, many authors have tried to address this issue by introducing 
different notions of quantitative semantics for temporal logic~\cite{pant2017smooth,filtering,rizk,donze2010robust,akazaki2015time}, where the binary 
satisfaction relation is replaced with a quantitative robustness degree function. 
The positive and negative sign of the robustness value indicates whether the formula is satisfied or violated, respectively.  This quantitative interpretation can be 
exploited in combination with several heuristics (e.g., ant colony, gradient ascent, statistical emulation) to
optimise the CPS design in order to satisfy or  falsify a given formal requirement~\cite{AbbasFSIG13tecs,AbbasWFJ14acc,annpureddy2011s,AnnapureddyF10iecon,DonzeKR09hscc,FainekosG03,Nghiem+Others/2010/Monte,YaghoubiF2017acc}.
\emph{Falsification analysis}~\cite{AbbasFSIG13tecs,AbbasWFJ14acc,annpureddy2011s,AnnapureddyF10iecon,FainekosG03,Nghiem+Others/2010/Monte,YaghoubiF2017acc}
refers to the activity of finding counterexamples in the space of CPS inputs
 by identifying those inputs that  minimise 
the robustness with respect to a given requirement.
Maximising the robustness can be used to tune the parameters (e.g., \emph{parameter synthesis})
that makes the system to satisfy a given formal requirement to a greater extent~\cite{tssl,stl-ps,eziotcs,BartocciL16,plos,DonzeKR09hscc,spatel,insulin}.

\item[From Monitoring to Control Synthesis.] 
  The use of formal logic-based languages has also enabled control
  engineers to build tools that automatically synthesise controllers
  starting from a given specification~\cite{belta2017formal}.
  Temporal logics such as Metric Temporal Logic
  (MTL)~\cite{Koymans1990}, and Signal Temporal Logic
  (STL)~\cite{maler2004monitoring} have been employed to specify
  time-dependent tasks and constraints in many control system
  applications~\cite{annpureddy2011s,stlce,WongpiromsarnTM12}.  
  In the context of Model Predictive Control
  (MPC)~\cite{BemporadM99,kim2017dynamic,pant2017smooth,raman2014model},
  the  monitoring of temporal logics constraints over the simulated traces
  of a plant model can be used to find iteratively the input that will
  optimize the robustness for the specification over a finite-horizon.

\end{description}

\subsection{Challenges}
\label{sec:hybrid-challenges}

Although specification-based monitoring of CPS is a well-established
research area, there are still many open challenges that need to be
addressed.  We now discuss some of the most important remaining
challenges.

\begin{challenge}{Autonomous CPS}
  \label{ch:autonomous-CPS}
  There is an increasing trend to equip CPS with Machine Learning and
  Artificial Intelligent (AI) components.  This creates a tremendous challenge
  to ensure the 
  safety and security properties of the system both at the design time
  and at its deployment time.  One promising research direction is to
  investigate how to use specification-based falsification analysis 
  techniques to test machine learning components such as neural 
  networks~\cite{DreossiDS17}. 
  
\end{challenge}

\begin{challenge}{From design-time to runtime}
  \label{ch:from-design}
  Specification languages for CPS typically assume a perfect
  mathematical world where the time is continuous and the state
  variables are all observable with infinite precision.  In contrast,
  runtime verification of such ideal specifications is not precise at
  runtime. One has to take into account that the CPS can be only
  observed at sampled points in time, that some state variables may
  not be observable and that the sensors may introduce noise and
  inaccuracies into measurements, including sampling noise.
  As a consequence, there is an urgent need to address these questions
  in the context of runtime verification of CPS.
\end{challenge}

\begin{challenge}{Limited resources}
CPS runtime monitors often need to be implemented on embedded devices 
with limited resources. It is hence essential to take into consideration the restrictions 
regarding the monitor's bandwidth as well as memory and time resources.
\end{challenge}

\begin{challenge}{From real-time to spatial and spectral specifications}
  Most of the existing work on runtime monitoring of CPS is focused on
  real-time temporal properties. However, CPS often consist of
  networked spatially distributed entities where timing constraints
  are combined with spatial relations between the components. In
  addition, many basic properties of continuous CPS entities are
  naturally definable in spectral (for instance frequency)
  domain~\cite{tffainekos,ftl}.  There is a necessity to study
  specification formalisms that gracefully integrate these important
  CPS aspects.
\end{challenge}

\begin{challenge}{Fault-localisation and explanation}  Detecting a fault while 
monitoring a CPS during its design or deployment time involves understanding 
and correcting the error. Complementing runtime verification methods 
with (semi) automated fault localisation~\cite{fault-stl} and explanation could significantly 
reduce the debugging efforts and help the engineer in building a safe and 
secure system.
\end{challenge}

%Important aspects and concepts: sampling, assumptions on the signals,
%noise, robustness
%
%Proposed logics: MTL (metric temporal logic), STL (signal temporal
%logic), TRE (timed regular expressions)
%
%Relevant

%!TEX root = ../main.tex

\section{Hardware}\label{sec:hardware}

{\em Hardware supported runtime verification} (HRV) has an immense potential
for runtime observation and can even allow the continuous assessment
of the behavior exhibited by the system.
The main idea is to move extensive analysis required for complex
hardware/software design from offline and limited data sets to an
online simultaneous non-intrusive analysis.
Observation and simultaneous correctness checking of system internals can reach a level of detail that is orders of magnitude better than today's tools and systems provide.

Online runtime verification hardware-based approaches may take
advantage of multiple technologies, for example, hardware description
languages and reconfigurable hardware.
Together the use of these technologies provides the means for
non-intrusiveness, observability, feasibility, expressiveness,
flexibility, adaptability and responsiveness of hardware-based
monitors that observe and monitor a target system and allow to react
to erroneous behavior. 
In addition, HRV can be used for other analysis, such as performance
monitoring.

Several solutions have been proposed that approach runtime verification (RV) differently, diverging on the methodologies used, goals and target system-induced limitations. 
Whether the monitor executes on external hardware or on-system, what
the monitor watches (that is, the meaningful events it cares about:
the events of interest), how it is connected to the system and what is
instrumented or not, are dependent on both the characteristics of the
system being monitored and the goals of the monitoring process.

\subsection{Context and Areas of Interest}

\subsubsection{Non-intrusiveness}

Ideally, observing and monitoring components should not interfere with
the normal behavior of the system being observed, thus negating what
is called ``the observer effect'' or ``the probe
effect''~\cite{Gait:86a}, in which the observing methodology hinders
the system behavior by affecting some of its functional or
non-functional (e.g., timeliness) properties.
Hardware-based approaches are inherently non-intrusive, while software-based solutions normally exhibit some degree of intrusiveness, even if minimal. 
Therefore, it is widely acknowledged that these approaches must be
used with care.

For example, the delays implicitly associated with the insertion of software-based probes may ill affect the timing and synchronisation characteristics of concurrent programs. 
Moreover, and perhaps less intuitively, the removal of such probes
from real-time embedded software which, in principle, leads to shorter
program/task execution times and may render a given task set
unschedulable due to changes in the corresponding cache-miss
profile~\cite{Lundqvist:99a,Nam:17a,Wilhelm:08a}.
Non-intrusiveness, i.e. the absence of interference may then be referred to as a RV constraint. RV constraints are not only relevant, but in fact fundamental, for highly critical systems~\cite{Pike:11a}. 

A comprehensive overview of various hardware (including on-chip), software and hybrid (i.e., a combination of hardware and software) methodologies for system observation, monitoring and verification of software execution in runtime is provided in~\cite{Watterson:07a}.

System observing solutions can be designed to be directly connected to some form of system bus, enabling information gathering regarding events of interest, such as data transfers and signalling taking place inside the computing platform, namely instruction fetch, memory read/write cycles and interrupt requests, with no required changes on the target system's architecture. 
Examples of such kind of hardware-based observation approaches are
proposed
in~\cite{Kane:15b,Pellizzoni:08a,RCPinto2014rv,Reinbacher:14b}.
 
% \begin{figure}[!bht]
% \centering
% \includegraphics[width=0.95\linewidth]{./3.hardware/Fig-Hardware-Diagram.pdf}
% \vspace*{-2mm}
% \caption{Non-intrusive observation and runtime verification.}
% \label{Fig:hardware-diagram}
% \end{figure}
 
As emphasized in~\cite{Watterson:07a} that observing mechanisms should: (1) be minimally intrusive, or preferably completely non-intrusive, so as to respect the RV constraint; (2) provide enough information about the target system so that the objectives of runtime verification can be met. 

\subsubsection{Observability}

Another important aspect raised in~\cite{Watterson:07a} is the sometimes limited observability of program execution with respect to its internal state and data information. In general, software-based monitoring may have access to extensive information about the operation of a complex system, in contrast to the limited information available to hardware probes~\cite{Watterson:07a}.
 
Thus, one first challenge is that hardware-based probes must be capable of observing enough information about the internal operation of the system to fulfil the purpose of the monitoring~\cite{Watterson:07a}. Gaining access to certain states or information is often problematic, since most systems do not provide access to system operation and software execution details. So, observability is sometimes limited to the data made available or accessible to observing components. Low observability of target system operation affects not only traditional hardware monitors, but also may jeopardize hybrid monitoring and may deem these observing and monitoring techniques ineffective.
 
\subsubsection{Feasibility} 
 
General purpose {\em Commercial Off-The-Shelf} (COTS) platforms offer
limited observing and monitoring capabilities.
For example, in those platforms based on Intel x86 architectures
observability is restricted to the Intel Control Flow
Integrity~\cite{Intel:16-CFI} and to the Intel Processor
Trace~\cite{Intel:16-PT} facilities.
Trying to enhance system observability through physical probing
implies either a considerable engineering effort~\cite{Kenny:07a} or
is restricted to specific behaviors, such as input/output
operations~\cite{Pellizzoni:08a}.

The trend to integrate the processing entities together with other
functional modules of a computing platform in an {\em Application Specific
Integrated Circuit} (ASIC), often known as {\em System on a Chip} (SoC), can
dramatically affect the overall system observability, depending on
whether or not special-purpose observers are also integrated.

The shortcomings and limitations of debug and trace resources
regarding runtime system observation is analysed
in~\cite{Lee:15a}, concluding that the deep integration of software
and hardware components within SoC-based devices hinders the use of
conventional analysis methods to observe and monitor the internal
state of those components. The situation is further exacerbated
whenever physical access to the trace interfaces is unavailable,
infeasible or cost prohibitive.

With the increased popularity of SoC-based platforms, one of the first
on-chip approaches to SoC observability was introduced in~\cite{Shobaki:01a}, where the
authors presented MAMon, a hardware-based probe-unit integrated within
the SoC and connected via a parallel-port link to a host-based
monitoring tool environment that performs both logic-level (e.g.,
interrupt request assertion detection) and system-level (e.g., system
call invocation) monitoring. This approach can either be passive (by
listening to logic- or system-level events) or activated by (minimally
intrusive) code instrumentation.

Many SoC designs integrate modules made from Intellectual Property
(IP) cores. An IP core design is pre-verified against its functional
specification, for example through assertion-based verification
methods. In hardware-based
designs, assertions are typically written in verification languages
such as the {\em Property Specification Language}
(PSL)~\cite{PSL-Standard:12a} and {\em System Verilog Assertions}
(SVA)~\cite{Verilog-Standard:12a}.
The pre-verification of IP core designs contributes to reduce the
effort placed in the debug and test of the system integration cycle.

The work~\cite{Todman:15a} presents an in-circuit RV solution that
targets the monitoring of the hardware itself rather than
software.
 Runtime verification is done by means of in-circuit
temporal logic-based monitors. Design specifications are separated
into compile-time and runtime properties, where runtime properties
cannot be verified at compile-time, since they depend on runtime
data. Compile-time properties are checked by symbolic
simulation. Runtime properties are verified by hardware monitors being
able to run at the same speed as the circuits they monitor.

System-wide observation of IP core functionality requires the
specification of a set of events to be observed and a set of
observation probes. The IP core designer will be the
best source of knowledge for determining which event probes can
provide the highest level of observability for each core. Such kind of
approach is followed in~\cite{Lee:11a}, for the specification of a
low-level hardware observability interface: a separate dedicated hardware observability bus is
used for accessing the hardware observation interface. 

The approach described in~\cite{Lee:11a} was further extended
in~\cite{Lee:15a} to include system level observations, achieved
through the use of processor trace interfaces. The solution discussed in~\cite{Lee:15a} introduces a
System-level Observation Framework (SOF) that monitors hardware and
software events by inserting additional logic within hardware cores
and by listening to processor trace ports. The proposed SOF provides
visibility for monitoring complex execution behavior of software
applications without affecting the system execution. Engineering and
evaluation of such approaches has resorted to FPGA-based
prototyping~\cite{Lee:11a,Lee:15a}.

Support for such kind of observation can be found also in modern processor architectures with multiple cores, implemented as single chip solutions and natively integrating embedded on-chip special-purpose observation resources, such as the ARM CoreSight~\cite{ARM-CoreSight:13a,Orme:08a}.

\subsubsection{Design approaches}
 
Nowadays there are two approaches for embedded multicore processor
observation. Software instrumentation is easy to use, but
very limited for debugging and testing (especially for integration
tests and higher levels). A more sophisticated approach and key
element in multicore observation are embedded trace based emulators. A
special hardware unit observes the processor's internal states,
compresses and outputs this information via a dedicated trace port. An
external trace device records the trace data stream and forwards the
data after the observation period to, e.g. a personal computer for
offline decompression and processing. 
Unfortunately, this approach still suffers from serious limitations in
trace data recording and offline processing:

\begin{itemize}
\item Trace trigger conditions are limited and fixed to the sparse
  functionality implemented in the ``embedded trace'' unit.
\item Because of the high trace data bandwidth it is impracticable on
  today's storage systems to save all the data obtained during an
  arbitrary long observation.
\item There is a discrepancy between trace data output bandwidth and
  trace data processing bandwidth, which is usually several orders of
  magnitude slower. This results in a very short observation period
  and a long trace data processing time, which 
  renders the debugging process inefficient.
\end{itemize}

Hardware supporting online runtime verification could
overcome these limitations. 
Trace data is not stored
before being pre-processed and verified, because both are done online. 
Debugging and runtime verification are accomplished without any
noticeable interference with the original system execution.
Verification is based on a given
specification of the system's correct behavior. In case a
misbehavior is detected, further complex processing steps are
triggered. This challenging solution enables an autonomous, arbitrary
enduring observation and brings out the highest possible observability
from ``embedded trace'' implementations.

Other solutions place the observation hardware inside the processing
units, which may, in some situations, require their modification. 
Some simple modifications may enable lower-level and finer-grained
monitoring, for example by allowing the precise instant of an
instruction execution to be observed. The choice of where to connect a
runtime verification hardware depends on the sort of verification one
aims to perform and at which cost, being a design challenge.
 
A {\em Non-Intrusive Runtime Verification} (NIRV) observer architecture for
real-time SoC-based embedded systems is presented
in~\cite{RCPinto2014rv}. The observer (also called {\em Observer Entity},
OE) synchronously monitors the SoC bus, comparing the values being
exchanged in the bus with a set of configured observation points, the
events of interest. 
Upon detection of an event of interest, the OE time-stamps the event
and sends it an external monitor.
This approach is extended in~\cite{Gouveia:2016a:NIRV} to enforce
system safety and security using a more precise observation of
programs execution, which are secured through the (non-intrusive)
observation of the buses between the processor and the L1 cache
sub-system.
%
% The availability of soft-processors and SoC designs has been opening
% room for novel monitoring and RV approaches, supporting non-intrusive
% hardware-based probing, in the basis of RV techniques for many systems
% and applications.

A wide spectrum of both functional and non-functional properties can
be targeted by these RV approaches, from timeliness to safety and
security, preventing misbehavior overall. 
The effectiveness of system observability is crucial for securing the
overall system monitoring.
Hardware-based observation is advantageous given its
non-intrusiveness, but software-based observation is more flexible,
namely with respect to capturing of context-related data.

\subsubsection{Flexibility: (self-)adaptability and reconfiguration}
  
Requirements for (self-)adaptability to different operational
conditions call for observers (and monitors) flexibility, which may be
characterized by a ready capability to adapt to new, different, or
changing needs.  
Flexibility implies that observing resources should be re-configurable
in terms of the types and nature of event triggers. 
This configurability may be defined via configuration files, supported
online by self-learning modules, or a combination of
both. Reconfigurable hardware implementations usually provide
sufficient flexibility to allow for changes of the monitored
specification without re-synthesising the hardware infrastructure.
This is a fundamental characteristic since logic synthesis is a very
time-consuming task and therefore unfit to be performed online.
Observer and monitor reconfigurability can be obtained in the
following ways:

\begin{itemize}
\item Using reconfiguration registers that can be changed
  online~\cite{RCPinto2014rv}, a flexible characteristic that supports
  simple to moderate adaptability capabilities. 
  Examples include to redefine the address scope for a function stack
  frame, upon its call, or to define function's calling addresses upon
  dynamic linking with shared object libraries.
\item Selecting an active monitor or a monitor specification from a
  predefined set of mutually exclusive
  monitors~\cite{Rufino:16a}. 
  This corresponds to a mode change in the operation of the
  system. 
  Mode changes needs to secure overall system stable
  operations~\cite{Phan:11a}.
\item Using a reconfigurable single monitor~\cite{Rahmatian:12a},
  which allows to update the monitor through the partial
  reconfiguration capabilities enabled by modern FPGAs.
\end{itemize}

The approach in~\cite{Rahmatian:12a} implements intrusion detection in
embedded systems by detecting behavioral differences between the
correct system and the malware. The system is implemented using FPGA
logic to enable the detection process to be regularly updated and
adapt to new malware and changing system behavior. The idea is to
protect against the execution of code that is different from the
correct code the system designer intends to execute. The technique
uses hardware support to enable attack detection in real time, using
finite state machines.

System adaptation triggered by non-intrusive RV techniques is
approached in~\cite{Rufino:16a} for complex systems, such as {\em Time- and
Space-Partitioned} (TSP) systems, where each partition hosts a
(real-time) operating system and the corresponding
applications. Special-purpose hardware resources provide support for:
partition scheduling, which are verified in runtime through (minimally
intrusive) RV software; process deadline violation monitoring, which
is fully non-intrusive while deadlines are fulfilled. Process level
exception handlers, defined the application programmer, establish the
actions to be executed by software components when a process
deadline violation is detected.  The monitoring component which
analyses the observed events (the trace data) may be a component
belonging to RV hardware itself, checking the system behavior as it
observes.

\subsubsection{Use case examples}

Given the numerous possibilities for implementing RV in hardware,
multiple contributions have been made that tackle the ongoing search
for improvement of hardware-based RV monitors.
Some solutions address monitoring and verification in a single
instance~\cite{Reinbacher:14b}. Here, the verification procedure is
mapped into soft-microcontroller units, embedded within the design,
and use formal languages such as past-time Linear Temporal Logic
(ptLTL).
An embedded CPU is responsible for checking ptLTL clauses in a
software-oriented fashion.

A System Health Management technique was introduced
in~\cite{Reinbacher:14a} which empowers real-time assessment of the
system status with respect to temporal-logic-based specifications and
also supports statistical reasoning to estimate its health at
runtime. By seamlessly intercepting sensor values through read-only
observations of the system bus and by on-boarding their platform
(rt-R2U2) aboard an existing FPGA already built into the standard UAS
(Unmanned Aerial Systems) design, system integration problems of
software instrumentation or added hardware were avoided, as well as
intrusiveness. 

A runtime verification architecture for
monitoring safety critical embedded systems which uses an external bus
monitor connected to the target system, is presented in~\cite{Kane:15a}. 
This architecture was designed for
distributed systems with broadcast buses and black-box components, a
common architecture in modern ground vehicles. 
This approach uses a passive external monitor which lines up well
against the constraints imposed by safety-critical embedded
systems. 
Isolating the monitor from the target system helps ensure that system
functionality and performance is not compromised by the inclusion of
the monitor.
 
The use of a hardware-based NIRV approach for mission-level adaptation
in unmanned space and aerial vehicles is addressed
in~\cite{Rufino:16b-OSPERT} with the goal to contribute to
mission/vehicle survivability.
For each phase of a flight, different schedules are defined to three
modes: normal, survival, recovery. The available processor time is
allocated to the different vehicle functions accordingly with its
relevance within each mode: normal implies the execution of the
activities defined for the mission; survival means the processor time
is mostly assigned to fundamental avionic functions; recovery foresees
also the execution of fault detection, isolation and recovery
functions.

Gouveia and Rufino~\cite{Gouveia:2016a:NIRV} attack the problem of
fine-grained memory protection in cyber-physical systems using a
hardware-based observation and monitoring entity are presented.
To ensure the security of the observer itself, the monitor is designed
as a black box, allowing it to be viewed in terms of its input and
output but not its internal functioning and thus preventing malicious
entities from hijacking its behavior.

No previous study concerning hardware-based observability has tackled
the problem of applying the concepts and techniques to the
non-intrusive observation and monitoring of programs in interpreted
languages, such as Python and Java bytecode, running on the
corresponding virtual machines.

\subsection{Challenges}

\begin{challenge}{Observability}
  There is no general results on defining which hardware entities
  (system bus, processor internal buses, IP core internals) of a
  system should be instrumented to guarantee the required
  observability and how to probe such entities. In general,
  observation at different levels of abstraction should be supported,
  from logic-level events (e.g., interrupt, request, assertion) up to
  system (e.g., system call invocation) and application levels (e.g.,
  value assigned to a given variable).
\end{challenge}

\begin{challenge}{Effectiveness}
  To ensure that hardware-based probing is able to provide effective
  system observability, meaning all the events of interest should be
  captured, while maintaining the complexity of hardware
  instrumentation in conformity with SWaP (Size, Weight and Power)
  constraints. This is especially important for observation and
  monitoring of hardware components, where the RV resources should
  have a much lower complexity than the observed infrastructure, but
  this results could also be applicable to the monitoring of software
  components.
\end{challenge}

\begin{challenge}{Feasibility and flexibility}
  To handle the potentially high volumes of trace data produced by 
  extensive system observation
  is  challenge. It includes confining the observed events of interest, and
  the use of advanced compression, pre-processing and runtime
  verification techniques to reduce the gap between trace data output
  and trace data processing capabilities. 
  Also, mapping of formal specification of system properties into actual
  observing and monitoring actions, making use of a minimal set of
  highly effective hardware/software probing components and
  monitors. 
  If applicable, provide support for flexible observation and
  monitoring, thus opening room for the integration of RV techniques
  in (self-)adaptable and reconfigurable systems.
\end{challenge}

\begin{challenge}{Hybrid approaches for observability}
Combining software-based instrumentation with
  hardware-based observability in a highly effective hybrid approach, to: 
  (1) Capture program execution flows and timing,
  without the need for special-purpose software hooks; (2)
  Observe fine-grained data, such as read/write accesses to global
  and local variables; (3) Monitor bulk data (e.g. arrays)
  through the observation of read/write accesses to individual
  members.
\end{challenge}

\begin{challenge}{Advanced system architectures}
  Extending hardware-based observability to advanced system
  architectures, such as processor and memory virtualisation,
  including time- and space-partitioning, and also to the execution of
  interpreted languages including bytecode that runs on virtual
  machines, like JVM.
\end{challenge}

%!TEX root = ../main.tex

\section{Security and  Privacy}\label{sec:security}

%\input{4.security/security_body}
%\input{4.security/privacy}

%\subsection{Introduction}
% Written by Gerardo
In the last years there has been a huge explosion in the availability of large
volumes of data. 
Large integrated datasets can
potentially provide a much deeper understanding of both nature and society and
open up many new avenues of research. 
These datasets are critical for addressing key societal
problems---from offering personalized services, improving public
health and managing natural resources intelligently to designing
better cities and coping with climate change.
More and more applications are deployed in our smart devices and used by our browsers in order to offer better services. 
However, this comes at a price: on one side most services are offered
in exchange of personal data, but on the other side the complexity of
the interactions of such applications and services makes it difficult
to understand and track what these applications have access to, and
what they do with the users' data.
Privacy and security are thus at stake.

Cybersecurity is not just a buzzword, as stated in the recent article ``All IT Jobs Are Cybersecurity Jobs Now''\footnote{\url{https://www.wsj.com/articles/all-it-jobs-are-jobs-now-1495364418}; The Wall Street journal.} where it is said that ``The rise of cyberthreats means that the people once assigned to setting up computers and email servers must now treat security as top priority''. Also, ``The largest ransom-ware infection in history''\footnote{\url{https://blog.comae.io/wannacry-the-largest-ransom-ware-infection-in-history-}  \url{f37da8e30a58}.} infected more than 200,000 computers. 
Referring to the event above, the Europol chief stated in a recent BBC interview that ``Cybersecurity should be a top line executive priority and you need to do something to protect yourself''\footnote{\url{http://www.bbc.com/news/technology-39913630}.}. 

Besides the above examples, which are well-known given their massive impact in the media and society, we know that security and privacy issues are present in our daily lives in different forms, including 
botnets,
%\cite{DBLP:conf/dimva/StinsonM07},
distributed denial-of-service attacks (DDoS), 
%\cite{REF},
hacking, 
%\cite{REF},
malware, 
%\cite{REF},
pharming, 
%\cite{REF},
phishing, 
%\cite{REF},
ransomware, 
%\cite{symantec16ist},
spam, 
%\cite{REF},
and numerous attacks leaking private information \cite{SER12sdl}.
The (global) protection starts with the protection of each single computer or
device connected to the Internet. 
However, nowadays only partial solutions can be done statically.
Runtime monitoring, verification and enforcement are thus
crucial to help in the fight against security and privacy threats.

\paragraph{Remark.} Given the breadth of the Security \& Privacy
domain, we do not present an exhaustive analysis of the different
application areas. 
We deliberately focus our attention on a small subset of the whole
research area, mainly privacy concerns from the EU General Data
Protection Regulation (GDPR), information flow, malware detection,
browser extensions, and privacy and security policies. 
Even within those specific areas, we present a subset of challenges
emerging from this areas.

\subsection{Context and Areas of Interest}

%\paragraph{Side Channels}
%In security, a {\em side-channel attack} is any attack based on information gained from the physical implementation of a system. 
%This includes timing information, power consumption, electromagnetic leaks,

%\subsection{GDPR (General Data Protection Regulation)}
\subsubsection{GDPR (General Data Protection Regulation)}
% Written by Gerardo
The European {\em General Data Protection Regulation}\footnote{\url{http://eur-lex.europa.eu/legal-content/en/ALL/?uri=CELEX:32016R0679}} (GDPR  ---EU--2016/679, adopted on 27 April 2016; entered into application on 25 May 2018)
 subjects companies, governmental organizations and any other data collector to
 stringent obligations when it comes to user privacy in their digital products
 and services. 
 Consequently, new systems need to be designed with privacy in mind
 ({\em privacy-by-design} \cite{cavoukian2009}) and existing
 systems have to provide evidence about their compliance with the new
 GDPR rules.  This is mandatory, and sanctions for data breaches are tough and costly.

 As an example, Article 5 of GDPR, related to the so-called {\em data
   minimization principle}, states: ``Personal data must be adequate,
 relevant, and limited to the minimum necessary in relation to the
 purposes for which they are processed''.  
 While determining what is ``adequate'' and ``relevant'' might seem
 difficult given the inherent imprecision of the terms, identifying
 what is ``minimum necessary in relation to the purpose'' is easier to
 define and reason about formally.

Independently on whether we are considering privacy by design or giving evidence about privacy compliance for already deployed systems, there are some issues to be considered. 
Not all the obligations stated in the regulations can be easily translated into
technical solutions, so there is a need to identify which regulations are
enforceable by technical means. 
For those rules or principles identified as being enforceable by
software, it is hard for engineers to assess and provide evidence of
whether a technical design is compliant with the law due to the gap
existing between a legal document written in natural language and a
technical solution in the form of a software system.

Consider again the data minimization principle.
One way to understand minimization is on how the data is {\em used},
that is we could consider ways to identify the \emph{purpose} for
which the input data collected is used in the program.
In this case we would need to look inside the program and track the
usage of the data by performing static analysis techniques like
tainting, def-use, information flow, etc. 
This, in turn, requires a precise definition of what ``purpose'' means
and a way to check that the intended purpose matches the real actions
that the program take to process the data at runtime.
Another aspect of minimization is related to when and how the data is
{\em collected} in order to limit the collection of data to what is
actually needed to perform the purpose of the program.
In this case we could consider that the purpose is given by the
specification of the program, which is the approach followed by
Antignac et al.~\cite{ASS17dm}.  
This results indicate that it may be possible to enforce data
minimization at runtime, at least in what concerns some of its
aspects. But other privacy principles are more difficult to tackle.

%\paragraph{Information Flow.}
\subsubsection{Information Flow}
% Written by Julien

In computer systems, it is often necessary to prevent some objects to
access specific data.
These permissions are usually defined through security policies, and
enforced using access control mechanisms. 
However, such mechanisms are typically insufficient in practice.
For instance, an application could require to access both private
data---such as the user contact list---and to connect to Internet but,
once the application is granted by the operating system's access
control policy, one would like to ensure that no data from the contact
list (assumed to be confidential) leaks to the Internet (a public
channel). 
Enforcing such fine-grained security policies require information flow
control mechanisms.
These mechanisms allow untrusted applications to access confidential
data as soon as they do not leak these data to public
channels. Denning's seminal work~\cite{denning76acm,denning77acm} in
that field proposed static verification techniques to ensure that a
program does not leak any confidential data. 
This property is usually called \emph{non-interference},
first formalized by Goguen and Meseguer~\cite{goguen82ssp}.
More generally, non-interference states that no private data leaks to
a public channel, either directly or indirectly. An indirect
non-secure flow may appear for instance when two different values of
some public data may be emitted on a public channel depending on some
private conditions. 
In this case, an observer can infer part of the private information
just by observing public data.
From the eighties to the early 2000's, many efforts have been put in
verifying non-interference properties
statically~\cite{sabelfeld03jsac,volpano96jcs}.

In 2004 Vachharajani \emph{et al}~\cite{vachharajani04micro} abandoned static
approaches and proposed \textsf{Rifle}, a runtime information flow security
system. 
After that, dynamic information flow approaches have been proposed for
different settings (\emph{e.g.}  JavaScript~\cite{austin10plas}, or
applied to databases~\cite{yang16pldi}).
The main advantage of dynamic information flow is its ability to deal
with dynamic languages and dynamic security policies. It is also
usually more permissive than static approaches with respect to
non-interference: dynamic approaches may accept secure flows that
would be rejected statically.
However, pure dynamic approaches have a major drawback: they cannot
take into account the branches uncovered by the examined executions
and so they may miss (indirect) insecure flows. In particular, Russo
and Sabelfeld~\cite{russo10cfs} demonstrated that pure dynamic
approaches cannot be sound with respect to flow-sensitive
non-interference, in the form of Hunt and
Sands~\cite{hunt06popl}. 
However flow-sensitivity is a very useful feature in practice, since
it is more permissive than flow-insensitivity by accepting that memory
locations store values of different security level.

In 2006 Le Guernic \emph{et al}~\cite{leguernic06asian} proposed a
hybrid approach that combines soundness of a static approach and
permissiveness of a dynamic approach. In recent years, hybrid
information flow has received a lot of attention, for instance for
languages such as C~\cite{barany17tap}, Haskell~\cite{buiras15icfp},
and JavaScript~\cite{fragososantos15tgc,hedin15csf}.  
To deal with the unsoundness of dynamic approaches, it is also
possible to consider multiple executions~\cite{devriese10sp} or
multiple facets~\cite{austin12popl}, the latter consisting in mapping
a variable to several values (or facets), each of them corresponding
to a particular security level.

Different variants of non-interference and ways of verifying them are
described by Hedin and Sabelfeld's~\cite{hedin12nato} and by Bielova
and Rezk~\cite{bielova16trust}.

%\paragraph{Cyber-physical systems (e.g. Internet of Things)}
%\todo{[MARTIN \& JOS\'E - SEE ALSO THE HARDWARE SECTION]}

%\paragraph{Malware Detection and Analysis.}
\subsubsection{Malware Detection and Analysis}
% Written by Gerardo
{\em Malware} refers to a malicious software specifically designed to disrupt, damage, or gain unauthorized access to a computer system. 
Malware usually exploits specific system vulnerabilities, such as a
programming bug in software (e.g., a browser application plugin) or a
bug in the underlying platform or OS.  Malware infiltration effects
range from simple disruption of the proper behavior of the system to
destruction or theft of private and sensitive data.  
The huge number of devices interconnected through the Internet has
turned the infection of malware a very serious threat, even more
with the current trend of digitizing almost all human activities,
notably economical transactions.

Malware {\em detection} is concerned with identifying software that is potentially malicious, ideally before the malware acts destructively. 
Malware {\em analysis} is about identifying the true intent and capabilities of malware by looking at some aspects of the code (statically) or by running it (dynamically).

Static analysis examines malware with or without viewing the actual code. The technical indicators gathered with basic static analysis can include file name, hashes, file type, file size and recognition by using  tools like antivirus. 
When it is possible to inspect the source code, static malware
analyzers try to detect whether the code has been intentionally
obfuscated or try to identify concrete well-known malicious lines
of code.  
Dynamic analysis, on the other hand, runs the malware in a controlled
environment to observe its behavior, in order to understand its
functionality and identify indicators of potential danger.
These indicators include domain names, IP addresses, file path
locations, and whether there are additional files located on the
system.
See \cite{JDF08bdm,BHSS08sia,SER12sdl,YLA+17smd} for surveys on
malware detection techniques.

%\paragraph{Monitoring and intrusion detection}
%\todo{[BERND, JOS\'E, NATASHA \& JUAN - SEE ALSO THE HARDWARE SECTION]}

%\paragraph{Browser Extensions.}
\subsubsection{Browser Extensions}
% Written by Gerardo
Browser extensions are small applications executed in a browser context in order to provide additional capabilities and enrich the user experience while surfing the web. The acceptance of extensions in current browsers is unquestionable. 
For instance, as of 2018, Chrome's official extension repository has
more than 140,000 applications, with some of these extensions having
more than 10 million users.
When an extension is installed, the browser often pops up a message
showing the permissions that this new extension requests and, upon
user approval, the extension is then installed and integrated within
the browser.  Extensions run through the JavaScript event listener
system. An extension can subscribe to a set of events associated with
the browser (e.g., when a new tab is opened or a new bookmark is
added) or the content (e.g., when a user clicks on an HTML element or
when the page is loaded). When a JavaScript event is triggered, the
event is captured by the browser engine and all extensions subscribed
to this event are executed.

Research on the understanding of browser extensions, detecting possible privacy and security threats, and mitigating them is on its infancy. 
The potential danger of extensions has been highlighted in
\cite{Heule15} where extensions were identified to be ``the most
dangerous code to user privacy'' in today's browsers.
Some recent works have focused on tracking the provenance of web
content at the level of DOM (Document Object Model)
elements~\cite{ArshadKR16}.
%, others XXXX

Another relevant issue is the order in which extensions are
executed.
When installed, extensions are pushed to an internal
stack within the browser, which implies that the last installed
extension is the last one that will be executed. 

Recent works~\cite{PSt18ayp} demonstrates empirically that this order
could be exploited by an unprivileged malicious extension (i.e., one
with no more permissions than those already assigned when accessing
web content) to get access to any private information that other
extensions have previously introduced.  
To the best of our knowledge, there still is no solution to this
problem.

Finally, there is the problem of collusion attacks, which occurs when
two or more extensions collaborate to extract more information from
the user based on the individual permissions of each extension.
Even tough in isolation they cannot do any harm, they can exercise an
additional power by collaborating and combining their
privileges. 
With few exceptions~\cite{SainiGLC16}, this is an unexplored area.

Given that extensions may subscribe to events after they have been installed (i.e., at runtime), there is no way to  statically detect potential attacks.\footnote{Extensions may statically declare to which events they want to subscribe, but there is nothing forbidding them to subscribe to new events later at runtime.}
One of the few works providing a runtime solution to information flow in browsers (Chromium in particular) is \cite{BauerCJPST15}.

Overall, there still are concerns regarding the effect of browser
extensions on security and privacy. 
Giving the limitations on what can be obtained by static analysis,
solutions to mitigate these issues must be accomplished by means of
runtime monitoring techniques.

%\paragraph{Privacy and Security Policies.}
\subsubsection{Privacy and Security Policies}
% Written by Gerardo
One way to mitigate security and privacy threats is to have suitable and powerful policies which are enforced statically or at runtime. This, however, is not easy for different reasons.
First, defining precisely a policy language requires to introduce its
syntax (what the policies can talk about), characterize its scope
(what are the limitations, i.e., what cannot be expressed/captured by
the language), and define an enforcement mechanism (how to implement
the mechanism that ensures the policies are to be respected). 
Getting a sound and complete result is too restrictive in general.
Second, static policies may be enforced only in very specific cases
and have to be done by designers and programmers at a very early stage
of the software development process. 
In some cases, this may be done at runtime when the code is
downloaded, but it requires to isolate the code to perform the
analysis, which is not always possible.
Last, security and privacy policies could be enforced at runtime: by
mitigating the attack right after it is detected.
This is not possible in general as we cannot foresee all possible
future threats and sometimes when an attack is detected, it is usually
too late.

%\subsection{Tools}
%\paragraph{Mocca}
%
%\paragraph{Jackstab}
%
%\subsection{Challenges}
%
%\paragraph{Side channels}
%
%Whether the monitor itself could be a side channel (probably yes, but what can we do).

\subsection{Challenges}

\begin{challenge}{Monitoring GDPR}  % Written by Gerardo
  One of the main challenges is to identify which privacy principles
  might be verified or enforced by using monitors. As the regulation
  is quite extensive, we advocate to start with the principle of {\em
    data minimization} as an example of the kind of challenges the
  community might face.
\end{challenge}

\begin{challenge}{Monitoring Data Minimization}
When considering how the data is {\em used}, 
%if not possible to analyse that statically, we will need to do this at runtime. A 
a challenge is that we will not being able to do runtime verification in a black box manner. 
%(as the usage of data is an internal activity) so we might need to consider invasive means for monitoring, using inlining monitoring. 
Getting access to proprietary code can be an issue. 
Concerning when and how the data is {\em collected}, we could do runtime verification in a black box manner, but
%The challenge is that this might be doable only for suitable notions of input domains, and the problem is that 
data minimization is not monitorable in general 
%but detection of its violation (checking for non-minimality) is under certain circumstances feasible 
\cite{PAS+17mdm,PSS18rvh}. 
For the more general notion of distributed data minimization, the property is not monitorable, 
%as it is a $\forall\forall\exists\exists$ property and 
therefore new techniques using {\em grey box} runtime verification might be needed \cite{BSS18mh}.
\end{challenge}

%\paragraph{Approximate monitorability of security and privacy policies}
%How to approximate the monitorability of such policies. 
%Knowing that some security properties are not monitorable how to perform runtime monitoring in a useful manner (reduce false negatives and positives).

\begin{challenge}{Hybrid Information Flow}
%\todo{[JULIEN}
  As mentioned earlier, it is not possible to have a sound yet
  permissive dynamic information flow analysis~\cite{russo10cfs}.
  Therefore, an important challenge for information flow monitoring is
  the design of a hybrid (static/dynamic) mechanism that is efficient
  yet permissive, and that can deal with real programs and security
  policy.
\end{challenge}

\begin{challenge}{Monitoring Declassification and Quantitative Information Flow}
Non-interference is often too strong a property. For instance, a
password checker usually leaks one bit of information: whether the
password is correct.
Declassification and quantitative information flow aim to solve this
issue, but verifying these properties is very hard. In spite of some
initial work on hybrid approaches~\cite{besson16csf}, monitoring these
properties remains an unresolved challenge.
\end{challenge}

\begin{challenge}{Generic Language for Information Flow}
  There are many variants and flavors of important properties like
  non-interference, but there is currently no mainstream accepted
  language that encompasses all these security policies, which are now
  recognized to be hyper-properties~\cite{clarkson10jcs}.
  The challenge is the design and adoption of a formalism for the
  hyperproperties of interest in information flow security and the
  thorough study of its monitoring algorithms and limitations.
\end{challenge}

%% Craft the right language or logic to express different notions of information flow and related security and privacy concerns. How to monitor and enforce such issues. 
%% Foundational aspects related to noninterference (Identify and formalize noninterference (hyper)properties.)
%% Monitorability.

%\paragraph{Security at different levels}
%Low level (OS, application level, etc)
%
%\paragraph{Cyber-physical systems (e.g. Internet of Things)}
%\todo{[MARTIN \& JOS\'E - SEE ALSO THE HARDWARE SECTION]}
%
%
%\paragraph{Malware detection and analysis}
%%Identify patterns \todo{[GILES \& JUAN]}
%% Written by Gerardo
%
%%Ye et al.~\cite{YLA+17smd} surveys different malware detection techniques using data mining. 
%
%
%\paragraph{Monitoring and intrusion detection}
%\todo{[BERND, JOS\'E, NATASHA \& JUAN - SEE ALSO THE HARDWARE SECTION]}

\begin{challenge}{Browser extensions}
% Written by Gerardo
%
  One challenge on the enforcement side is how to ensure that
  malicious extensions do not expose private information from a user's
  homepage. 
  This private leakage might be done by an external entity or by
  another extension which may aggregate this information with the
  information the extension has already collected, eventually
  performing a collusion attack.  
  A related issue has to do with implementation: a robust runtime
  enforcement mechanism might need to modify the core of the browser
  (e.g., Chromium), which is quite invasive and requires a high level
  of expertise.
\end{challenge}

\begin{challenge}{Privacy and security policies}
  One challenge is how to define security and privacy policy languages
  to write policies about concrete known threats. 
  Also, this challenge involves the use of runtime monitoring
  techniques in order to detect potential and real threats, log that
  information and give this to an offline analyzer to identify
  patterns in order to generalize existing policies, or create new
  ones.  A related challenge is how to learn the policies at
  runtime. This could be done by learning them from the attacker
  models (e.g., as in \cite{AA16lpc}), and improve the precision
  taking feedback from the runtime monitors.
\end{challenge}

%
%\subsection*{References}are~\cite{kinder10proactive,kinder10precise,kinder12towards,song13ltl,babic12recognizing,dallapreda15unveiling,dallapreda14analyzing,naval15employing,macedo13mining}
%

%!TEX root = ../main.tex

\section{Reliable Transactional Systems}
\label{sec:transactions}

%\subsection{Introduction}

The human society is increasingly dependent on computing systems, even
in areas like entertainment (e.g., Netflix), social (e.g., Facebook)
and economic interactions (e.g., Amazon). 
The ubiquity of computer systems, and the large scale at which they
operate, make hardware and software failures both common and
inevitable.
At first glance it might seem that the majority of systems should not experience failures as frequently because they do not serve a world-scale user base. But with the advent of Infrastructure as a Service (IaaS) products (e.g., Amazon EC2) small and medium-sized companies are deploying their systems over IaaS offerings~\cite{armbrust:2010:cacm}, which are supported by fault prone large-scale clusters~\cite{wired:2011}.
This setting exploits modern hardware systems features to
provide fault tolerance while keeping the software systems running
efficiently, correctly, and with ease to develop and use, hence
building computer systems with improved reliability and resilience and
lower energy consumption.
% \CC{feels like a gap here but unsure what the story line is}

Database systems have successfully exploited parallelism for decades,
both at software and hardware levels.  
Databases can improve their performance by issuing many queries
simultaneously and by running those queries on multiple computing
systems in parallel, while preserving the same programming model as if
the queries were executed one at a time in a single computing system.
Transactions are at the core of most database systems.  
A transaction is an abstraction that specifies a program semantics
where computations behave as if they are executing one at a time with
exclusive access to the database.
Transactional systems implement a {\em serializable} model.
This means that even if the system allows multiple transactions to
execute concurrently, the final result of their execution must be
indistinguishable from executing one after the other (in some total
order).
Consequently, a transaction is a sequence of actions that appear to
execute instantaneously as a single, indivisible, operation. 
The system manages concurrency between transactions automatically, and
is free to execute transactions concurrently as long as the result is
equivalent to some serial execution of the transactions.

State machine replication
(SMR)~\cite{lamport:1978:cacm,schneider:1990:cacm} is the standard way
to build such fault-tolerant systems. 
An SMR system maintains multiple replicas that keep a copy of the
system's data, and coordinates the execution of operations on each of those
data replicas. 
Since replicas also execute every operation submitted to the system,
the system can continue operating as long as a majority of correct
replicas execute the operations. 
When requests to execute operations arrive, an ``agree-execute''
protocol keeps replicas synchronized: they first agree on an order to
execute the incoming operations, and then execute the operations
sequentially in the agreed order, driving all replicas to the same
final state.
However, to take advantage of contemporary hardware systems, one should
use all the available processor cores to execute multiple operations
at the same time. 
That said, this concurrent execution of operations is at odds with the
``agree-execute'' protocol because concurrent execution is inherently
non-deterministic so replicas may arrive at different final states and
the system could become inconsistent.

Improving SMR's efficiency and performance can be achieved by
exploiting multi-core processors, while still preserving determinism
and correctness.
This, however, requires to have operations that can be expressed as
serializable transactions, and that the concurrency control protocol
ensures that the concurrent execution of transactions respects the
order replicas have agreed upon.

In a typical SMR setting, a set of clients concurrently submit
requests to the system.
The system, made of replicas, runs an agreement protocol, e.g.,
Paxos~\cite{lamport:1998:PP:acmtcs}, that totally orders the incoming
requests. 
Each replica executes the requests sequentially in the agreed order,
driving all the (correct) replicas to the same final
state. Essentially, we can divide state machine replication in two
phases. 
First, the {\em agreement phase}, where replicas agree on an order
for all requests. This is then followed by the {\em execution phase}, where replicas
execute the requested operations in the agreed order.  When using
SMR there is a clear tension between the fact that the replicas have
multi-core processors and the requirement that replicas execute the
operations in a specific order.

Recovery and reparations in transactional systems \cite{survey} are
multi-layered: 
when recovering within a transaction which may still succeed,
reparations may be expressed in a \emph{try-catch} fashion. 
However, if the action is considered to have failed, then any
previously completed parts of the transaction need to be rolled back.
This is done to preserve the atomicity of the transaction, i.e.,
either the transaction entirely succeeds or entirely fails.
The problem arises when it is not possible to isolate a transaction
with the result that its actions affect other parts of the system
before the transaction is committed. 
This usually happens due to the long-life nature of the transaction ---making it
infeasible to lock the relevant resources for a long duration.

\subsection{Context and Areas of Interest}

\subsubsection{Dependable Storage Systems}

Main database vendors, such as IBM and Oracle, have business solutions
for high-performant dependable storage systems.  Innovative approaches
to such dependable storage systems are based on state machine
replication, either in
KV-stores~\cite{bessani:2013:usenixatc,bolosky:2011:nsdi,rao:2011:vldb},
filesystems~\cite{castro:2002:acmtcs,liskov:1991:sosp}, or
transactional storages~\cite{elnikety:2005:srds,garcia:2011:eurosys}.
These systems are frequently used to build business-critical (and
sometimes even life-critical) systems and must be constantly monitored
to assess the correct behavior of the storage system.  Monitoring
these systems, specially those involving SMR, both in terms of the
architecture of monitoring system itself and of the information to be
collects to reason upon, is an open and very interesting challenge.

\subsubsection{Coordination services}

Concurrent operations on distributed applications frequently need to be coordinated to ensure system correctness. These services are often provided by a small database, which stores configuration data to implement resource locking, leader election, message ordering, etc.  
%Such coordination systems are later used in more complex solutions, such as Google's Chubby distributed lock service~\cite{burrows:2006:osdi}, which is used by Bigtable (now in production in Google Analytics and other products).
%%
%In Ceph~\cite{weil:2006:osdi}, as part of the monitor processes to agree which storage units are up and in the cluster; or in the Clustrix distributed SQL database for distributed transaction resolution.
%
Such coordination systems have been recently used in more complex solutions, for example in: 
%Such coordination systems are later used in more complex solutions, for example in: 
i) Google’s Chubby distributed lock service~\cite{burrows:2006:osdi}, which is used by Bigtable (now in production in Google Analytics and other products); ii) the Ceph storage system~\cite{weil:2006:osdi}, where the coordination system is part of the monitor processes to agree which OSDs are up and in the cluster;  iii) the Clustrix distributed SQL database, which leverages on a coordination system for distributed transaction resolution.

\subsubsection{Network Services}

{\em Software-Defined Networks} (SDNs) are a step towards the separation of the network control plane from the data plane, aiming at improving the manageability, programmability and extensibility of networks.
In these SDNs, the controller should neither be a bottleneck nor a single point of failure.  State machine replication is a natural answer to such fault-tolerance requirements.
For example, the Ananta distributed load
balancer~\cite{patel:2013:sigcomm} uses Paxos for maintaining
high-availability in its manager component and serves thousands of
data flows per day in the Windows Azure cloud.

\subsubsection{Main Memory Contention Management}

The transactional model as used by database systems can be of use to
manage the contention to shared data residing in main memory.
This was first observed by Lomet in~1977~\cite{lomet:1977:sigsoft},
and proposed as a hardware solution by Herlihy and Moss in
1993~\cite{herlihy:1993:sigarch}, and by Herlihy et al in
2003~\cite{herlihy:2003:podc} as the first practical software only
solution.

\subsection{Challenges}

\begin{challenge}{Low-overhead monitoring} 
A step towards the reconciliation of SMR with the current computer processor architecture, i.e. multicore processors, is to devise new concurrency control protocols that explore pre-ordered transactions to ensure the correctness of a SMR system where individual replicas execute the local operations concurrently~\cite{vale:2016:taaco}.
The correctness of such new concurrency protocols must be assessed by intensive testing and monitoring of the system behavior.  Any deviations to the specification must be fully diagnosed and corrected.  Understanding what is happening at the level of the concurrency protocol itself (including the algorithm internal state and the ordering of concurrent events) plays an important role in such process and must be supported by lightweight (non-intrusive) monitoring techniques, so that the errors are not masked when monitoring is active.
\end{challenge}

\begin{challenge}{Reduction of the conflicting window}
When using the typical API to declare transactions (e.g., begin, read, write, and commit) the system is blind to the application's semantics, i.e., how values read are used by the application. 
Since transactional speculation is only effective when it succeeds,
there is also the need to reduce the number of conflicting
transactions by introducing variations in the typical API to declare
transactions.
The allows clients to express more clearly the intended semantics of
the program while executing over an abstract replica state, resulting
in fewer conflicts and thus more successful speculative executions.
How to reduce both the interactions with the remote database nodes (replicas) and to the ``conflicting window'' for transactions? %\CC{could not understand this question}
Some work has been done on delaying read accesses to the database using futures~\cite{baker:1977:sigplan} and double barriers and epochs~\cite{ravichandran:2014:pact}. 
Such concepts are still not mainstream in monitoring and logging of
transactional systems.
Another alternative would be to increase the expressiveness of the
transactional API to better express the application semantics and
hence improving transactional performance in SMR.
\end{challenge}

\begin{challenge}{Expressiveness of logs}
  The performance of concurrency control protocols depends on whether
  concurrent transactions conflict with each other. 
  The decision of whether two transactions conflict depends on how
  aware of the concurrency control protocol is of the transactions'
  semantics.
How to do the automatic translation of existing applications into the new transactional SMR infrastructure and how to ensure the new application (using the new transactional API) is functionally equivalent to the original? 
%\CC{This question, as the one in the previous challenge seems to be disjointed from the rest of the text... perhaps this could be put in italics as the main question of this challenge?}
%
Any changes to the protocol will create a new transactional infrastructure and any changes to the API will create a new application.  
In both cases, the new system must be backwards compatible with the
original system.
Such backward compatibility must be assessed by observing the dynamic behavior of both systems and reason over the collected information to detect any deviations of the new system to the expected behavior.
In addition to the huge logs, this challenge raises another question on expressiveness of the logs: What information is registered and how does it express the semantics of the intended transactional operations.
\end{challenge}

\begin{challenge}{Unification of multiple system huge logs} 
Observing long living distributed computations such as transactional systems replicated using SMR, may be a main requirement to automatically decompose transactions~\cite{xie:2015:sosp} and/or ensure that the workload is safe~\cite{xie:2014:sosp}.
In these cases, if the workload changes or new operations are created, the whole system must be monitored, re-analyzed and re-deployed.
In such a distributed setting, possibly many huge logs are collected (one per processor or one per replica) that must be dealt with (see Section~\ref{sec:huge}) and possibly unified into a single log, raising issues on resources' usage and consistency of the multiple observations.
\end{challenge}

\begin{challenge}{Expressing reparations in transactional systems}
  In non-transactional applications monitors typically need to have
  their own reparation code that executes in case the monitor flags a
  problem.
  In the case of transactional application monitoring, reparations are
  readily available and the monitor simply needs to trigger them.
  While this is more of an opportunity, the challenge lies in how to
  improve upon current practices and express the behavior of
  reparations formally and succinctly in a specification
  language---similarly to the way monitors are defined.
There have been several works in this
regard~\cite{colombo13monitor,colombo14comprehensive,colombo12safer}
for example through the use of {\em compensating automata}. However, future
work can focus on further simplifying the specification language and
perhaps providing a library of ready-made constructs which developers
can use directly.
\end{challenge}

\begin{challenge}{Management of historic data to be used in the reparations}
  From a more pragmatic point of view, compensations and rollbacks
  present the challenge of managing historic data values to be used in
  the reparation code.
  In this respect runtime monitors can be useful in the same way
  software monitors are typically stateful.
  Reparations can be parametrized through the monitors' state,
  avoiding complex wiring to pass the data around. To the best of our
  knowledge this approach has not been implemented.
\end{challenge}

\begin{challenge}{Monitoring transactional memory}
The time-scale for transactional memory is orders of magnitude smaller than transactional databases.  In transactional memory, each access to a shared memory location must be handled by the transactional monitor and considered for the success or failure of the memory transaction.  
Any additional probing or logging introduced by a monitoring system
may influence the scheduling and have a strong impact in a
malfunctioning transactional memory application, by changing the
serialization order of the transactions, possibly masking or hiding
previously observed errors.  
Researchers have partially addressed this challenge in the
past~\cite{hvc:2013:dias,dias:2012:ecoop,lourenco:2009:padtad,orosa:2016:euromicro}
aiming at both correctness and performance.
\end{challenge}

%!TEX root = ../main.tex

\section{Contracts and Policies}\label{sec:contracts}
%\todo[inline]{Add an intro}

The term {\em contract} is overloaded in computer science, so it may
be understood in different ways depending on the community:
\begin{enumerate}[(i)] 
\item \emph{Conventional contracts} are legally binding documents,
  establishing the rights and obligations of different signatories, as in
  traditional, judicial and commercial, activities. 
\item {\em Normative documents} are a generalization of the notion of legal contracts.  
  The main feature is the inclusion of certain normative notions such
  as {\it obligations}, {\it permissions}, and {\it prohibitions}, either directly, or by representing them indirectly. 
  These include legal documents, regulations, terms of services, contractual agreements and workflow descriptions.
\item {\it Electronic contracts} are machine-oriented, and may be written directly in a formal specification language, or translated from a conventional contract. In this context, the signatories of a contract may be objects, agents, web services, etc.
\item \emph{Behavioral interfaces} are considered to be contracts between different components specifying the history of interactions between different agents (participants, objects, principals, entities, etc.). Rights and obligations are thus determined by ``legal'' (sets of) traces which are permissible.
\item The term ``contract'' is sometimes used for specifying the interaction between communicating entities (agents, objects, etc.). It is common to talk then about a \emph{contractual protocol}.
\item \emph{Programming by contract} or \emph{design by contract} is an influential methodology popularized first in the context of the programming language Eiffel  \cite{meyer:eiffel}.  
  ``Contract'' here means a relation between pre- and post-conditions of routines, method calls, etc.
  This concept of contract is also used in approaches such as the Ke\kern-0.1emY program verification tool \cite{key}.
\item In the context of web services, ``contracts'' may be understood as
  {\it service-level agreements} usually written in an XML-like language
  like IBM's Web Service Level Agreement (WSLA \cite{wsla}).
\item More recently, the term ``contract'' is used in the context of {\em
    blockchain} and other \emph{distributed ledger technologies} as
  programs that ensure certain properties concerning
  transactions. 
  These programs are called {\em smart contracts} \cite{szabo}, as popularized by the Ethereum platform \cite{Buterin}.
\end{enumerate}

In this section we focus on the use of the term in the computational
domain but with a richer interpretation than just a specification or property.
In particular, we consider two types of contracts: normative documents
(including conventional contracts and their electronic versions as
described above), and smart contracts.
In both cases, we refer to ``full
contracts''~\cite{FOP+08bsc,pace09challenges}, that is agreements between
different entities regulating not only the normal interactive
behaviors, but also exceptional ones.
A common aspect of such contracts is that they should express not only
the sequence and causality of events, but also what obligations,
permissions and prohibitions the participating entities have (basic modalities studied in deontic logic~\cite{VW51dl}), as well as the associated penalties in case of violations.

The specification of such contracts requires a formal language rich
enough to capture these deontic notions, temporal and dynamic aspects,
real-time issues such as deadlines, the handling of actions (events)
and exception mechanisms.
The main aim is not only to specify such contracts, but to analyze
them using techniques like model checking and runtime verification.
Clearly, the use of contracts is only meaningful if there is a
mechanism to validate their fulfillment.

A related concept is that of {\em policies}. 
At a certain level of abstraction, policies can be seen as contracts in the sense that they prescribe behavior. 
Since the term policy is also very generic with a broad scope, we
concentrate on {\em privacy policies} (or privacy settings) and more
specifically in the context of Online Social Networks (OSN) like Facebook and Twitter.
Though the formalization of such policies can be quite different from
that of contracts ---for example, it requires to use {\em epistemic} instead
of {\em deontic} logic--- from a runtime verification perspective, the two
are very similar.

\subsection{Context and Areas of Interest}

\subsubsection{Contracts: Normative documents}

The complete specification of full contracts ---normative texts
which include tolerated exception, and which enable reasoning about
the contracts themselves--- can be achieved using a combination of
temporal and deontic concepts \cite{FOP+08bsc}.
Formalizing such contracts requires operators and combinators for
choice, obligations over sequences, contrary-to-duty obligations, and
the representation of how internal and external decisions may be
incorporated in an action- or state-based language for specifying contracts.
There have been several interpretations and approaches for the
development of such a logic~\cite{pace09challenges}, including modal
extensions of logics and automata in order to address the issue of how
contracts can be formalized and reasoned about. 
See, for
example~\cite{APS+16ca,CRS18wbt,GGV17nrl,LS03dis,MS13ecc,PS13fan,PS12ddl,Wyner15},
just to mention a few.\footnote{The literature is quite vast and the
  list of citations is not exhaustive.  The main conferences,
  workshops and journals in the area include JURIX \cite{jurix}, DEON
  \cite{deon}, RuleML \cite{ruleml}, and the Journal of Artificial
  Intelligence and Law \cite{jail}.}
%
%Some of these approaches are more amenable to monitorability, while
%others take more global view and are more appropriate for whole-system
%verification.

Why is there a need for a logic or some other formal language? 
One of the aims of formalizing contracts is not simply to use them as
specification, but also to be able to prove properties about the
contracts themselves, to perform queries on the contracts (like what
each party is agreeing to), and ultimately to ensure at runtime that
the contract is satisfied (or alternatively to detect for violations). 
An alternative approach is to use {\em machine learning} (or other artificial intelligence techniques). 
For instance, one may avoid the use of formal methods by using {\em natural
language processing} (NLP) combined with machine learning to directly
perform queries on the textual representation. 
While this is feasible in certain cases, it is well known that the
state of the art in NLP is still far from being able to deliver fully automatic and sufficiently reliable techniques.
Moreover, performing semantic queries or running simulations still require a formal representation.
This is an important and interesting research area in
itself, but here we are concerned not with the problems of obtaining
such normative documents but with the specific issue of monitoring their satisfaction or violation.

In terms of monitoring of contracts, most of the current work start
from some form of formal semantics.
There are various outstanding questions of what subsets of deontic
logics are tractably and practically monitorable.
For example, are more standard logics, like classic or temporal
logics, enough? How important is to get full complex semantics
(e.g., based on Kripke semantics) for the logic?
For a full representation and analysis of contracts, Kripke semantics
might be necessary, but for monitoring purposes a much simpler approach considering trace semantics seems to be sufficient.

Concerning monitoring, an ideal goal is to automatically extract a
monitor from the document's formal representation, but this is, in general, not feasible.
We assume then that we obtain the monitors from a given contract manually or semi-automatically.
This is still not an easy task, as there is no standard, easy and direct way to extract a model from a document in natural language.

The use of \emph{controlled natural languages} (CNL)
\cite{DBLP:journals/corr/Kuhn15a} has been proposed in different works
in order to facilitate bridging the gap between the natural language
description of the original document and a more formal representation
in the form of a formal language~\cite{ACS13fca,CRS18wbt,CPS14cnl,LCC+17pvr}.
In a legal specification setting, there is initial work in this
direction, but we are still far from reaching this goal~\cite{DBLP:conf/cnl/BunzliH10,DBLP:conf/cnl/CamilleriPR10,CPS14cnl}.

\subsubsection{Smart contracts}
If the computer science community borrowed the notion of contracts by remarking on the similarity between specifications and legal agreements, the legal community saw an opportunity in viewing computer code as a form of executable enforcement or enactment of agreements or legislation.  
The notion that executable code regulates the behavior of different
parties very much in the same manner that legal code does was proposed
by Lessig~\cite{lessig} in 1999.
The dual view, that the use of executable smart contracts can enforce
compliance as an integral part of the behavior, was argued earlier by
Szabo in 1996 \cite{szabo}.

The introduction of blockchain \cite{Nakamoto2008} and other
distributed ledgers technologies, which enable the automated
management of digital assets, has changed the way in which computer
systems can regulate the interaction between real-world parties.
In particular, these technologies have enabled the deployment of
Szabo's notion of smart contracts in a distributed setting, without
the participation of trusted central authorities or resource managers.
For instance, the Ethereum~\cite{ethereum} blockchain supports smart
contracts which can be expressed using a Turing-complete programming
model, to be executed on the Ethereum Virtual Machine (EVM) and
typically programmed using one of a number of languages supporting a
higher level of abstraction.

Smart contracts are executable specifications of the way the contract will update the state of the underlying system. 
Although specifications can be executable or not
(see~\cite{DBLP:journals/iee/Fuchs92}
and~\cite{Hayes:1989:SE:84458.84466}), it is generally accepted that
executable specifications must elucidate \emph{how} to achieve the
desired state of affairs, while non-executable specifications simply
characterize properties that the desired state should satisfy.
The former is substantially more complex, which is why the fields of
validation and verification arose to explore ways in which executable
specifications (code) can be verified against non-executable ones
(properties).

This gives rise to a challenge: that of verifying that smart
contracts indeed perform as they should.
Although one can argue that the challenge behind verification of such executable code is no different from that of verifying standard programs, there are a number of issues which are particular to smart contracts. 
There has been little work yet addressing the special idiosyncrasies
of smart contracts.
Static analysis techniques for the verification of smart contracts has been proposed in \cite{Bhargavan:2016go}, via a translation from smart contracts into another language (F* in this case) for verification. See \cite{APS18sc} for a discussion on some challenges concerning the verification of smart contracts using deductive verification techniques.
From a runtime perspective, there has been some work on using
blockchain technology to regulate distributed systems (see
\cite{DBLP:conf/bpm/Garcia-Banuelos17,DBLP:conf/ruleml/GovernatoriR10,DBLP:journals/corr/Prybila0HW17,DBLP:conf/bpm/WeberXRGPM16}),
but the focus of this work is not on the verification of the smart
contracts themselves.
Initial attempts to address runtime verification of smart contracts and building tools to automate this have started to appear~\cite{contractlarvaisola,contractlarva}, but many challenges remain to be addressed~\cite{contractlarvatutorial}.

One particular aspect that presents specific challenges is that these smart contracts are typically mainly concerned with the movement of digital assets, with built-in notions of failing transactions and computation roll-back to handle failure. 
Although this has been investigated in the domain of financial system
verification~\cite{colombo12safer,DBLP:conf/cade/PassmoreI17}, there
is a major difference. 
Before the rise of cryptocurrencies, all such systems were deployed on
a central trusted system, typically residing within the infrastructure
of the payment institution.
In contrast, in the context of distributed ledgers, the storage and
computation are, by their very nature, distributed, and particularly
runtime verification require the instrumentation and deployment to
take this into consideration.

There is a major difference with regular financial transaction software
deployed on, or interacting with, payment institutions. That is that given the
critical nature of such systems (payment applications have been built
using a strict validation process) ensuring compliance to legislation
and adherence to specifications.
However, with what has been hailed as the democratization of currency
systems, came the popularization of payment application development,
with many smart contracts being developed without the necessary care
and responsibility.
This approach has suffered a number of huge financial losses due to
bugs \cite{DBLP:conf/post/AtzeiBC17}. 
The need for lightweight runtime validation of such systems, whether
inbuilt in the execution of the smart contracts or inherent in the
blockchain or alternative distributed ledger technology is essential
to ensure user safety.

Turing-complete environments for smart contracts suffer from the possibility of non-termination or excessively long computation. 
Rather than limit the power of the programming language, the solution
adopted in systems such as Ethereum was that of introducing the notion
of \emph{gas} ---a resource required to enable computation and that
has to be paid for using other digital assets, typically the
underlying cryptocurrency.
Although efficiency of computation has always been an important issue
in computing, it has typically been detached from functional
correctness issues addressed by formal methods. 
With the notion of gas, the direct correlation between execution steps
and financial cost is a new challenge for runtime verification.
As a direct corollary, additional computation to check for correctness will directly induce additional cost. 
However, there is also the issue that gas affects computation, in that
once gas runs out, computation is reverted, which has been exploited
in a number of smart contract attacks.
Finally, the use of gas throughout the computation may justify qualitative dynamic analysis to measure the extent of satisfaction or violation using a distance metric to detect failure due to lack of gas.

Finally, the multitude of contracts and interaction platforms provided by the underlying distributed technology is likely to give increased importance to contract comparison and negotiation. 
We envision a scenario, in which one may negotiate for increased
dependability (e.g. by monitoring additional logic) against a stake
paid by the developer or provider of the contract.
At a more complex level, one can have a system where different or
additional functionalities are negotiated upon setting up a smart
contract. 
In both cases, the process is a form of meta-contract which regulates
how the parties may interact to negotiate and agree upon a contract
which will be set up.

See \cite{APS18sc} and references therein for a discussion on the verification of smart contracts, as well as papers in \cite{isola18rsc} for recent advances and a discussion on open issues in the area. 

%\subsubsection{About policies (privacy policies for OSNs).}
\subsubsection{Privacy policies for OSNs}
Policies may be understood, at a certain level of abstraction, as
contracts: they prescribe what actions are allowed or not.
The term policy is generic and may be applied to many different cases
or applications.
We focus here on privacy policies, and in particular on privacy
policies for Online Social Networks (OSNs).
OSNs provide an opportunity for interaction between people in different ways
depending on the kind of relationship that links them.
One of the aims of OSNs is to be flexible in the way one shares
information, being as permissive as possible in how people communicate
and disseminate information.
While preserving the spirit of OSNs, users would like to be sure that
their privacy is not compromised.
One way to do so is by providing users with means to define 
privacy policies and provide them with guarantees that their requested policy will be respected.

For defining policies one might use simple checkbox privacy settings
(as it is the case in most OSNs today), or allow user to define more
richer policies using expressive formal languages or logics.
Given means to specify privacy policies is not enough, as these
policies must be enforced at runtime.
Enforcement of checkbox privacy settings is rather well-understood, at
least for most of the kind of policies currently implemented in
existing OSNs.
However, if one wants to allow the definition of richer policy
languages, the challenge goes beyond identifying an appropriately
expressible language to the problem of automatically extracting a
runtime monitor to act as an enforcement mechanism.
This is currently beyond the state of the art and no concrete
solutions exist.

Furthermore, the state of the art today is focused on static policies.
For instance, in Facebook users can state polices like \emph{``Only my
friends can see a post on my timeline''} or \emph{``Whenever I am tagged, the
picture should not be shown on my timeline unless I approve it''.}
However, no current OSN provides the possibility of defining and
enforcing {\em evolving} (dynamic) privacy policies.
Policies may evolve due to explicit changes done by the users (e.g., a
user may change the audience of an intended post to make it more
restrictive), or because the privacy policy is dynamic per se.
Consider for instance: \emph{``Co-workers cannot see my posts while I
  am not at work, and only family can see my location while I am at
  home'',} \emph{``Only up to 3 posts disclosing my location are
  allowed per day on my timeline'',} \emph{``My boss cannot know my
  location between 20:00-23:59 every day'',} and \emph{``Only my
  friends can know my location from Friday at 20:00 till Monday at
  08:00''.}
No current OSN addresses the specification and enforcement of such
policies.
Formal languages are needed to express such time and event-dependent
recurrent policies, and suitable enforcement mechanisms need to be
defined.
This could be done by defining real-time extensions of epistemic
logic, or combining existing static privacy policy languages with
automata, as done for instance in~\cite{PCP+16aba,PKS+16sep,PSS18tek}.

\subsection{Challenges}
%In conclusion, we identify some general challenges concerning the formalization and runtime verification of normative documents, smart contracts and privacy policies for OSNs.

\begin{challenge}{Formalizing natural language contracts} 
%Since for any realistic real-world contract, it is unfeasible and error-prone to carry out its formalisation manually, a major challenge is the identification of techniques to extract a formal model from a normative document in an automatic manner. 
%  
%Natural language processing is a major stumbling block in using legislation, contracts and privacy policies as specifications for runtime verification and formal analysis. Many existing tools are not yet adapted to the legal domain and terminology, even if work based on training directly on legal corpora is becoming more common e.g. \cite{DBLP:conf/lrec/PetersW16,DBLP:conf/rweb/AthanGPPW15}. 
%  
%Another direction which has started being investigated, and holds promise is that of the use of controlled natural languages --- constrained versions of natural language in order to support easier understanding and analysis. One of the original drives which led to controlled natural languages was, in fact, the need to be able to go from natural language requirements to formal specifications. In a legal specification setting, there is initial work looking into this \cite{DBLP:conf/cnl/CamilleriPS14,DBLP:conf/cnl/CamilleriPR10,DBLP:conf/cnl/BunzliH10}, but we are still far from being achieved.
A major challenge is the identification of techniques to extract a formal model from a normative document in an automatic manner. In particular, the challenge is to adapt NLP techniques and use machine learning techniques to (semi-)automatically translate natural language text into a suitable CNL.
\end{challenge}

\begin{challenge}{Formal reasoning about legal documents} 
A challenge in the formalization of legal documents is the choice of the right formal language adequate for the type of analysis required, as there is a trade-off between expressiveness and tractability. 
In particular, the notion of \emph{permission} (and \emph{rights})
poses challenges in monitoring, since one party's permission to
perform an action typically entails an obligation on the other party
to allow the action, and this obligation may not be observable unless
the right is exercised.
\end{challenge}

\begin{challenge}{Operationalization of legal documents} 
  Most legal texts are written in a declarative style, and typically
  require to be operationalized for automated analysis. 
  Furthermore, parts of these texts may refer to events or attributes
  which are not observable and thus not monitorable.  Most runtime
  monitoring and verification approaches for legal texts interpret
  the term \emph{runtime} to refer to the time during which the legal
  text regulates. Another possible interpretation is that of
  monitoring the process of drafting of a contract or legislation, or
  the negotiation of a contract. A monitoring regime could be useful
  in this setting.
\end{challenge}

\begin{challenge}{Smart contract monitoring and verification} 
  How to adapt dynamic verification to smart contract monitoring is
  unclear, particularly because once a problem arises, it is not
  always possible to take reparatory action to recover. An open
  question is how enforcement, verification and reparation can be
  combined in a single formalism and framework.
\end{challenge}

\begin{challenge}{Monitoring gas in smart contracts} 
  Another challenge is the use of the notion of `gas' to justify
  computation on ledger systems such as Ethereum, although it is
  unclear how dynamic analysis can be used effectively to track such a
  non-functional property.
  Furthermore, the introduction of runtime verification overheads in
  terms of gas poses new challenges for monitoring.
\end{challenge}

\begin{challenge}{Compliance between legal and smart contracts} 
%The relation between the underlying legal document and smart contracts is still to be addressed. While the former regulates agreements between parties, having to revert to external enforcement if required, the latter uses code to describe, enforce and execute the agreed upon behaviour. However, one can envisage situations in which a legal contract can regulate the parties' behaviour with respect to a smart contract e.g. a party may be obliged to use certain services in a smart contract, or a legal contract may regulate at a meta-level the behaviour expected from a smart contract. On the other hand, a smart contract may trigger clauses in a legal contract e.g. a transfer of assets may require one of the parties to report (outside the blockchain) certain information to the other party. 
The relation between the underlying legal document and smart contracts is still to be addressed.
The challenge here is how to monitor compliance between both versions
of the contract, and relate violations in the execution of the smart
contract with the corresponding clause in the real legal contract.
\end{challenge}

\begin{challenge}{Policy monitoring and verification} 
  The challenges we identified for contracts also apply to
  policies. In particular, there might be a need to combine the
  enforcement mechanism with machine learning techniques and with
  natural language processing.  For instance, a post might contain a
  sentence like \emph{``I am here with John drinking a glass of
    wine'',} where \emph{``here''} clearly refers to a place which
  might be inferred from the location associated with the post. This
  kind of inference is difficult to do automatically by machine.
\end{challenge}

\begin{challenge}{Policy monitoring in OSNs} 
  For Online Social Networks (OSNs), the use of epistemic logic to
  reason about whether and how explicit (and derived) knowledge of
  users adhere to policies has been explored. 
  However, the operationalization of such policies and the extraction
  of monitors from policies have proved to be particularly difficult.
\end{challenge}

\begin{challenge}{Policy monitoring and verification} 
  The evolution of policies due to specific events or timeouts also
  poses a number of challenges. Some initial work has been recently
  done on the specification side with a proof of concept
  implementation. 
  The work in~\cite{PKS+16sep,PSS18tek} presents an approach based on
  extending a privacy language with real-time, while~\cite{PCP+16aba}
  proposes a combination of static privacy policy language with
  automata. 
  However, a general working solution to this challenge is still
  missing.
\end{challenge}

\section{Huge Data and Approximate Monitoring}
\label{sec:huge}

This section describes runtime verification challenges related to the
analysis of very large logs or streams of events from the system under
observation.
The general goal when dealing with huge data streams is to develop
algorithms that offer scalability, specification language
expressiveness, precision, and utility.
Below we discuss the advances made along each of these dimensions and
some of the remaining challenges.

\subsection{Context and Areas of Interest}

\subsubsection{Scalability}
In runtime verification, the focus to date has mainly been on
efficiency, expressiveness, and correctness, and less so on
scalability to {\em Big Data} in realistic scenarios. 
A few exceptions exist and are summarized below, which mostly address
offline monitoring.

Barre et al. \cite{barre_mapreduce_2012} and Hall\'e and
Soucy-Boivin~\cite{halle_mapreduce_2015} use Hadoop's MapReduce
framework to scale up the monitoring of propositional LTL properties
using parallelization. In their experiments, they used event
logs with more than nine million entries. In these approaches,
formulas are processed bottom up using multiple MapReduce iterations.
While the evaluation in the map phase is completely parallelized for
different time points from the event log, the results of the map phase
for a subformula for the whole log are collected and processed by a
single reducer. In a single iteration there are as many reducers as there are
independent subformulas with the same height. 
The reducers, therefore, become bottlenecks that limit the
scalability.

Bianculli et al.~\cite{bianculli_trace_2014} extend this approach to
the offline monitoring of large traces, for properties expressed in
MTL with aggregation operators.
Similarly to the aforementioned approaches, the memory consumption of
the reducers limits the scalability of this approach. More
specifically, reducers (that implement the semantics of temporal and
aggregate operators) need to keep track of the positions relevant to
the time window specified in the formula: the more time points there
are the denser the time window becomes, with a consequent increase in
memory usage.  Bersani et al.~\cite{bersani_efficient_2016} worked
around this problem by considering an alternative semantics for MTL,
called the \emph{lazy semantics}. This semantics evaluates temporal
formulas and Boolean combinations of temporal-only formulas at any
arbitrary time instant.  
It is more expressive than the point-based semantics and supports the
sound rewriting of any MTL formula into an equivalent one with
smaller, bounded time intervals.
The lazy semantics has the drawback that basic logical properties do
not hold anymore. 
%
% For example, the formula \emph{true} only holds at
% points where there is an event, and is false otherwise. 
%
This disallows formula simplifications and complicates the
formalization of properties given in natural language, since familiar
concepts have a different meaning. 
%
% Another limitation of this approach is that determining the most
% appropriate value for the decomposition parameter is not automated,
% and relies on manual experimentation with the configuration of the
% available infrastructure. 
%
Unlike the previous approaches, Bersani et al. implemented the monitor
on top of the Apache Spark framework~\cite{zaharia_spark:_2010} that
is optimized for iterative distributed computations.

Parametric trace slicing~\cite{chen_parametric_2009,RegerR-RV2015} is
a technique for monitoring a parametric LTL property by grounding it
to several plain LTL properties. In this approach logged events are
grouped into slices based on the values of the parameters. A slice is
created for each parameter value or for each combination of values
depending on the number of parameters. The individual slices are then
processed by a propositional LTL monitor unaware of the parameters.
The initial main goal of this approach was not scalability, but rather
monitoring the more expressive parametric LTL specification language.
However, the approach is also relevant for scalability since
it easily lends itself to parallelization.

Another line of work~\cite{basin_scalable_2014,basin_scalable_2016}
similarly splits the logged events into slices, but it avoids
grounding first-order properties altogether.
This is enabled by using a more powerful monitor, 
MonPoly~\cite{RV-CuBES2017:MonPoly,basin_monpoly:_2011,basin_monitoring_2015,basin_monitoring_2015-1},
to process the slices. Overall, the approach allows for scalable 
offline monitoring of properties expressed in {\em Metric First-Order Temporal Logic} (MFOTL).
%A log with several billion events was distribed the workload among 1000
%machines~\cite{basin_scalable_2016}.
%
The core idea in this work is to split the log into multiple slices and check
the same formula on each slice independently.
This allows the solution to scale, by handling one slice on a single
computer. 
The key component is a log-splitting framework used to distribute the
log to different parallel monitors based on data and time.
The framework takes as input the formula and a splitting strategy and
splits the log ensuring soundness and completeness.
The approach was implemented in Google's MapReduce framework where
the log-splitting framework is executed in the map phase. 
The approach is, however, limited to offline monitoring since it uses
MapReduce. 
Parallelization is not limited as in the previous
approaches, but it is potentially wasted, since to ensure correctness,
the log splitting framework may completely duplicate the original log
into some of the individual slices.  
Another limitation is that the slicing framework relies on a domain
expert to supply a splitting strategy manually. 
For example, if a monitored property involves events parametrized with
``servers'' and ``clients'', one could split the log along the
different ``servers'', along the different ``clients'', or along both.

Loreti et al.~\cite{loreti2017:distributed-com} discuss two MapReduce
architectures to tame scalability in the context of compliance
monitoring of business processes, using the SCIFF
framework~\cite{Alberti:2008:VAI:1380572.1380578}. Such a framework
provides a logic-based proof procedure for checking declarative
constraints on sequences of events, in terms of expectations and
happened events. The two MapReduce architectures proposed in this work
were adapted from similar ideas in process
mining~\cite{10.1007/978-3-642-28872-2_1} and distinguish between
\emph{vertical} and \emph{horizontal} distribution.  
In the vertical distribution all nodes receive the complete
specification and a subset of the complete log.
During the map phase, the log is split across the various nodes such
that all the events of a trace are sent to the same node.
In the reduce phase, each node checks the conformance of each log
fragment to the specification.  
In horizontal distribution both the specification and the logs are
partitioned across the nodes.
Each node checks a partial specification on a fragment of the log that
contains only the events used in the partial specification.
The results of all the nodes are then merged together with a logical
AND.
The limitation of the approach is the expressiveness of the SCIFF
logic programming framework that cannot handle parametric
specification.

Yu et al.~\cite{yu2017:verifying-tempo} propose an approach for
parallel runtime verification of programs written in the {\em Modeling,
Simulation and Verification Language} (MSVL), with properties expressed
in {\em Propositional Projection Temporal Logic} (PPTL). 
The approach divides each program trace into several
segments, which are verified in parallel by threads running on
under-utilized CPU cores.
The verification results of all segments are then merged and further analyzed
to produce a verdict.  

\subsubsection{Expressiveness}
Most of the works on runtime verification borrow logics from static 
verification approaches and focus on designing algorithms that either (1)
generate a monitor that can analyze a trace online, or (2) can process
dumps of traces offline.
Optionally, one could use a general programming language or a domain-specific language to write the queries that process the input traces
online or offline.
In both cases, we would like to monitor Big Data with a highly expressive specification language.
More expressive logics naturally require more computation resources for monitoring. Thus, a worthwhile research question is:
\emph{What are the limits of the specification language expressiveness to achieve scalable monitoring of Big Data?} Below we discuss some directions of how expressive specification languages could look like.

{\em Complex Event Processing} (CEP) and {\em Data Stream Management Systems}
(DSMS), for example, can serve as specialized languages for building
stream processors (see~\cite{margara_processing_2011} for a recent
survey).
The query languages of DSMS are mostly extensions of SQL (e.g., with
window operators~\cite{arasu_cql_2006}), and thus typically much
weaker than logics such as MFOTL due to the absence of proper negation
and more limited capabilities for expressing temporal
relationships. 
Moreover, DSMS tend to focus on efficient query execution at the
expense of sacrificing a clean semantics of the property
specifications.
The reference model of DSMS has been defined in the seminal work on
the {\em Continuous Query Language} (CQL)~\cite{arasu_cql_2006}. 
In CQL, the processing of streams is split in three steps.
i) Stream-to-relation operators---that is, windows---select a portion
of each stream thus implicitly creating static database table. 
ii) The actual computation takes place on these tables, using
relation-to-relation (mostly SQL) operators. 
iii) Finally,
relation-to-stream operators generate new streams from tables, after
data manipulation. Several variants and extensions have been proposed,
but they all rely on the same general processing abstractions defined
above.

CEP~\cite{luckham_power_2005,margara_processing_2011} systems are
closely related to DSMS. CEP systems analyze timestamped data streams
by recognizing composite events consisting of multiple atomic events
from the original stream that adhere to certain patterns. The user of
a CEP system controls the analysis by specifying such patterns of
interest. The predominant specification languages for patterns are
descendants of SQL~\cite{gyllstrom_sase:_2006}. An alternative is
given by rule-based languages, such as
Etalis~\cite{anicic_rule-based_2010}, which resembles Prolog. Although
CEP systems improve the ease of specification of temporal
relationships between events over DSMS, they are still significantly
less expressive than MFOTL due to their restricted support for
parametrization of events and lack of quantification over parameters.
Interestingly, some CEP systems use interval timestamps. In this
model, each data element is associated with two points in time that
define the first and the last moment in time in which the data element
is
valid~\cite{Schultz-Moller:2009:DCE:1619258.1619264,White:2007:NEP:1265530.1265567}.

For logical specification languages such as LTL a recent trend has
been to incorporate regular-expression-like constructs in the
logic. 
This gave rise to the industrially standardized {\em Property Specification
Language} (PSL)~\cite{eisner_practical_2006}, the development of
{\em Regular Linear Temporal Logic}
(RLTL)~\cite{leucker07regular,sanchez10regular} and its more recent
incarnation in the form of {\em (Parametric) Linear Dynamic Logic} ((P)LDL)
\cite{giacomo_linear_2013,faymonville_parametric_2014} and its metric
counterpart (MDL)~\cite{basin-monitoring-mdl-2017}. 
Due to the extension with regular expressions, those languages are
more expressive than LTL in that they capture all $\omega$-regular
languages.  
Vardi~\cite{vardi_church_2008} observed that these
extensions were essential for the practical usage of PSL in many
industrial application settings.
% \begin{quote}
% The conclusion was that LTL is not expressive enough for industrial usage. In particular, many properties that are expressible in [...] are not expressible in LTL. Thus, it turned out that the theoretical considerations regarding the expressiveness of LTL, i.e., its lack of $\omega$-regularity, had practical significance.
% \end{quote}
First-order extensions of languages like PSL, RLTL, (P)LDL, and MDL,
which should be more expressive than MFOTL, have not yet been
considered for monitoring.

However, to keep things manageable for Big Data, it may be necessary
to restrict or even remove features from our property specification
languages. The usage of negation is a candidate for restriction while
the first-order aspect of MFOTL is a candidate for  removal (or for replacement with freeze quantifiers).
Many works~\cite{basin_scalable_2016,basin_monitoring_2012,basin_monitoring_2015,basin_monitoring_2015-1,basin-monitoring-out-of-order-2017} had to define (efficiently) monitorable fragments using similar restrictions. A syntactic restriction (e.g., of the allowed occurrences of negation) is preferable over a modification of the semantics as seen on the example of negation in many data stream management systems (DSMS). The user of a specification language with a syntactic restriction can at least rely on the familiar semantics. Moreover, properties outside of the monitorable fragment can be often automatically rewritten into equivalent formulas within the fragment.

\subsubsection{Precision}
Compromising on soundness is
not a common approach in runtime verification. 
However, when faced with very large logs (or streams) of data and hard
real-time constraints on providing verdicts, this can become a
very useful compromise.
In some cases, sound algorithms cannot be used in
practice. For example, a sound algorithm that determines the number of 
distinct elements in a data stream must use space linear in the
cardinality it estimates, which is impractical.
Determining cardinality is a large component of many practical
monitoring tasks such as detecting worm propagations, denial of
service (DoS) attacks, or link-based spam.
Ideally, tradeoffs between monitoring efficiency and precision of the
provided verdicts should be formulated as an additional input to the 
monitor.
We call such an extension \emph{approximate monitoring}.

Approximate monitoring deals with the issue of providing approximate
(or imprecise) results to the standard monitoring problem,  with
guarantees on the ``distance'' to the actual (precise) results from
the provided ones.

%Dmitriy: unclear: should be rephrased or dropped

%These guarantees can be global, i.e., applicable to
%all produced verdicts, or local, i.e., provided on an individual verdict
%basis.

% Approximate monitoring vs
% Monitoring probabilistic properties vs
% Monitoring uncertain data (imprecise/disagreeing/incomplete logs + errors) vs
% probabilistic monitoring of dynamically evolving state
One should make a clear distinction between approximate
monitoring and monitoring probabilistic properties. The latter deals
with monitoring specification languages that
can express probabilistic and statistical properties of data
streams. However, it still provides verdicts with absolute precision
given the semantics of the specification language.
A related facet is the monitoring of uncertain data, which deals with
the problems of data collection and data reliability, and it often
carries over to monitoring by invalidating certain assumptions on the
data stream.
There are many sources of uncertainties in the monitored
data: timestamps can be imprecise due to clock skew, logs may be
incomplete due to outages, or even disagree when coming from various
sources. Uncertainty can come from the monitored systems themselves
which can exhibit stochastic and faulty behavior.
Another related field is state inference of the monitored
system using probabilistic approaches where a belief state is
maintained and updated during monitoring. Although these approaches
provide probabilistic guarantees as part of the resulting belief
state, they perform a specific monitoring task.

Existing work on approximate monitoring stems from the fields of
databases~\cite{Babcock:2002:MID:543613.543615}, streaming
algorithms~\cite{DBLP:journals/crossroads/Nelson12}, and property
testing~\cite{DBLP:conf/propertytesting/2010}.  All approaches can be
classified based on two criteria: the specific
queries %(or query languages)
they approximate and the resources they optimize.  Commonly
approximated queries in the literature are cardinality
estimation~\cite{Flajolet07hyperloglog:the}, top-k
items~\cite{Babcock:2003:DTM:872757.872764}, frequent items (heavy
hitters)~\cite{Keralapura:2006:CDM:1142473.1142507,Manjhi:2005:FFI:1053724.1054115,Yi:2013:OTD:3118739.3118856},
quantiles~\cite{Cormode:2005:HAN:1066157.1066161,Yi:2013:OTD:3118739.3118856},
frequency moments~\cite{Cormode:2005:SST:1083592.1083598,Cormode:2011:ADF:1921659.1921667},
entropy~\cite{Arackaparambil:2009:FMW:1577399.1577411}, other
non-linear functions over (possibly distributed) streams, and distance
queries~\cite{Agarwal.2465355}.  Orthogonally, the approaches either
optimize memory consumption, communication
cost, %(e.g., number of exchanged messages),
execution time, or the monitor's overhead.

% Optimizing memory
\emph{Optimizing memory consumption}
has led Morris to develop his well-known
approximate algorithm for 
counting~\cite{Morris:1978:CLN:359619.359627}.
% Consider a stream of $n$ 
% increments: instead of incrementing a counter $c$ upon every received
% increment and using up to $\lceil log_2 n\rceil$ memory, the counter
% is incremented each time with a probability $2^c$. The counter then
% basically  holds an ``order of magnitude estimate'', which is much more
% memory efficient, while the expected value of $2^c$ is $n+2$ after $n$
% updates~\cite{Flajolet:1985:ACD:3765.3780}.
%
% Hyperloglog
The HyperLogLog algorithm~\cite{Flajolet07hyperloglog:the} tackles the
cardinality estimation problem mentioned in the example above.
% To
% avoid using liner space in the number of distinct elements in the
% stream (which is the size of the stream in the worst case), it
% estimates the number approximately. Using an auxiliary memory of $m$
% units it performs a single pass over the data and produces an estimate
% of the cardinality such that the relative accuracy (the standard
% error) is typically about $\frac{1.04}{\sqrt{m}}$.
% It can estimate cardinalities beyond $10^9$ with an accuracy of 2\%
% while using a memory of only 1.5 kilobytes.  
%
% frequency count (heavy hitters)
Counting the most frequent items in a stream is a very common
query. In fact there has been an ample amount of work in devising
good approximation algorithms. One of the oldest streaming algorithms
for detecting frequent items is the MJRTY algorithm~\cite{Boyer1991}
and its
generalizations~\cite{Demaine:2002:FEI:647912.740658,Karp:2003:SAF:762471.762473,Misra:1982:FRE:867576}.
% It guarantees to find an item that appears in the stream $0<\varepsilon\leq
% 1$ of the time, for a given parameter $\varepsilon$.
% For a stream of $m$ elements coming from
% the domain $D=\{1,\ldots,n\}$,
% The algorithm creates groups of
% $k=\lceil 1/\epsilon\rceil$ different elements and as soon as a group
% is completed (i.e., reaches cardinality $k$) it is discarded. Any
% value with more then $m\epsilon$ occurrences in the stream will be
% present in the last incomplete group.
% the algorithm uses $O(\lceil 1/\varepsilon\rceil \mathsf{log}(m+n))$ bits %
% of memory.

% Optimizing communication
\emph{Optimizing communication cost} is a common problem in the field
of streaming databases. 
Consider $k$ data streams and a monitor that consisting of $k+1$
distributed components ---one for every stream and an additional
central coordinator.
Components are only allowed to send messages to the central
coordinator. 
The goal is to track a (reasonably precise) value of a function
defined over the data in all $k$ streams at the central coordinator,
while minimizing the number of messages sent.
This problem is a good abstraction of many network monitoring tasks
where the goal is to detect global properties of routed data.
The communication cost is the primary measure of
complexity of a tracking algorithm.
Initial work dealt with optimizing the top-$k$ items
query~\cite{Babcock:2003:DTM:872757.872764}; it was then extended to
non-temporal
functions~\cite{Cormode:2011:ADF:1921659.1921667,Woodruff:2012:TBD:2213977.2214063}.
Temporal queries are facilitated by introducing various types of
windows, and the approximation is achieved by maintaining a uniform
sample of events per window at the
coordinator~\cite{Cormode:2011:CDM:2031792.2031793,Cormode:2012:CSD:2160158.2160163}.

\emph{Optimizing execution time} using approximation methods 
involves ignoring parts of the input, predicated on strong statistical
guarantees on precision of the output. This is enabled by sampling
techniques~\cite{Cormode:2012:SMD:2344400.2344401} that are shown to
work for specific queries. These techniques are often referred to as
Approximate Query Processing (AQP) and they are implemented by many
existing 
systems~\cite{Agarwal.2465355,Mozafari:2017:AQE:3035918.3056098,mozafarisnappydata,spoth:2017:cidr:adaptive}.
When sampling, a random sample is a ``representative'' subset
of the data, obtained via some stochastic mechanism. Samples are
quicker to obtain, smaller than the data itself and are hence used to
 answer queries more efficiently.
A histogram summarizes the data by grouping its values into
subsets (or ``buckets'') and then computing a small set
of summary statistics for each bucket.
These statistics allow to approximately reconstruct the data in each
bucket.
Wavelets are techniques by which a dataset is viewed as a set of M 
elements in a vector, i.e., a function defined on the set
$\{0,1,2,\dots,M - 1\}$. Such a function can be seen as a weighted sum of
some carefully chosen wavelet ``basis functions''.
Coefficients that are close to zero in magnitude can then be ignored,
with the remaining small set of coefficients serving as the data
summary.
Sketches are particularly well-suited to streaming data. Linear
sketches view a numerical dataset as a matrix, and multiply the data
by  some fixed matrix. Such sketches are massively
parallelizable and used to successfully estimate answers to
set cardinality, union and sum queries, as well as top-$k$ or min-$k$
queries.

\emph{Optimizing monitoring overhead} is a problem often encountered
in runtime verification.
When optimizing overhead,
one must consider the monitored system in addition to the event stream.
In this setting, computing resources (time, memory, and network) are 
shared by the monitored system and the monitor. Overhead can be seen as
the percentage of the resources used by the monitor.
Bartocci at
al.~\cite{DBLP:conf/rv/BartocciGKSSZS12,KalajdzicBSSG13,Stoller2012} use
dynamic knowledge about the monitored system to control the amount of 
resources that are allocated for monitoring. More precisely they enable
and disable monitoring of certain events as needed. This
can be seen as sampling, however the stochastic mechanism is informed
by the probabilistic model of the monitored system.
Given how likely it is that an event will participate in a violation of
a given temporal property, the system decides to include it in the
monitored stream. The aforementioned approaches all differ in the
probabilistic formalism used to model the monitored
system~\cite{EPTCS124.1}.

\subsubsection{Utility}
Another important dimension is the usefulness (or utility) of the monitoring output.
The expected output of the monitoring problem is often
underspecified and usually different approaches employ different
assumptions derived from the implementation details of the monitoring
algorithms.
Yet, the underlying time and space complexity of the monitoring 
problem highly depends on its precise output specification.

%Single boolean vs a stream of booleans
For instance, some monitoring algorithms output a single Boolean
verdict stating that, overall, the trace satisfies or violates the
monitored property.
Other monitoring algorithms solve a strictly harder 
problem - they output a stream of Boolean verdicts attesting to the
satisfaction of the monitored property for every prefix of the trace
(or stream).
While the complexity of the former variants have been studied for
various specification
languages~\cite{bundala_complexity_2014,DBLP:conf/dlt/FengLQ15,kuhtz_ltl_2009},
the latter have mostly been ignored.
%

% Verdicts vs violations
An interesting distinction to make is between outputting a stream composed only of
violations, {\em versus} giving a (more general) stream of verdicts that
includes satisfactions of the monitored property as well.

% In-order vs out-of-order
Traditional monitoring algorithms for temporal logics with
future operators, scale poorly when subjected to high-velocity event
streams. One reason is that the monitor is constrained to produce outputs
strictly in the order defined by the incoming events. It can be shown
that this ordering constraint, although providing more usable output,
makes for a more complex monitoring problem. 
An interesting special case of monitors producing out-of-order output
are monitors that output violations as soon as possible, i.e., as soon
as they have enough information from the input to pinpoint some
violation. Monitors that produce ordered output violate this seemingly
natural monitoring requirement.

% All vs some violations
Orthogonally, in contrast to reporting all violations of a property,
there are many valid use cases where monitors report only some (most
relevant) violations. Examples include reporting only the first, or
the last (most recent) violation. However, the impact of these choices on the
monitoring complexity is unclear.

% Boolean vs non-boolean output
It is also possible to design algorithms that produce non-Boolean
verdicts, for example using {\em Stream Runtime
Verification}~\cite{DAngelo05time}, which allows to compute streams
from arbitrary domains.
Other system use verdicts that target specific (potentially relaxed)
output requirements and may or may not contain enough information to
reconstruct the standard Boolean verdict output.
For example, % ,aerial_mdl_2017
Basin et al.~\cite{aerial_mtl_2017} proposed the so-called equivalence verdicts that state that the
monitor does not know the Boolean verdict at a particular point in the
event stream, but it knows that the verdict will be equal to another,
also presently unknown, verdict at a different point. The equivalence
verdicts carry enough information to reconstruct a stream of Boolean
verdicts. To do so, one must reorder
the verdicts reported in the output stream and propagate Boolean
verdicts to the equivalent ones. 
%
% But even without this post-processing, such a nonstandard output is
% useful, while the corresponding monitoring algorithm outperforms
% algorithms that output purely Boolean verdicts.

% More detailed satisfaction/violation explanations
All output variations mentioned so far compromise utility for the
sake of scalability. However, sometimes starting from a stream of
verdicts, it is quite nontrivial to understand why  a complex
property is satisfied (or violated) at some point in the trace.
One can increase the utility of the monitors by
replacing the stream of Boolean verdicts with a stream of proof
objects that encode the explanations as to why property has been
satisfied or violated. The proof objects can take the forms
of minimal-size proof trees~\cite{basinoptimal},
or a compressed
summary trace capturing the essentials of the original trace that
contribute to a violation.

%not sure if one wants to talk about robustness of the
%satisfaction/violation here

% Here are some itemized challenges to be incorporated in the special section:
% \begin{itemize}
% \item To study the complexity of the monitoring algorithms
% outputting streams of Boolean verdicts
% \item Related to the above challenge, study the variants of the
% monitoring problem that outputs out-of-order or non-boolean verdicts 
% \item Adopt a new model of computation suitable for analyzing
%   complexity of online monitoring algorithms (i.e., the algorithms
%   that deal with unbounded input streams), since the current model
%   of computation (a Turing machine) assumes finite input. A starting
%   point can be persistent or interactive Turing machines.
% \item Study possible proof objects that can represent explanations
% \item Develop algorithms that provide forms of explanations for the
%   satisfaction or violation of the monitoring property.
% \end{itemize}

\subsection{Challenges}

\begin{challenge}{Combining Horizontal and Vertical Parallelization}
  \label{ch:huge:combination}
  The different approaches to parallelize monitoring algorithms have
  different advantages and limitations. Horizontal parallelization as
  in Barre et al.~\cite{barre_mapreduce_2012} and Hall\'e and
  Soucy-Boivin~\cite{halle_mapreduce_2015} does not dependent on the
  actual events but is limited by the formula's structure. Vertical
  parallelization as in Basin et al.~\cite{basin_scalable_2016} or
  parametric trace slicing~\cite{chen_parametric_2009,RegerR-RV2015}
  offers an {\em a priori} unbounded amount of parallelization but may lead
  to data duplication for certain formulas. A combination of the
  approaches may achieve the best of both worlds and is worth 
  investigating.
\end{challenge}

\begin{challenge}{Scalable Monitoring in Online Setting}
  \label{ch:huge:online}
  Most of the described approaches rely on MapReduce as a technical
  solution for distributed fault tolerant computation. However,
  its batch-processing nature restricts monitoring to the
  offline setting, in which the complete log of events is given as
  input to the monitor at once. More recently, 
  systems research has moved towards a proper streaming paradigm, as
  witnessed by widely adopted streaming frameworks such as Apache
  Flink~\cite{carbone-flink2015} or Timely
  Dataflow~\cite{mcsherry-naiad2013}. These frameworks can be used to 
  achieve scalability in the online setting, in which individual events
  steadily arrive at the monitor. The challenge thereby is to adapt the
  offline approaches (both horizontal and vertical)  to the online
  setting.
\end{challenge}

\begin{challenge}{Adaptive Scalability}
  \label{ch:huge:adaptive}
  A related challenge that arises only in the online setting is
  adaptivity. To retain scalability, a parallel monitor, and in
  particular its log slicing component, may need to adapt to changes in
  behavior of the monitored system. For example, an 
  event-rate increase or change in the occurrence distribution of some
  system events.
  Detecting such changes and adequately reacting to them are both
  challenging. In particular, the latter will most likely require a
  reshuffling of the parallel monitors' states in a way that maintains
  a consistent global state, that is, it does not compromise the
  soundness of monitoring.
\end{challenge}

\begin{challenge}{Automatically Synthesizing Splitting Strategies}
  Log slicing techniques, like Basin et al.~\cite{basin_scalable_2016}
  rely on a domain expert to supply a splitting strategy. 
  An open challenge is how to synthesize such a splitting strategy
  automatically, based on the monitored property and some formalized
  domain knowledge, for example, statistics on types of events in the
  log.
  The holy grail would be an algorithm that picks the
  \emph{optimal} splitting strategy, i.e., one that minimizes the
  amount of duplicated data between the slices and creates balanced
  slices that require equal computational effort to monitor.
  % A more tractable first step would be to identify
  % the assumptions on the event stream needed to establish the
  % existence of such an optimal splitting strategy for a fixed property
  % specification language.
\end{challenge}

\begin{challenge}{Expressive Specification Languages}
  \label{ch:expressive}
  Richer specification languages allow to capture more sophisticated
  properties. For example, hyperproperties allow to express relational
  properties (essentially properties that relate different traces). 
  These traces can come from a single large trace that is
  processed offline.
  For example, a specification can relate two traces, which are
  extracted from the large trace as requests coming from different
  users or different requests performed at different points in time.
  This richer language would allow to express properties like
  differential SLA that are beyond the expressiveness of the
  specification formalisms currently used.
  Another family of specification languages that allow to express rich
  properties is stream runtime verification languages. 
  Currently, these languages only have online and offline evaluation
  algorithms for small traces, in the sense of traces that can be
  stored in a single computer.
  A challenge is then to come up with parallel algorithms for large
  traces.
\end{challenge}

\begin{challenge}{Richer Verdicts and Concise Model Witnesses}
  \label{ch:verdicts}
  Classical specification formalisms from runtime verification,
  borrowed from behavioral languages used in static verification,
  generate Boolean outcomes from a given trace, which indicate whether
  the trace observed is a model of the specification.
  One challenge is to compute richer outcomes of the monitoring
  process.
  Examples include computing quantitative verdicts, like for example
  how robustly was the specification satisfied or computing statistics
  from the input trace, like the average number of retransmissions or
  the worst-case response time.
  A related challenge is the computation of witnesses of the
  satisfaction or violation of the property for offline traces.
  The main goal is that the monitoring algorithm computes the verdict
  and, as by-product, a compressed summary trace, where irrelevant
  information has been omitted and consolidated.
  Algorithms will have to be created to (1) check that the summary
  trace is indeed a summary of the input trace, and (2) that the
  summary trace has the claimed verdict against the specification.
  This process, if successful, will allow to check fast and
  independently that the runtime verification process was correctly
  performed.
\end{challenge}

\begin{challenge}{Approximate monitoring}
  \label{ch:approximate:monitoring}
The monitoring setting should provide a systematic and explicit way
to specify tradeoffs between the resources the monitoring algorithms may utilize (e.g., maximum memory consumption or running time) and the
precision of the verdicts they provide.
Existing work provides such tradeoffs for a few fixed
monitored properties (usually involving aggregations), however,
support for complete language fragments is an open problem.
\end{challenge}

\begin{challenge}{Impact of utility on monitoring complexity}
  \label{ch:utility:monitoring}
The existing work on the complexity of
monitoring~\cite{bundala_complexity_2014,DBLP:conf/dlt/FengLQ15,kuhtz_ltl_2009}
(called path checking in this context) only considers the problem of
providing a single Boolean verdict in an offline manner. Tight
complexity bounds for the online monitoring problem or other variants
of the problem with different output utility (e.g., a verdict stream) have not yet been established. The impact of the different kinds of
verdicts on
the complexity of the resulting monitoring problem needs to be better understood.
\end{challenge}

%%% Local Variables:
%%% mode: latex
%%% TeX-master: "../main"
%%% End:

\section{Conclusion}
\label{sec:conclusion}

Runtime verification techniques have been traditionally applied to
software in order to monitor programs.
One of the missions of the EU COST Action IC1402 ({\em Runtime Verification
beyond Monitoring}) was to identify application domains where runtime
verification and monitoring could be applied, and describe the
challenges that these domains would entail.
This paper has explored seven selected areas of application, namely,
distributed systems, hybrid systems, hardware monitoring, security and
privacy, transactional systems, contracts and policies and monitoring
large and unreliable traces.
For each of these seven domains, we survey the state-of-the-art
focusing on monitoring techniques in these areas, and finally
presented some of the most important challenges (collecting a total of
\thechallenge{} challenges) to be addressed by the runtime
verification research community in the next years.

\subsection*{Acknowledgements}
This research has been supported by the European ICT COST Action IC1402 ({\it
  Runtime Verification beyond Monitoring (ARVI)}). 
The authors would like to thank Fonenantsoa Maurica and Pablo Picazo-Sanchez for their feedback
on parts of a preliminary version of this document.

\bibliographystyle{abbrv}
\bibliography{refs,%
  1.distributed/refs,%
  2.hybrid/refs,%
  3.hardware/refs,%
  4.security/refs,%
  5.transactions/refs,%
  6.contracts/refs,%
  7.huge/refs%
}

\begin{thebibliography}{100}

\bibitem{AA16lpc}
M.~Abadi and D.~G. Andersen.
\newblock Learning to protect communications with adversarial neural
  cryptography.
\newblock Technical Report CoRR abs/1610.06918, arXiv.org, 2016.

\bibitem{AbbasFSIG13tecs}
H.~Abbas, G.~E. Fainekos, S.~Sankaranarayanan, F.~Ivancic, and A.~Gupta.
\newblock Probabilistic temporal logic falsification of cyber-physical systems.
\newblock {\em ACM Transactions on Embedded Computing Systems},
  12(s2):95:1--95:30, 2013.

\bibitem{AbbasWFJ14acc}
H.~Abbas, A.~Winn, G.~E. Fainekos, and A.~A. Julius.
\newblock Functional gradient descent method for metric temporal logic
  specifications.
\newblock In {\em Proc. of the American Control Conference (ACC'14)}, pages
  2312--2317. IEEE, 2014.

\bibitem{AcetoAFI17:FSTTCS}
L.~Aceto, A.~Achilleos, A.~Francalanza, and A.~Ing{\'{o}}lfsd{\'{o}}ttir.
\newblock Monitoring for silent actions.
\newblock In {\em Proc. of the 37th {IARCS} Annual Conf. on Foundations of
  Software Technology and Theoretical Computer Science, (FSTTCS'17)}, volume~93
  of {\em LIPIcs}, pages 7:1--7:14. Schloss Dagstuhl - Leibniz-Zentrum fuer
  Informatik, 2017.

\bibitem{AcetoAFI18}
L.~Aceto, A.~Achilleos, A.~Francalanza, and A.~Ing{\'{o}}lfsd{\'{o}}ttir.
\newblock A framework for parameterized monitorability.
\newblock In {\em Proc. of the 21st Int'l Conf. on Foundations of Software
  Science and Computation Structures (FOSSACS'18)}, volume 10803 of {\em LNCS},
  pages 203--220. Springer, 2018.

\bibitem{AcetoAFIK17:CIAA}
L.~Aceto, A.~Achilleos, A.~Francalanza, A.~Ing{\'{o}}lfsd{\'{o}}ttir, and
  S.~{\"{O}}. Kjartansson.
\newblock {O}n the complexity of determinizing monitors.
\newblock In {\em Proc. of the 22nd Int'l Conf. on Implementation and
  Application of Automata (CIAA'17)}, volume 10329 of {\em LNCS}, pages 1--13.
  Springer, 2017.

\bibitem{Agarwal.2465355}
S.~Agarwal, B.~Mozafari, A.~Panda, H.~Milner, S.~Madden, and I.~Stoica.
\newblock Blinkdb: Queries with bounded errors and bounded response times on
  very large data.
\newblock In {\em Proc. of the 8th ACM European Conf. on Computer Systems
  (EuroSys '13)}, pages 29--42. ACM, 2013.

\bibitem{APS18sc}
W.~Ahrendt, G.~J. Pace, and G.~Schneider.
\newblock Smart contracts---a killer application for deductive source code
  verification.
\newblock In {\em Festschrift on the Occasion of Arnd Poetzsch-Heffter's 60th
  Birthday (ARND'18)}, 2018.
\newblock To appear.

\bibitem{akazaki2015time}
T.~Akazaki and I.~Hasuo.
\newblock Time robustness in mtl and expressivity in hybrid system
  falsification.
\newblock In {\em Proc. of the 27th Int'l Conf. on Computer Aided Verification
  (CAV'15)}, volume 9207 of {\em LNCS}, pages 356--374. Springer, 2015.

\bibitem{alagar01techniques}
S.~Alagar and S.~Venkatesan.
\newblock Techniques to tackle state explosion in global predicate detection.
\newblock {\em IEEE Transactions on Software Engineering (TSE)},
  27(8):704--714, 2001.

\bibitem{Alberti:2008:VAI:1380572.1380578}
M.~Alberti, F.~Chesani, M.~Gavanelli, E.~Lamma, P.~Mello, and P.~Torroni.
\newblock Verifiable agent interaction in abductive logic programming: The
  sciff framework.
\newblock {\em ACM Transactions on Computational Logic}, 9(4):29:1--29:43, Aug.
  2008.

\bibitem{Althoff2013}
M.~Althoff.
\newblock Reachability analysis of nonlinear systems using conservative
  polynomialization and non-convex sets.
\newblock In {\em Proc. the 16th Int'l Conf. on Hybrid Systems Computation and
  Control (HSCC'13)}, pages 173--182. ACM, 2013.

\bibitem{alur96thebenefits}
R.~Alur, T.~Feder, and T.~A. Henzinger.
\newblock The benefits of relaxing punctuality.
\newblock {\em Journal of the ACM}, 43(1):116--146, 1996.

\bibitem{ACS13fca}
K.~Angelov, J.~J. Camilleri, and G.~Schneider.
\newblock A framework for conflict analysis of normative texts written in
  controlled natural language.
\newblock {\em {Journal of Logic and Algebraic Programming}}, 82(5-7):216--240,
  July-October 2013.

\bibitem{anicic_rule-based_2010}
D.~Anicic, P.~Fodor, S.~Rudolph, R.~St\"uhmer, N.~Stojanovic, and R.~Studer.
\newblock A rule-based language for complex event processing and reasoning.
\newblock In {\em Proc. of the 4th Int'l Conf. on Web Reasoning and Rule
  Systems (RR'10)}, volume 6333 of {\em LNCS}, pages 42--57. Springer, 2010.

\bibitem{annpureddy2011s}
Y.~Annapureddy, C.~Liu, G.~Fainekos, and S.~Sankaranarayanan.
\newblock S-taliro: A tool for temporal logic falsification for hybrid systems.
\newblock In {\em Proc. of the 17th Int'l Conf/ on Tools and Algorithms for the
  Construction and Analysis of Systems (TACAS'11)}, volume 6605 of {\em LNCS},
  pages 254--257. Springer, 2011.

\bibitem{AnnapureddyF10iecon}
Y.~S.~R. Annapureddy and G.~E. Fainekos.
\newblock Ant colonies for temporal logic falsification of hybrid systems.
\newblock In {\em Proc. of the 36th Annual Conf. on IEEE Industrial Electronics
  Society (IECON'10)}, pages 91--96. IEEE, 2010.

\bibitem{ASS17dm}
T.~Antignac, D.~Sands, and G.~Schneider.
\newblock {Data Minimisation: A Language-Based Approach}.
\newblock In {\em Proc. of the 32nd IFIP TC Int'l Conf. on ICT Sys. Security \&
  Privacy Protection (IFIP SEC'17)}, volume 502 of {\em IFIP Advances in
  Information and Communication Technology (AICT)}, pages 442--456. Springer,
  2017.

\bibitem{Arackaparambil:2009:FMW:1577399.1577411}
C.~Arackaparambil, J.~Brody, and A.~Chakrabarti.
\newblock Functional monitoring without monotonicity.
\newblock In {\em Proc. of the 36th Int'l Colloquium on Automata, Languages and
  Programming: Part I (ICALP'09)}, volume 5555 of {\em LNCS}, pages 95--106.
  Springer, 2009.

\bibitem{arasu_cql_2006}
A.~Arasu, S.~Babu, and J.~Widom.
\newblock The {CQL} continuous query language: semantic foundations and query
  execution.
\newblock {\em The VLDB Journal}, 15(2):121--142, 2006.

\bibitem{ARM-CoreSight:13a}
ARM, Cambridge, England.
\newblock {\em ARM CoreSight Architecture Specification}, 2.0 edition, Sept.
  2013.

\bibitem{armbrust:2010:cacm}
M.~Armbrust, A.~Fox, R.~Griffith, A.~D. Joseph, R.~H. Katz, A.~Konwinski,
  G.~Lee, D.~A. Patterson, A.~Rabkin, I.~Stoica, and M.~Zaharia.
\newblock A view of cloud computing.
\newblock {\em Communications of the ACM}, 53(4):50--58, 2010.

\bibitem{ArshadKR16}
S.~Arshad, A.~Kharraz, and W.~Robertson.
\newblock Identifying extension-based ad injection via fine-grained web content
  provenance.
\newblock In {\em Proc. of the 19th Int'l Symp. on Research in Attacks,
  Intrusions, and Defenses (RAID'16)}, volume 9854 of {\em LNCS}, pages
  415--436. Springer, 2016.

\bibitem{asarin02timed}
E.~Asarin, P.~Caspi, and O.~Maler.
\newblock Timed regular expressions.
\newblock {\em Journal of the ACM}, 49(2):172--206, 2002.

\bibitem{AMP95rad}
E.~Asarin, O.~Maler, and A.~Pnueli.
\newblock Reachability analysis of dynamical systems having piecewise-constant
  derivatives.
\newblock {\em Theoretical Computer Science}, 138(1):35--65, 1995.

\bibitem{AMP+07}
E.~Asarin, V.~Mysore, A.~Pnueli, and G.~Schneider.
\newblock Low dimensional hybrid systems - decidable, undecidable, don't know.
\newblock {\em Information and Computation}, 211:138--159, 2012.

\bibitem{ASY01drp}
E.~Asarin, G.~Schneider, and S.~Yovine.
\newblock On the decidability of the reachability problem for planar
  differential inclusions.
\newblock In {\em 4th International Workshop on Hybrid Systems: Computation and
  Control (HSCC'01)}, number 2034 in LNCS, pages 89--104. Springer-Verlag,
  2001.

\bibitem{ASY07tcs1}
E.~Asarin, G.~Schneider, and S.~Yovine.
\newblock {Algorithmic Analysis of Polygonal Hybrid Systems. Part I:
  Reachability}.
\newblock {\em Theoretical Computer Science}, 379(1-2):231--265, 2007.

\bibitem{AttardF16}
D.~P. Attard and A.~Francalanza.
\newblock A monitoring tool for a branching-time logic.
\newblock In {\em Proc. of the 16th Int'l Conf. on Runtime Verification
  (RV'16)}, volume 10012 of {\em {LNCS}}, pages 473--481. Springer, 2016.

\bibitem{AttardF17:SEFM}
D.~P. Attard and A.~Francalanza.
\newblock {T}race partitioning and local monitoring for asynchronous
  components.
\newblock In {\em Proc. of the 15th Int'l Conf. on Software Engineering and
  Formal Methods (SEFM'17)}, volume 10469 of {\em LNCS}, pages 219--235.
  Springer, 2017.

\bibitem{attiya04distributed}
H.~Attiya and J.~L. Welch.
\newblock {\em Distributed computing: fundamentals, simulations and advanced
  topics}.
\newblock Wiley, 2004.

\bibitem{DBLP:conf/post/AtzeiBC17}
N.~Atzei, M.~Bartoletti, and T.~Cimoli.
\newblock A survey of attacks on ethereum smart contracts ({SoK}).
\newblock In {\em Proc. of the 6th Int'l Conf. on Principles of Security and
  Trust (POST'17)}, volume 10204 of {\em LNCS}, pages 164--186. Springer, 2017.

\bibitem{austin10plas}
T.~H. Austin and C.~Flanagan.
\newblock {Permissive Dynamic Information Flow Analysis}.
\newblock In {\em Proc. of the Workshop on Programming Languages and Analysis
  for Security (PLAS'10)}, pages 1--12. ACM, 2010.

\bibitem{austin12popl}
T.~H. Austin and C.~Flanagan.
\newblock Multiple facets for dynamic information flow.
\newblock In {\em Proc. of the 39th ACM SIGPLAN-SIGACT Symp, on Principles of
  Programming Languages (POPL'12)}, pages 165--178. ACM, Jan. 2012.

\bibitem{tssl}
E.~Aydin-Gol, E.~Bartocci, and C.~Belta.
\newblock A formal methods approach to pattern synthesis in reaction diffusion
  systems.
\newblock In {\em Proc. of the the 53rd {IEEE} Conference on Decision and
  Control (CDC'14)}, pages 108--113. IEEE, 2014.

\bibitem{contractlarvatutorial}
S.~Azzopardi, J.~Ellul, and G.~J. Pace.
\newblock Monitoring smart contracts: Contractlarva and open challenges beyond.
\newblock In {\em Proc. of the 18th Int'l Conf. on Runtime Verification
  (RV'18)}, LNCS. Springer, 2018.
\newblock To appear.

\bibitem{APS+16ca}
S.~Azzopardi, G.~J. Pace, F.~Schapachnik, and G.~Schneider.
\newblock {Contract Automata: An Operational View of Contracts Between
  Interactive Parties}.
\newblock {\em Artificial Intelligence and Law}, 24(3):203--243, September
  2016.

\bibitem{Babcock:2002:MID:543613.543615}
B.~Babcock, S.~Babu, M.~Datar, R.~Motwani, and J.~Widom.
\newblock Models and issues in data stream systems.
\newblock In {\em Proc. of the 21st ACM SIGMOD-SIGACT-SIGART Symposium on
  Principles of Database Systems (PODS'02)}, pages 1--16. ACM, 2002.

\bibitem{Babcock:2003:DTM:872757.872764}
B.~Babcock and C.~Olston.
\newblock Distributed top-k monitoring.
\newblock In {\em Proc. of the 2003 ACM SIGMOD Int'l Conf. on Management of
  Data (SIGMOD'03)}, pages 28--39. ACM, 2003.

\bibitem{baker:1977:sigplan}
H.~C. Baker, Jr. and C.~Hewitt.
\newblock The incremental garbage collection of processes.
\newblock {\em SIGPLAN Not.}, 12(8):55--59, Aug. 1977.

\bibitem{barany17tap}
G.~Barany and J.~Signoles.
\newblock {Hybrid Information Flow Analysis for Real-World C Code}.
\newblock In {\em Proc. of the 11th Int'l Conf. on Tests and Proofs (TAP'17)},
  LNCS, pages 23--40. Springer, July 2017.

\bibitem{barre_mapreduce_2012}
B.~Barre, M.~Klein, S.-M. Boivin, P.-A. Ollivier, and S.~Hall\'e.
\newblock {MapReduce} for parallel trace validation of {LTL} properties.
\newblock In {\em Proc. of the 17th Int'l Conf. on Runtime Verification
  (RV'12)}, volume 7687 of {\em LNCS}, pages 184--198. Springer, 2012.

\bibitem{Bartocci2016TNCS}
E.~Bartocci, E.~Aydin-Gol, I.~Haghighi, and C.~Belta.
\newblock A formal methods approach to pattern recognition and synthesis in
  reaction diffusion networks.
\newblock {\em IEEE Transactions on Control of Network Systems}, PP(99):1--12,
  2016.

\bibitem{Bartocci17memocode}
E.~Bartocci, L.~Bortolussi, M.~Loreti, and L.~Nenzi.
\newblock Monitoring mobile and spatially distributed cyber-physical systems.
\newblock In {\em Proc. of the 15th ACM-IEEE International Conference on Formal
  Methods and Models for System Design (MEMOCODE'17)}, pages 146--155. ACM,
  2017.

\bibitem{sstl2}
E.~Bartocci, L.~Bortolussi, D.~Milios, L.~Nenzi, and G.~Sanguinetti.
\newblock Studying emergent behaviours in morphogenesis using signal
  spatio-temporal logic.
\newblock In {\em Proc. of the Fourth Int'l Workshop on Hybrid Systems and
  Biology (HSB'15)}, volume 9271 of {\em LNCS}, pages 156--172. Springer, 2015.

\bibitem{stl-ps}
E.~Bartocci, L.~Bortolussi, and L.~Nenzi.
\newblock A temporal logic approach to modular design of synthetic biological
  circuits.
\newblock In {\em Proc. of the 11th Int'l Conf. on Computational Methods in
  Systems Biology (CMSB'13)}, volume 8130 of {\em LNBI}, pages 164--177.
  Springer, 2013.

\bibitem{eziotcs}
E.~Bartocci, L.~Bortolussi, L.~Nenzi, and G.~Sanguinetti.
\newblock System design of stochastic models using robustness of temporal
  properties.
\newblock {\em Theoretical Computer Science}, 587:3--25, 2015.

\bibitem{Bartocci2009}
E.~Bartocci, F.~Corradini, M.~R.~D. Berardini, E.~Entcheva, S.~A. Smolka, and
  R.~Grosu.
\newblock Modeling and simulation of cardiac tissue using hybrid {I/O}
  automata.
\newblock {\em Theoretical Computer Science}, 410(33-34):3149--3165, 2009.

\bibitem{BartocciDDFMNS18}
E.~Bartocci, J.~V. Deshmukh, A.~Donz{\'{e}}, G.~E. Fainekos, O.~Maler,
  D.~Nickovic, and S.~Sankaranarayanan.
\newblock Specification-based monitoring of cyber-physical systems: {A} survey
  on theory, tools and applications.
\newblock In {\em Lectures on Runtime Verification - Introductory and Advanced
  Topics}, volume 10457 of {\em LNCS}, pages 135--175. Springer, 2018.

\bibitem{DBLP:series/lncs/10457}
E.~Bartocci and Y.~Falcone, editors.
\newblock {\em Lectures on Runtime Verification - Introductory and Advanced
  Topics}, volume 10457 of {\em Lecture Notes in Computer Science}.
\newblock Springer, 2018.

\bibitem{fault-stl}
E.~Bartocci, T.~Ferr\'ere, N.~Manjunath, and D.~Nickovic.
\newblock Localizing faults in {S}imulink/{S}tateflow models with {STL}.
\newblock In {\em Proc. of the 21st ACM Int'l Conf. on Hybrid Systems
  Computation and Control (HSCC'18)}, pages 197--206. ACM, 2018.

\bibitem{EPTCS124.1}
E.~Bartocci and R.~Grosu.
\newblock Monitoring with uncertainty.
\newblock In {\em Proc. 3rd Int'l Workshop on Hybrid Autonomous Systems},
  volume 124 of {\em Theoretical Computer Science}, pages 1--4. Open Publishing
  Association, 2013.

\bibitem{DBLP:conf/rv/BartocciGKSSZS12}
E.~Bartocci, R.~Grosu, A.~Karmarkar, S.~A. Smolka, S.~D. Stoller, E.~Zadok, and
  J.~Seyster.
\newblock Adaptive runtime verification.
\newblock In {\em {RV} 2012}, pages 168--182, 2012.

\bibitem{BartocciL16}
E.~Bartocci and P.~Li{\`{o}}.
\newblock Computational modeling, formal analysis, and tools for systems
  biology.
\newblock {\em PLoS Computational Biology}, 12(1), 2016.

\bibitem{basinoptimal}
D.~Basin, B.~Bhatt, and D.~Traytel.
\newblock Optimal proofs for linear temporal logic on lasso words.
\newblock In {\em Proc. of the 16th Int'l Symp. on Automated Technology for
  Verification and Analysis (ATVA'18)}, volume 11138 of {\em LNCS}. Springer,
  2018.

\bibitem{basin_scalable_2016}
D.~Basin, G.~Caronni, S.~Ereth, M.~Harvan, F.~Klaedtke, and H.~Mantel.
\newblock Scalable offline monitoring of temporal specifications.
\newblock {\em Formal Methods in System Design}, Mar. 2016.

\bibitem{Basin-inconsistencies-2013}
D.~Basin, F.~Klaedtke, S.~Marinovic, and E.~Z{\u{a}}linescu.
\newblock Monitoring compliance policies over incomplete and disagreeing logs.
\newblock In {\em Proc. of the 4th Int'l Conf. on Runtime Verification
  (RV'13)}, volume 8174 of {\em LNCS}, pages 151--167. Springer, 2013.

\bibitem{RV-CuBES2017:MonPoly}
D.~Basin, F.~Klaedtke, and E.~Zalinescu.
\newblock The {MonPoly} monitoring tool.
\newblock In {\em An International Workshop on Competitions, Usability,
  Benchmarks, Evaluation, and Standardisation for Runtime Verification Tools
  (RV-CuBES 2017)}, volume~3 of {\em Kalpa Publications in Computing}, pages
  19--28. EasyChair, 2017.

\bibitem{basin-monitoring-mdl-2017}
D.~Basin, S.~Krsti\'c, and D.~Traytel.
\newblock Almost event-rate indepedent monitoring of metric dynamic logic.
\newblock In {\em Proc. of the 17th Int'l Conf. on Runtime Verification
  (RV'17)}, volume 10548 of {\em LNCS}, pages 85--102. Springer, 2017.

\bibitem{aerial_mtl_2017}
D.~A. Basin, B.~N. Bhatt, and D.~Traytel.
\newblock Almost event-rate independent monitoring of metric temporal logic.
\newblock In {\em Proc. of the 23rd Int'l Conf. on Tools and Algorithms for the
  Construction and Analysis of Systems (TACAS'17): Part {II}}, volume 10206 of
  {\em LNCS}, pages 94--112. Springer, 2017.

\bibitem{basin_scalable_2014}
D.~A. Basin, G.~Caronni, S.~Ereth, M.~Harvan, F.~Klaedtke, and H.~Mantel.
\newblock Scalable offline monitoring.
\newblock In {\em Proc. of 14th Int'l. Conf. on Runtime Verification (RV'14)},
  volume 8734 of {\em {LNCS}}, pages 31--47. Springer, 2014.

\bibitem{basin_monpoly:_2011}
D.~A. Basin, M.~Harvan, F.~Klaedtke, and E.~Zalinescu.
\newblock {MONPOLY}: Monitoring usage-control policies.
\newblock In {\em Proc. of the Second Int'l Conf. on Runtime Verification
  (RV'11)}, volume 7186 of {\em LNCS}, pages 360--364. Springer, 2011.

\bibitem{basin_monitoring_2012}
D.~A. Basin, F.~Klaedtke, S.~Marinovic, and E.~Zalinescu.
\newblock Monitoring compliance policies over incomplete and disagreeing logs.
\newblock In {\em Proc. of the Third Int'l Conf. on Runtime Verification
  (RV'12)}, volume 7687 of {\em LNCS}, pages 151--167. Springer, 2012.

\bibitem{basin_monitoring_2015}
D.~A. Basin, F.~Klaedtke, S.~Marinovic, and E.~Zalinescu.
\newblock Monitoring of temporal first-order properties with aggregations.
\newblock {\em Formal Methods in System Design}, 46(3):262--285, 2015.

\bibitem{basin_monitoring_2015-1}
D.~A. Basin, F.~Klaedtke, S.~Müller, and E.~Zalinescu.
\newblock Monitoring metric first-order temporal properties.
\newblock {\em Journal of the ACM}, 62(2), 2015.

\bibitem{basin15failure}
D.~A. Basin, F.~Klaedtke, and E.~Zalinescu.
\newblock Failure-aware runtime verification of distributed systems.
\newblock In {\em Proc. of the 35th IARCS Annual Conf. on Foundations of
  Software Technology and Theoretical Computer Science (FSTTCS'15)}, volume~45
  of {\em Leibniz International Proceedings in Informatics (LIPIcs)}, pages
  590--603, Dagstuhl, Germany, 2015. Schloss Dagstuhl--Leibniz-Zentrum fuer
  Informatik.

\bibitem{basin-monitoring-out-of-order-2017}
D.~A. Basin, F.~Klaedtke, and E.~Zalinescu.
\newblock Runtime verification of temporal properties over out-of-order data
  streams.
\newblock In {\em Proc. of the 29th Int'l Conf. on Computer Aided Verification
  (CAV'17)}, volume 10426 of {\em LNCS}, pages 356--376. Springer, 2017.

\bibitem{bauer12decentralised}
A.~K. Bauer and Y.~Falcone.
\newblock Decentralised {LTL} monitoring.
\newblock In {\em Proc. of the 18th Int'l Symp. on Formal Methods (FM'12)},
  volume 7436 of {\em LNCS}, pages 85--100. Springer, 2012.

\bibitem{bauer16decentralised}
A.~K. Bauer and Y.~Falcone.
\newblock Decentralised {LTL} monitoring.
\newblock {\em Formal Methods in System Design}, 48(1--2):49--93, 2016.

\bibitem{Bauer:ltl}
A.~K. Bauer, M.~Leucker, and C.~Schallhart.
\newblock Runtime verification for {LTL} and {TLTL}.
\newblock {\em ACM Transactions on Software Engineering Methodology},
  20:14:1--14:64, 2011.

\bibitem{BauerCJPST15}
L.~Bauer, S.~Cai, L.~Jia, T.~Passaro, M.~Stroucken, and Y.~Tian.
\newblock Run-time monitoring and formal analysis of information flows in
  chromium.
\newblock In {\em Proc. of the 22nd Annual Network and Distributed System
  Security Symp. (NDSS'15)}. The Internet Society, 2015.

\bibitem{belta2017formal}
C.~Belta, B.~Yordanov, and E.~A. Gol.
\newblock {\em Formal Methods for Discrete-Time Dynamical Systems}.
\newblock Springer, 2017.

\bibitem{BemporadM99}
A.~Bemporad and M.~Morari.
\newblock Control of systems integrating logic, dynamics, and constraints.
\newblock {\em Automatica}, 35(3):407--427, 1999.

\bibitem{bersani_efficient_2016}
M.~M. Bersani, D.~Bianculli, C.~Ghezzi, S.~Krstic, and P.~S. Pietro.
\newblock Efficient large-scale trace checking using mapreduce.
\newblock In {\em Proc. of the 38th Int'l Conf. on Software Engineering
  (ICSE'16)}, pages 888--898. ACM, 2016.

\bibitem{bessani:2013:usenixatc}
A.~Bessani, M.~Santos, J.~{a}o Felix, N.~Neves, and M.~Correia.
\newblock On the efficiency of durable state machine replication.
\newblock In {\em Proc. of the 2013 USENIX Conf. on Annual Technical Conference
  (ATC'13)}, pages 169--180. USENIX Association, 2013.

\bibitem{besson16csf}
F.~Besson, N.~Bielova, and T.~P. Jensen.
\newblock Hybrid monitoring of attacker knowledge.
\newblock In {\em Proc. of the IEEE 29th Computer Security Foundations
  Symposium (CSF'16)}, pages 225--238. IEEE, June 2016.

\bibitem{Bhargavan:2016go}
K.~Bhargavan, N.~Swamy, S.~Zanella-B{\'e}guelin, A.~Delignat-Lavaud,
  C.~Fournet, A.~Gollamudi, G.~Gonthier, N.~Kobeissi, N.~Kulatova, A.~Rastogi,
  and T.~Sibut-Pinote.
\newblock Formal verification of smart contracts.
\newblock In {\em Proc. of the 2016 ACM Workshop on Programming Languages and
  Analysis for Security (PLAS'16)}, pages 91--96. ACM Press, 2016.

\bibitem{bianculli_trace_2014}
D.~Bianculli, C.~Ghezzi, and S.~Krstic.
\newblock Trace checking of {M}etric {T}emporal {L}ogic with aggregating
  modalities using {MapReduce}.
\newblock In {\em Proc. of the 12th Int'l Conf. on Software Engineering and
  Formal Methods (SEFM'14)}, volume 8702 of {\em LNCS}, pages 144--158.
  Springer, 2014.

\bibitem{bielova16trust}
N.~Bielova and T.~Rezk.
\newblock A taxonomy of information flow monitors.
\newblock In {\em Proc. of the 7th Int'l Conf. on Principles of Security and
  Trust (POST'16)}, volume 9635 of {\em LNCS}, pages 46--67. Springer, Apr.
  2016.

\bibitem{DBLP:conf/isola/BonakdarpourFRT16}
B.~Bonakdarpour, P.~Fraigniaud, S.~Rajsbaum, and C.~Travers.
\newblock Challenges in fault-tolerant distributed runtime verification.
\newblock In {\em Proc. of the 7th Int'l Symp. on Leveraging Applications of
  Formal Methods, Verification and Validation: Foundational Techniques
  (ISoLA'16): Part {II}}, volume 9952 of {\em LNCS}, pages 363--370. Springer,
  2016.

\bibitem{BSS18mh}
B.~Bonakdarpour, C.~S\'anchez, and G.~Schneider.
\newblock {Monitoring Hyperproperties by Combining Static Analysis and Runtime
  Verification}.
\newblock In {\em Proc. of the 8th Int'l Symp. on Leveraging Applications of
  Formal Methods, Verification and Validation (ISoLA'18); Track: A Broader View
  on Verification: From Static to Runtime and Back}, LNCS. Springer, 2018.
\newblock To appear.

\bibitem{Bortolussi2015}
L.~Bortolussi, D.~Milios, and S.~Guido.
\newblock {U-Check}: Model checking and parameter synthesis under uncertainty.
\newblock In {\em Proc. of the 12th Int'l. Conf. on Quantitative Evaluation of
  Systems (QEST'15)}, volume 9259 of {\em LNCS}, pages 89--104. Springer, 2015.

\bibitem{Boyer1991}
R.~S. Boyer and J.~S. Moore.
\newblock {\em Automated Reasoning: Essays in Honor of Woody Bledsoe}, chapter
  MJRTY---A Fast Majority Vote Algorithm, pages 105--117.
\newblock Springer, 1991.

\bibitem{buiras15icfp}
P.~Buiras, D.~Vytiniotis, and A.~Russo.
\newblock Hlio: Mixing static and dynamic typing for information-flow control
  in haskell.
\newblock In {\em Proc. of the 20th ACM SIGPLAN Int'l Conf. on Functional
  Programming (ICFP'15)}, pages 289--301. ACM, 2015.

\bibitem{bundala_complexity_2014}
D.~Bundala and J.~Ouaknine.
\newblock On the complexity of temporal-logic path checking.
\newblock In {\em Proc. of 41st Int'l Colloquium on Automata, Languages, and
  Programming (ICALP'14). Part II}, volume 8573 of {\em LNCS}, pages 86--97.
  Springer, 2014.

\bibitem{DBLP:conf/cnl/BunzliH10}
A.~B{\"{u}}nzli and S.~H{\"{o}}fler.
\newblock Controlling ambiguities in legislative language.
\newblock In {\em Proc. of the Int'l Workshop on Controlled Natural Language
  (CNL'10)}, volume 7175 of {\em LNCS}, pages 21--42. Springer, 2010.

\bibitem{burrows:2006:osdi}
M.~Burrows.
\newblock The chubby lock service for loosely-coupled distributed systems.
\newblock In {\em Proc. of the 7th Symp. on Operating Systems Design and
  Implementation (OSDI'06)}, pages 335--350. USENIX Association, 2006.

\bibitem{Buterin}
V.~Buterin.
\newblock {A Next Generation Smart Contract \& Decentralized Application
  Platform}.
\newblock Ethereum White Paper. Available online from
  \url{https://github.com/ethereum/wiki/wiki/White-Paper}, 2017.

\bibitem{CRS18wbt}
J.~J. Camilleri, M.~R. Haghshenas, and G.~Schneider.
\newblock {A Web-Based Tool for Analysing Normative Documents in English}.
\newblock In {\em Proc. of the the 33rd ACM/SIGAPP Symposium On Applied
  Computing--Software Verification and Testing track (SAC-SVT'18)}. ACM, 2018.

\bibitem{DBLP:conf/cnl/CamilleriPR10}
J.~J. Camilleri, G.~J. Pace, and M.~Rosner.
\newblock Controlled natural language in a game for legal assistance.
\newblock In {\em Proc. of the 2nd Int'l Workshop on Controlled Natural
  Language (CNL'10)}, volume 7175 of {\em LNCS}, pages 137--153. Springer,
  2010.

\bibitem{CPS14cnl}
J.~J. Camilleri, G.~Paganelli, and G.~Schneider.
\newblock A cnl for contract-oriented diagrams.
\newblock In {\em Proc. of the 4th Int'l Workshop on Controlled Natural
  Language (CNL'14)}, volume 8625 of {\em LNCS}, pages 135--146. Springer,
  2014.

\bibitem{carbone-flink2015}
P.~Carbone, A.~Katsifodimos, S.~Ewen, V.~Markl, S.~Haridi, and K.~Tzoumas.
\newblock Apache flink{\texttrademark}: Stream and batch processing in a single
  engine.
\newblock {\em {IEEE} Data Engineering Bulleting}, 38(4):28--38, 2015.

\bibitem{detecterRV15}
I.~Cassar and A.~Francalanza.
\newblock {R}untime {A}daptation for {A}ctor {S}ystems.
\newblock In {\em Proc. of the 6th Int'l Conf. on Runtime Verification
  (RV'15)}, volume 9333 of {\em LNCS}, pages 38--54. Springer, 2015.

\bibitem{cassar16implementing}
I.~Cassar and A.~Francalanza.
\newblock On implementing a monitor-oriented programming framework for actor
  systems.
\newblock In {\em Proc. of the 12th Int'l Conf. on Integrated Formal Methods
  {iFM'16}}, volume 9681 of {\em LNCS}, pages 176--192, Berlin, 2016. Springer.

\bibitem{CasFS:15}
I.~Cassar, A.~Francalanza, and S.~Said.
\newblock Improving runtime overheads for detecter.
\newblock In {\em Proc. of the 12th Int'l Workshop on Formal Engineering
  approaches to Software Components and Architectures (FESCA'15)}, volume 178
  of {\em EPTCS}, pages 1--8, 2015.

\bibitem{castro:2002:acmtcs}
M.~Castro and B.~Liskov.
\newblock Practical byzantine fault tolerance and proactive recovery.
\newblock {\em ACM Transactions on Computing Systems}, 20(4):398--461, Nov.
  2002.

\bibitem{cavoukian2009}
A.~Cavoukian.
\newblock Privacy by design: The 7 foundational principles.
\newblock {\em Information and Privacy Commissioner of Ontario, Canada}, 2009.

\bibitem{tffainekos}
A.~Chakarov, S.~Sankaranarayanan, and G.~E. Fainekos.
\newblock Combining time and frequency domain specifications for periodic
  signals.
\newblock In {\em Proc. of the 2nd Int'l Conf. on Runtime Verification
  (RV'11)}, volume 7186 of {\em LNCS}, pages 294--309. Springer, 2011.

\bibitem{Chang:92:ALP}
E.~Chang, Z.~Manna, and A.~Pnueli.
\newblock Characterization of temporal property classes.
\newblock In {\em Proc. of the 19th Int'l Colloquium on Automata, Languages and
  Programming (ICALP'92)}, volume 623 of {\em LNCS}, pages 474--486. Springer,
  1992.

\bibitem{chauhan13distributed}
H.~Chauhan, V.~K. Garg, A.~Natarajan, and N.~Mittal.
\newblock A distributed abstraction algorithm for online predicate detection.
\newblock In {\em Proc. of the 32nd {IEEE} Symp. on Reliable Distributed
  Systems, (SRDS'13)}, pages 101--110. IEEE, 2013.

\bibitem{chen_parametric_2009}
F.~Chen and G.~Roşu.
\newblock Parametric trace slicing and monitoring.
\newblock In {\em Proc. of the 15th Int'l Conf. on Tools and Algorithms for the
  Construction and Analysis of Systems (TACAS'09)}, volume 5505 of {\em
  {LNCS}}, pages 246--261. Springer, 2009.

\bibitem{chen2013flow}
X.~Chen, E.~{\'A}brah{\'a}m, and S.~Sankaranarayanan.
\newblock Flow*: An analyzer for non-linear hybrid systems.
\newblock In {\em Proc. of the 25th Int'l Conf. on Computer Aided Verification
  (CAV'13)}, volume 8044 of {\em LNCS}, pages 258--263. Springer, 2013.

\bibitem{clarkson10jcs}
M.~R. Clarkson and F.~B. Schneider.
\newblock Hyperproperties.
\newblock {\em Journal of Computer Security}, 18(6):1157--1210, Sept. 2010.

\bibitem{contractlarvaisola}
C.~Colombo, J.~Ellul, and G.~J. Pace.
\newblock Contracts over smart contracts: Recovering from violations
  dynamically.
\newblock In {\em Proc. of the 8th Int'l Symp. on Leveraging Applications of
  Formal Methods, Verification and Validation (ISoLA'18)}, LNCS. Springer,
  2018.

\bibitem{colombo16organising}
C.~Colombo and Y.~Falcone.
\newblock Organising {LTL} monitors over distributed systems with a global
  clock.
\newblock {\em Formal Methods in System Design}, 49(1--2):109--158, 2016.

\bibitem{polyLarva-CFMP12}
C.~Colombo, A.~Francalanza, R.~Mizzi, and G.~J. Pace.
\newblock poly{L}arva: Runtime verification with configurable resource-aware
  monitoring boundaries.
\newblock In {\em Proc. of the 10th Int'l Conf. on Software Engineering and
  Formal Methods (SEFM'12)}, volume 7504 of {\em LNCS}, pages 218--232.
  Springer, 2012.

\bibitem{colombo13monitor}
C.~Colombo and G.~J. Pace.
\newblock Monitor-oriented compensation programming through compensating
  automata.
\newblock {\em ECEASST}, 58:1--12, 2013.

\bibitem{survey}
C.~Colombo and G.~J. Pace.
\newblock Recovery within long running transactions.
\newblock {\em ACM Computing Surveys}, 45, 2013.

\bibitem{colombo14comprehensive}
C.~Colombo and G.~J. Pace.
\newblock Comprehensive monitor-oriented compensation programming.
\newblock In {\em Proc. of the 11th Int'l Workshop on Formal Engineering
  approaches to Software Components and Architectures (FESCA'14)}, volume 147
  of {\em EPTCS}, page 4761. Open Publishing Association, 2014.

\bibitem{colombo12safer}
C.~Colombo, G.~J. Pace, and P.~Abela.
\newblock Safer asynchronous runtime monitoring using compensations.
\newblock {\em Formal Methods in System Design}, 41(3):269--294, 2012.

\bibitem{cooper91consistent}
R.~Cooper and K.~Marzullo.
\newblock Consistent detection of global predicates.
\newblock In {\em Proc. of the ACM/ONR Workshop on Parallel and Distributed
  Debugging (PADD '91)}, pages 163--173. ACM, 1991.

\bibitem{Cormode:2011:CDM:2031792.2031793}
G.~Cormode.
\newblock Continuous distributed monitoring: A short survey.
\newblock In {\em Proc. of the First Int'l Workshop on Algorithms and Models
  for Distributed Event Processing (AlMoDEP'11)}, volume 585 of {\em ACM
  International Conference Proceeding Series}, pages 1--10, New York, NY, USA,
  2011. ACM.

\bibitem{Cormode:2005:SST:1083592.1083598}
G.~Cormode and M.~Garofalakis.
\newblock Sketching streams through the net: Distributed approximate query
  tracking.
\newblock In {\em Proc. of the 31st Int'l Conf. on Very Large Data Bases
  (VLDB'05)}, pages 13--24. VLDB Endowment, 2005.

\bibitem{Cormode:2012:SMD:2344400.2344401}
G.~Cormode, M.~Garofalakis, P.~J. Haas, and C.~Jermaine.
\newblock Synopses for massive data: Samples, histograms, wavelets, sketches.
\newblock {\em Foundations and Trends in Databases}, 4(1--3):1--294, 2012.

\bibitem{Cormode:2005:HAN:1066157.1066161}
G.~Cormode, M.~Garofalakis, S.~Muthukrishnan, and R.~Rastogi.
\newblock Holistic aggregates in a networked world: Distributed tracking of
  approximate quantiles.
\newblock In {\em Proc. of the 2005 ACM SIGMOD Int'l Conf. on Management of
  Data (SIGMOD'05)}, pages 25--36, New York, NY, USA, 2005. ACM.

\bibitem{Cormode:2011:ADF:1921659.1921667}
G.~Cormode, S.~Muthukrishnan, and K.~Yi.
\newblock Algorithms for distributed functional monitoring.
\newblock {\em ACM Transactions on Algorithms}, 7(2):21:1--21:20, 2011.

\bibitem{Cormode:2012:CSD:2160158.2160163}
G.~Cormode, S.~Muthukrishnan, K.~Yi, and Q.~Zhang.
\newblock Continuous sampling from distributed streams.
\newblock {\em Journal of the ACM}, 59(2):10:1--10:25, 2012.

\bibitem{coulouris11distributed}
G.~Coulouris.
\newblock {\em Distributed Systems: Concepts and Design}.
\newblock Addison-Wesley, 2011.

\bibitem{ddm04}
T.~Dang, A.~Donz{\'e}, and O.~Maler.
\newblock Verification of analog and mixed-signal circuits using hybrid system
  techniques.
\newblock In {\em Proc. of the 5th Int'l Conf. on Formal Methods in
  Computer-Aided Design (FMCAD'04)}, volume 3312 of {\em LNCS}, pages 21--36.
  Springer, 2004.

\bibitem{DangGM09}
T.~Dang, C.~{Le~{G}uernic}, and O.~Maler.
\newblock Computing reachable states for nonlinear biological models.
\newblock In {\em Proc. of the 7th Int'l Conf. on Computational Methods in
  Systems Biology (CMSB'09)}, volume 5688 of {\em LNCS}, pages 126--141.
  Springer, 2009.

\bibitem{DAngelo05time}
B.~D'Angelo, S.~Sankaranarayanan, C.~Sanchez, W.~Robinson, B.~Finkbeiner,
  H.~Sipma, S.~Mehrotra, and Z.~Manna.
\newblock {LOLA}: runtime monitoring of synchronous systems.
\newblock In {\em Proc. of the 12th International Symposium on Temporal
  Representation and Reasoning (TIME'05)}, pages 166--174. IEEE, 2005.

\bibitem{dangelo05lola}
B.~D'Angelo, S.~Sankaranarayanan, C.~S\'anchez, W.~Robinson, B.~Finkbeiner,
  H.~B. Sipma, S.~Mehrotra, and Z.~Manna.
\newblock {LOLA}: Runtime monitoring of synchronous systems.
\newblock In {\em Proc. of the 12th Int'l Symp. of Temporal Representation and
  Reasoning (TIME'05)}, pages 166--174. IEEE CS Press, 2005.

\bibitem{Demaine:2002:FEI:647912.740658}
E.~D. Demaine, A.~L\'{o}pez-Ortiz, and J.~I. Munro.
\newblock Frequency estimation of internet packet streams with limited space.
\newblock In {\em Proc. of the 10th Annual European Symp. on Algorithms
  (ESA'02)}, volume 2461 of {\em LNCS}, pages 348--360. Springer, 2002.

\bibitem{denning76acm}
D.~E.~R. Denning.
\newblock A lattice model of secure information flow.
\newblock {\em Communications of the ACM}, 19(5):236--243, 1976.

\bibitem{denning77acm}
D.~E.~R. Denning and P.~J. Denning.
\newblock Certification of programs for secure information flow.
\newblock {\em Communications of the ACM}, 20(7):504--513, 1977.

\bibitem{deon}
{DEON Conferences}.
\newblock \url{https://deon2018.sites.uu.nl/}.

\bibitem{online_journal}
J.~V. Deshmukh, A.~Donz{\'{e}}, S.~Ghosh, X.~Jin, J.~Garvit, and S.~A. Seshia.
\newblock Robust online monitoring of signal temporal logic.
\newblock {\em Formal Methods in System Design}, 1/2017, 2017.

\bibitem{devriese10sp}
D.~Devriese and F.~Piessens.
\newblock Noninterference through secure multi-execution.
\newblock In {\em Proc. of the 31st IEEE Symposium on Security and Privacy
  (SP'10)}, pages 109--124. IEEE, 2010.

\bibitem{hvc:2013:dias}
R.~J. Dias, V.~Pessanha, and J.~M. Lourenço.
\newblock Precise detection of atomicity violations.
\newblock In {\em Hardware and Software: Verification and Testing}, volume 7857
  of {\em LNCS}, pages 8--23. Springer, Nov. 2013.
\newblock HVC 2012 Best Paper Award.

\bibitem{dias:2012:ecoop}
R.~J. D.~D. Distefano, J.~C. Seco, and J.~M. Lourenço.
\newblock Verification of snapshot isolation in transactional memory java
  programs.
\newblock In J.~Noble, editor, {\em Proc. of the 26th European Conf. on
  Object-Oriented Programming (ECOOP'12)}, volume 7313 of {\em LNCS}, pages
  640--664. Springer, 2012.

\bibitem{dokhanchi14online}
A.~Dokhanchi, B.~Hoxha, and G.~E. Fainekos.
\newblock On-line monitoring for temporal logic robustness.
\newblock In {\em Proc. of the 5th Int'l Conf. on Runtime Verification
  (RV'14)}, LNCS, pages 231--246. Springer, 2014.

\bibitem{breach}
A.~Donz{\'{e}}.
\newblock {B}reach, a toolbox for verification and parameter synthesis of
  hybrid systems.
\newblock In {\em Proc. of the 22nd Int'l Conf. on Computer Aided Verification
  (CAV'10)}, volume 6174 of {\em LNCS}, pages 167--170. Springer, 2010.

\bibitem{plos}
A.~Donz{\'{e}}, E.~Fanchon, L.~M. Gattepaille, O.~Maler, and P.~Tracqui.
\newblock Robustness analysis and behavior discrimination in enzymatic reaction
  networks.
\newblock {\em PLoS ONE}, 6(9):e24246, 09 2011.

\bibitem{donze13efficient}
A.~Donz\'{e}, T.~Ferr\`{e}re, and O.~Maler.
\newblock Efficient robust monitoring for {STL}.
\newblock In {\em Proc. of the 25th Int'l Conf. on Computer Aided Verification
  (CAV'13)}, volume 8044 of {\em LNCS}, pages 264--279. Springer, 2013.

\bibitem{DonzeKR09hscc}
A.~Donz{\'{e}}, B.~Krogh, and A.~Rajhans.
\newblock Parameter synthesis for hybrid systems with an application to
  simulink models.
\newblock In {\em Proc. of the 12th Int'l Conf. on Hybrid Systems: Computation
  and Control (HSCC'09)}, volume 5469 of {\em LNCS}, pages 165--179. Springer,
  2009.

\bibitem{donze2010robust}
A.~Donz{\'e} and O.~Maler.
\newblock Robust satisfaction of temporal logic over real-valued signals.
\newblock In {\em Proc. of the 8th Int'l Conf. on Formal Modeling and Analysis
  of Timed Systems (FORMATS'10)}, volume 6246 of {\em LNCS}, pages 92--106.
  Springer, 2010.

\bibitem{ftl}
A.~Donz{\'{e}}, O.~Maler, E.~Bartocci, D.~Ni\v{c}kovi\'{c}, R.~Grosu, and S.~A.
  Smolka.
\newblock On temporal logic and signal processing.
\newblock In {\em Proc. of the 10th Int'l Symp. on Automated Technology for
  Verification and Analysis (ATVA'12)}, volume 7561 of {\em LNCS}, pages
  92--106. Springer, 2012.

\bibitem{DreossiDS17}
T.~Dreossi, A.~Donz{\'{e}}, and S.~A. Seshia.
\newblock Compositional falsification of cyber-physical systems with machine
  learning components.
\newblock In {\em Proc. of the 9th Int'l Symp. on {NASA} Formal Methods
  (NFM'17)}, volume 10227 of {\em LNCS}, pages 357--372. Springer, 2017.

\bibitem{eisner_practical_2006}
C.~Eisner and D.~Fisman.
\newblock {\em A Practical Introduction to PSL}.
\newblock Series on Integrated Circuits and Systems. Springer, 2006.

\bibitem{El-HokayemF17a}
A.~El-Hokayem and Y.~Falcone.
\newblock Monitoring decentralized specifications.
\newblock In {\em Proc. of the 26th ACM SIGSOFT Int'l Symp. on Software Testing
  and Analysis (ISSTA'17)}, pages 125--135. ACM, 2017.

\bibitem{corr/abs-1808-02692}
A.~El{-}Hokayem and Y.~Falcone.
\newblock On the monitoring of decentralized specifications semantics,
  properties, analysis, and simulation.
\newblock {\em CoRR}, abs/1808.02692, 2018.

\bibitem{contractlarva}
J.~Ellul and G.~J. Pace.
\newblock Runtime verification of ethereum smart contracts.
\newblock In {\em Proc. of the 1st Int'l Workshop on Blockchain Dependability,
  in conjunction with EDCC'18}, IEEE, 2018.

\bibitem{elnikety:2005:srds}
S.~Elnikety, W.~Zwaenepoel, and F.~Pedone.
\newblock Database replication using generalized snapshot isolation.
\newblock In {\em Proc. of the 24th IEEE Symp. on Reliable Distributed Systems
  (SRDS'05)}, pages 73--84. IEEE Computer Society, 2005.

\bibitem{FainekosG03}
G.~E. Fainekos and K.~C. Giannakoglou.
\newblock Inverse design of airfoils based on a novel formulation of the ant
  colony optimization method.
\newblock {\em Inverse Problems in Engineering}, 11(1):21--38, 2003.

\bibitem{FainekosGP06formats}
G.~E. Fainekos, A.~Girard, and G.~J. Pappas.
\newblock Temporal logic verification using simulation.
\newblock In {\em Proc. of the 4th Int'l Conf. on Formal Modelling and Analysis
  of Timed Systems (FORMATS'06)}, volume 4202 of {\em LNCS}, pages 171--186.
  Springer, 2006.

\bibitem{fainekos09robustness}
G.~E. Fainekos and G.~J. Pappas.
\newblock Robustness of temporal logic specifications for continuous-time
  signals.
\newblock {\em Theoretical Computer Science}, 410(42):4279--4291, 2009.

\bibitem{falcone14efficient}
Y.~Falcone, T.~Cornebize, and J.~Fernandez.
\newblock Efficient and generalized decentralized monitoring of regular
  languages.
\newblock In {\em Proc. of 34th IFIP Int'l Conf. on Formal Techniques for
  Distributed Objects, Components, and Systems (FORTE'14)}, volume 8461 of {\em
  LNCS}, pages 66--83. Springer, 2014.

\bibitem{FalconeFM12:STTT}
Y.~Falcone, J.~Fernandez, and L.~Mounier.
\newblock What can you verify and enforce at runtime?
\newblock {\em Int'l Journal on Software Tools for Technology Transfer (STTT)},
  14(3):349--382, 2012.

\bibitem{falcone15runtime}
Y.~Falcone, M.~Jaber, T.-H. Nguyen, M.~Bozga, and S.~Bensalem.
\newblock Runtime verification of component-based systems in the {BIP}
  framework with formally-proved sound and complete instrumentation.
\newblock {\em Software and System Modeling}, 14(1):173--199, 2015.

\bibitem{faymonville_parametric_2014}
P.~Faymonville and M.~Zimmermann.
\newblock Parametric linear dynamic logic.
\newblock In {\em Proc. of the 5th Int'l Symp. on Games, Automata, Logics and
  Formal Verification (GandALF'14)}, volume 161 of {\em EPTCS}, pages 60--73,
  2014.

\bibitem{FOP+08bsc}
S.~Fenech, J.~Okika, G.~J. Pace, A.~P. Ravn, and G.~Schneider.
\newblock On the specification of full contracts.
\newblock {\em Proc. of the 6th Int'l Workshop on Formal Engineering Approaches
  to Software Components and Architectures (FESCA'09)}, 253(1):39--55, Mar.
  2009.

\bibitem{DBLP:conf/dlt/FengLQ15}
S.~Feng, M.~Lohrey, and K.~Quaas.
\newblock Path checking for {MTL} and {TPTL} over data words.
\newblock In I.~Potapov, editor, {\em Proc. of the 19th Int'l Conf. on
  Developments in Language Theory (DLT'15)}, volume 9168 of {\em LNCS}, pages
  326--339. Springer, 2015.

\bibitem{FerrereMN15}
T.~Ferr{\`{e}}re, O.~Maler, and D.~Nickovic.
\newblock Trace diagnostics using temporal implicants.
\newblock In {\em Proc. of the 13th International Symposium on Automated
  Technology for Verification and Analysis (ATVA'15)}, volume 9364 of {\em
  LNCS}, pages 241--258. Springer, 2015.

\bibitem{measures}
T.~Ferr\`{e}re, O.~Maler, D.~Ni\v{c}kovi\'{c}, and D.~Ulus.
\newblock Measuring with timed patterns.
\newblock In {\em Proc. of the 27th Int'l Conf. on Computer Aided Verification
  (CAV'15)}, volume 9207 of {\em LNCS}, pages 322--337. Springer, 2015.

\bibitem{Flajolet07hyperloglog:the}
P.~Flajolet, E.~Fusy, O.~Gandouet, and F.~Meunier.
\newblock Hyperloglog: The analysis of a near-optimal cardinality estimation
  algorithm.
\newblock In {\em Proc. of the 2007 Int'l Conf. on Analysis of Algorithms
  (AOFA'07)}. DMTCS, 2007.

\bibitem{fraigniaud14onthenumber}
P.~Fraigniaud, S.~Rajsbaum, and C.~Travers.
\newblock On the number of opinions needed for fault-tolerant run-time
  monitoring in distributed systems.
\newblock In {\em Proc. of the 5th Int'l Conf. on Runtime Verification
  (RV'14)}, volume 8734 of {\em LNCS}, pages 92--107. Springer, 2014.

\bibitem{Fra16:Fossacs}
A.~Francalanza.
\newblock A theory of monitors - (extended abstract).
\newblock In {\em Proc. of the 19th Int'l Conf. on Foundations of Software
  Science and Computation Structures (FOSSACS'16)}, volume 9634 of {\em LNCS},
  pages 145--161. Springer, 2016.

\bibitem{Fra17:Concur}
A.~Francalanza.
\newblock Consistently-detecting monitors.
\newblock In R.~Meyer and U.~Nestmann, editors, {\em Proc. of the 28th Int'l
  Conf. on Concurrency Theory (CONCUR'17)}, volume~85 of {\em Leibniz
  International Proceedings in Informatics (LIPIcs)}, pages 8:1--8:19,
  Dagstuhl, Germany, 2017. Schloss Dagstuhl--Leibniz-Zentrum fuer Informatik.

\bibitem{FraAI17:FMSD}
A.~Francalanza, L.~Aceto, and A.~Ingolfsdottir.
\newblock {M}onitorability for the {H}ennessy--{M}ilner {L}ogic with
  {R}ecursion.
\newblock {\em Formal Methods in System Design}, pages 1--30, 2017.

\bibitem{francalanza13distributed}
A.~Francalanza, A.~Gauci, and G.~J. Pace.
\newblock Distributed system contract monitoring.
\newblock {\em Journal of Logic and Algebraic Programming}, 82(5--7):186--215,
  2013.

\bibitem{FrancalanzaH07:JLP}
A.~Francalanza and M.~Hennessy.
\newblock A theory for observational fault tolerance.
\newblock {\em Journal of Logic and Algebraic Programmming}, 73(1-2):22--50,
  2007.

\bibitem{FH08:InfComp}
A.~Francalanza and M.~Hennessy.
\newblock A {T}heory of {S}ystem {B}ehaviour in the presence of {N}ode and
  {L}ink failure.
\newblock {\em Information and Computation}, 206(6):711 -- 759, 2008.

\bibitem{FraMezTuo18:DAIS}
A.~Francalanza, C.~A. Mezzina, and E.~Tuosto.
\newblock Reversible choreographies via monitoring in erlang.
\newblock In {\em Proc. of the 18th {IFIP} {WG} 6.1 Int'l Conf. on Distributed
  Applications and Interoperable Systems (DAIS'18)}, volume 10853 of {\em
  LNCS}, pages 75--92. Springer, 2018.

\bibitem{FrancalanzaPS18}
A.~Francalanza, J.~A. P{\'{e}}rez, and C.~S{\'{a}}nchez.
\newblock Runtime verification for decentralised and distributed systems.
\newblock In E.~Bartocci and Y.~Falcone, editors, {\em Lectures on Runtime
  Verification - Introductory and Advanced Topics}, volume 10457 of {\em
  Lecture Notes in Computer Science}, pages 176--210. Springer, 2018.

\bibitem{Fra:Sey:13}
A.~Francalanza and A.~Seychell.
\newblock Synthesising correct concurrent runtime monitors (extended abstract).
\newblock In {\em Proc. of the 4th Int'l Conf. on Runtime Verification
  (RV'13)}, volume 8174 of {\em LNCS}, pages 112--129. Springer, 2013.

\bibitem{FraSey15}
A.~Francalanza and A.~Seychell.
\newblock Synthesising {C}orrect concurrent {R}untime {M}onitors.
\newblock {\em Formal Methods in System Design}, 46(3):226--261, 2015.

\bibitem{FranzleHTRS07}
M.~Fr{\"{a}}nzle, C.~Herde, T.~Teige, S.~Ratschan, and T.~Schubert.
\newblock Efficient solving of large non-linear arithmetic constraint systems
  with complex boolean structure.
\newblock {\em Journal on Satisfiability, Boolean Modeling and Computation},
  1(3-4):209--236, 2007.

\bibitem{Fre08}
G.~Frehse.
\newblock Phaver: algorithmic verification of hybrid systems past hytech.
\newblock {\em International Journal on Software Tools for Technology
  Transfer}, 10(3):263--279, 2008.

\bibitem{helicopter}
G.~Frehse, C.~{Le~{G}uernic}, A.~Donz{\'e}, S.~Cotton, R.~Ray, O.~Lebeltel,
  R.~Ripado, A.~Girard, T.~Dang, and O.~Maler.
\newblock Space{E}x: Scalable verification of hybrid systems.
\newblock In {\em Proc. of the 23rd Int'l Conf. on Computer Aided Verification
  (CAV'11)}, volume 6806 of {\em LNCS}, pages 379--395. Springer, 2011.

\bibitem{DBLP:journals/iee/Fuchs92}
N.~E. Fuchs.
\newblock Specifications are (preferably) executable.
\newblock {\em Software Engineering Journal}, 7(5):323--334, 1992.

\bibitem{Gait:86a}
J.~Gait.
\newblock A probe effect in concurrent programs.
\newblock {\em Software - Practise and Experience}, 16(3):225–233, Mar. 1986.

\bibitem{GGV17nrl}
F.~Gandon, G.~Governatori, and S.~Villata.
\newblock Normative requirements as linked data.
\newblock In {\em Proc. of the 30th Annual Conference on Legal Knowledge and
  Information Systems (JURIX'17)}, volume 302 of {\em Frontiers in Artificial
  Intelligence and Applications}, pages 1--10. IOS Press, 2017.

\bibitem{garcia:2011:eurosys}
R.~Garcia, R.~Rodrigues, and N.~Pregui\c{c}a.
\newblock Efficient middleware for byzantine fault tolerant database
  replication.
\newblock In {\em Proc. of the 6th Conf. on Computer Systems (EuroSys'11)},
  pages 107--122. ACM, 2011.

\bibitem{DBLP:conf/bpm/Garcia-Banuelos17}
L.~Garc{\'{\i}}a{-}Ba{\~{n}}uelos, A.~Ponomarev, M.~Dumas, and I.~Weber.
\newblock Optimized execution of business processes on blockchain.
\newblock In {\em Proc. of the 15th Int'l Conf. on Business Process Management
  (BPM'17)}, volume 10445 of {\em LNCS}, pages 130--146. Springer, 2017.

\bibitem{GargJD11}
D.~Garg, L.~Jia, and A.~Datta.
\newblock Policy auditing over incomplete logs: theory, implementation and
  applications.
\newblock In Y.~Chen, G.~Danezis, and V.~Shmatikov, editors, {\em Proc. of the
  18th {ACM} Conf. on Computer and Communications Security, (CCS'11)}, pages
  151--162. ACM, 2011.

\bibitem{garg02elements}
V.~K. Garg.
\newblock {\em Elements of Distributed Computing}.
\newblock Wiley-IEEE Press, 2002.

\bibitem{giacomo_linear_2013}
G.~D. Giacomo and M.~Y. Vardi.
\newblock Linear temporal logic and linear dynamic logic on finite traces.
\newblock In {\em Proc. of the 23rd Int'l Joint Conf. on Artificial
  Intelligence (IJCAI'13)}, pages 854--860. IJCAI/AAAI, 2013.

\bibitem{wired:2011}
S.~Gilbertson.
\newblock https://www.wired.com/2011/04/lessons-amazon-cloud-failure/, 2011.

\bibitem{goguen82ssp}
J.~A. Goguen and J.~Meseguer.
\newblock Security policies and security models.
\newblock {\em IEEE Symposium on Security and Privacy}, pages 11--20, 1982.

\bibitem{DBLP:conf/propertytesting/2010}
O.~Goldreich, editor.
\newblock {\em Property Testing - Current Research and Surveys}, volume 6390 of
  {\em LNCS}. Springer, 2010.

\bibitem{Gouveia:2016a:NIRV}
I.~Gouveia and J.~Rufino.
\newblock Enforcing safety and security through non-intrusive runtime
  verification.
\newblock In {\em Proc. 1st Workshop on Security and Dependability of Critical
  Embedded Real-Time Systems}, pages 19--24, Porto, Portugal, Dec. 2016. IEEE.

\bibitem{DBLP:conf/ruleml/GovernatoriR10}
G.~Governatori and A.~Rotolo.
\newblock Norm compliance in business process modeling.
\newblock In {\em Proc. of the 4th International Web Rule Symposium: Research
  Based and Industry Focused (RuleML'10)}, pages 194--209, 2010.

\bibitem{GrigoreK:18:Concur}
R.~Grigore and S.~Kiefer.
\newblock Selective monitoring.
\newblock In {\em Proc. 29th Int'l Conf. on Concurrency Theory (CONCUR'18)},
  volume 118 of {\em LIPIcs}, pages 20:1--20:16. Schloss Dagstuhl -
  Leibniz-Zentrum fuer Informatik, 2018.

\bibitem{gkr04}
S.~Gupta, B.~H. Krogh, and R.~A. Rutenbar.
\newblock Towards formal verification of analog designs.
\newblock In {\em Proc. of the Int'l Conf. on Computer-Aided Design
  (ICCAD'04)}, pages 210--217. IEEE CS Press, 2004.

\bibitem{gyllstrom_sase:_2006}
D.~Gyllstrom, E.~Wu, H.-J. Chae, Y.~Diao, P.~Stahlberg, and G.~Anderson.
\newblock {SASE}: Complex event processing over streams.
\newblock {\em CoRR}, abs/cs/0612128, 2006.

\bibitem{spatel}
I.~Haghighi, A.~Jones, Z.~Kong, E.~Bartocci, R.~Grosu, and C.~Belta.
\newblock {S}pa{T}e{L}: a novel spatial-temporal logic and its applications to
  networked systems.
\newblock In {\em Proc. of the 18th Int'l Conf. on Hybrid Systems: Computation
  and Control (HSCC'15)}, pages 189--198. IEEE, 2015.

\bibitem{halle_mapreduce_2015}
S.~Hall{\'e} and M.~Soucy-Boivin.
\newblock {MapReduce} for parallel trace validation of {LTL} properties.
\newblock {\em Journal of Cloud Computing}, 4(1):8, 2015.

\bibitem{havelund05verify}
K.~Havelund and A.~Goldberg.
\newblock Verify your runs.
\newblock In {\em Proc. of the First IFIP TC 2/WG 2.3 Conference on Verified
  Software: Theories, Tools, Experiments (VSTTE'05)}, volume 4171 of {\em
  LNCS}, pages 374--383. Springer, 2005.

\bibitem{Hayes:1989:SE:84458.84466}
I.~Hayes and C.~B. Jones.
\newblock Specifications are not (necessarily) executable.
\newblock {\em Software Engineering Journal}, 4(6):330--338, Nov. 1989.

\bibitem{hedin15csf}
D.~Hedin, L.~Bello, and A.~Sabelfeld.
\newblock Value-sensitive hybrid information flow control for a
  {J}avascript-like language.
\newblock In {\em Proc. of the IEEE 28th Computer Security Foundations
  Symposium (CSF'15)}, pages 351--365. IEEE, 2015.

\bibitem{hedin12nato}
D.~Hedin and A.~Sabelfeld.
\newblock A perspective on information-flow control.
\newblock In {\em Software Safety and Security}, volume~33 of {\em {NATO}
  Science for Peace and Security Series - {D:} Information and Communication
  Security}, pages 319--347. IOS Press, 2012.

\bibitem{Henzinger95}
T.~A. Henzinger, P.~W. Kopke, A.~Puri, and P.~Varaiya.
\newblock What's decidable about hybrid automata?
\newblock In {\em Journal of Computer and System Sciences}, pages 373--382. ACM
  Press, 1995.

\bibitem{herlihy:2003:podc}
M.~Herlihy, V.~Luchangco, M.~Moir, and W.~N. {Scherer, III}.
\newblock Software transactional memory for dynamic-sized data structures.
\newblock In {\em Proc. of the 22nd ACM Symp. on Principles of Distributed
  Computing (PODC'03)}, pages 92--101. ACM, 2003.

\bibitem{herlihy:1993:sigarch}
M.~Herlihy and J.~E.~B. Moss.
\newblock Transactional memory: architectural support for lock-free data
  structures.
\newblock {\em SIGARCH Computer Architecture News}, 21(2):289--300, May 1993.

\bibitem{Heule15}
S.~Heule, D.~Rifkin, A.~Russo, and D.~Stefan.
\newblock The most dangerous code in the browser.
\newblock In {\em 15th Workshop on Hot Topics in Operating Systems (HotOS XV)},
  Kartause Ittingen, Switzerland, 2015. USENIX Association.

\bibitem{hunt06popl}
S.~Hunt and D.~Sands.
\newblock On flow-sensitive security types.
\newblock In {\em Proc. of the 33rd ACM SIGPLAN-SIGACT Symp. on Principles of
  Programming Languages (POPL'06)}, pages 79--90. ACM, 2006.

\bibitem{Verilog-Standard:12a}
IEEE.
\newblock {\em IEEE P1800/D6, IEEE Approved Draft Standard for System Verilog -
  Unified Hardware Design, Specification, and Verification Language}, Aug.
  2012.

\bibitem{PSL-Standard:12a}
IEEE Standards Association.
\newblock {\em IEC 62531, IEEE Std 1850 Standard for Property Specification
  Language (PSL)}, 2012.

\bibitem{Intel:16-CFI}
Intel Corporation.
\newblock {\em Control-flow Enforcement Technology Preview}, June 2016.

\bibitem{isola18rsc}
{Reliable Smart Contracts: State-of-the-art, Applications, Challenges and
  Future Directions}.
\newblock \url{http://www.isp.uni-luebeck.de/Isola2018-SmartContracts}.
\newblock ISoLA'18 track
  (\url{http://www.isola-conference.org/isola2018/tracks.html}).

\bibitem{bolosky:2011:nsdi}
W.~J.~B. J., D.~Bradshaw, R.~B. Haagens, N.~P. Kusters, and P.~Li.
\newblock Paxos replicated state machines as the basis of a high-performance
  data store.
\newblock In {\em Proc. of the 8th USENIX Conf. on Networked Systems Design and
  Implementation (NSDI'11)}, pages 141--154. USENIX Association, 2011.

\bibitem{JDF08bdm}
G.~Jacob, H.~Debar, and E.~Filiol.
\newblock Behavioral detection of malware: from a survey towards an established
  taxonomy.
\newblock {\em Journal in Computer Virology}, 4(3):251--266, 2008.

\bibitem{jail}
{Journal of Artificial Intelligence and Law}.
\newblock \url{https://link.springer.com/journal/10506}.

\bibitem{fpga}
S.~Jaksic, E.~Bartocci, R.~Grosu, R.~Kloibhofer, T.~Nguyen, and
  D.~Ni\v{c}kovi\'{c}.
\newblock From signal temporal logic to {FPGA} monitors.
\newblock In {\em Proc. of the the 13th {ACM/IEEE} International Conference on
  Formal Methods and Models for Codesign (MEMOCODE'15)}, pages 218--227.
  {IEEE}, 2015.

\bibitem{wed}
S.~Jaksic, E.~Bartocci, R.~Grosu, and D.~Nickovic.
\newblock Quantitative monitoring of {STL} with edit distance.
\newblock In {\em Proc. of the 16th Int'l Conf. on Runtime Verification
  (RV'16)}, volume 10012 of {\em LNCS}, pages 201--218. Springer, 2016.

\bibitem{abs-1802-03775}
S.~Jaksic, E.~Bartocci, R.~Grosu, and D.~Nickovic.
\newblock An algebraic framework for runtime verification.
\newblock {\em CoRR}, abs/1802.03775, 2018.

\bibitem{DBLP:journals/tocs/JoyceLSU87}
J.~Joyce, G.~Lomow, K.~Slind, and B.~W. Unger.
\newblock Monitoring distributed systems.
\newblock {\em ACM Transactions on Computational Systems}, 5(2):121--150, 1987.

\bibitem{jurix}
{JURIX Conferences}.
\newblock \url{http://jurix.nl}.

\bibitem{KalajdzicBSSG13}
K.~Kalajdzic, E.~Bartocci, S.~A. Smolka, S.~D. Stoller, and R.~Grosu.
\newblock Runtime verification with particle filtering.
\newblock In {\em Proc. of the 4th Int'l Conf. on Runtime Verification
  (RV'13)}, volume 8174 of {\em LNCS}, pages 149--166. Springer, 2013.

\bibitem{Kane:15a}
A.~Kane.
\newblock {\em Runtime Monitoring for Safety-Critical Embedded Systems}.
\newblock PhD thesis, Carnegie Mellon University, USA, Feb. 2015.

\bibitem{Kane:15b}
A.~Kane, O.~Chowdhury., A.~Datta, and P.~Koopman.
\newblock A case study on runtime monitoring of an autonomous research vehicle
  ({ARV}) system.
\newblock In {\em Proc. of the 15th Int. Conf. on Runtime Verification
  (RV'15)}, volume 9333 of {\em LNCS}, pages 102--117. Springer, 2015.

\bibitem{Karp:2003:SAF:762471.762473}
R.~M. Karp, S.~Shenker, and C.~H. Papadimitriou.
\newblock A simple algorithm for finding frequent elements in streams and bags.
\newblock {\em ACM Transactions on Database Systems}, 28(1):51--55, 2003.

\bibitem{Kenny:07a}
J.~R. Kenny and B.~Mackin.
\newblock {FPGA} coprocessing in multi-core architectures for {DSP}.
\newblock Altera Corporation Application Note, Sept. 2007.

\bibitem{Keralapura:2006:CDM:1142473.1142507}
R.~Keralapura, G.~Cormode, and J.~Ramamirtham.
\newblock Communication-efficient distributed monitoring of thresholded counts.
\newblock In {\em Proc. of the 2006 ACM SIGMOD Int'l Conf. on Management of
  Data (SIGMOD'06)}, pages 289--300. ACM, 2006.

\bibitem{key}
{KeY}.
\newblock \url{https://www.key-project.org/applications/program-verification},
  2017.

\bibitem{kim2017dynamic}
E.~S. Kim, S.~Sadraddini, C.~Belta, M.~Arcak, and S.~A. Seshia.
\newblock Dynamic contracts for distributed temporal logic control of traffic
  networks.
\newblock In {\em Prc. of the IEEE 56th Annual Conf. on Decision and Control
  (CDC'17)}, pages 3640--3645. IEEE, 2017.

\bibitem{kong2015dreach}
S.~Kong, S.~Gao, W.~Chen, and E.~Clarke.
\newblock d{R}each: $\delta$-reachability analysis for hybrid systems.
\newblock In {\em Proc. of the 21st Int'l Conf. on Tools and Algorithms for the
  Construction and Analysis of Systems (TACAS'15)}, volume 9035 of {\em LNCS},
  pages 200--205. Springer, 2015.

\bibitem{Koymans1990}
R.~Koymans.
\newblock Specifying real-time properties with metric temporal logic.
\newblock {\em Real-Time Systems}, 2(4):255--299, 1990.

\bibitem{DBLP:journals/corr/Kuhn15a}
T.~Kuhn.
\newblock A survey and classification of controlled natural languages.
\newblock {\em Journal of Computational Linguistics}, 40(1):121--170, 2014.

\bibitem{kuhtz_ltl_2009}
L.~Kuhtz and B.~Finkbeiner.
\newblock {LTL} path checking is efficiently parallelizable.
\newblock In {\em Proc. of the 36th Int'l Colloquium on Automata, Languages and
  Programming (ICALP'09): Part II}, volume 5556 of {\em LNCS}, pages 235--246.
  Springer, 2009.

\bibitem{kuhtz_efficient_2012}
L.~Kuhtz and B.~Finkbeiner.
\newblock Efficient parallel path checking for linear-time temporal logic with
  past and bounds.
\newblock {\em Logical Methods in Computer Science}, 8(4), 2012.

\bibitem{lamport:1978:cacm}
L.~Lamport.
\newblock Time, clocks, and the ordering of events in a distributed system.
\newblock {\em Communications of the ACM}, 21(7):558--565, July 1978.

\bibitem{lamport:1998:PP:acmtcs}
L.~Lamport.
\newblock The part-time parliament.
\newblock {\em ACM Transactions on Computer Systems}, 16(2):133--169, May 1998.

\bibitem{leguernic06asian}
G.~{Le Guernic}, A.~Banerjee, T.~P. Jensen, and D.~A. Schmidt.
\newblock {Automata-based confidentiality monitoring}.
\newblock In {\em Proc. of the 11th Asian Computing Science Conference - Secure
  Software and Related Issues (ASIAN'06)}, volume 4435 of {\em LNCS}. Springer,
  Dec. 2006.

\bibitem{Lee:11a}
J.~C. Lee, A.~S. Gardnerd, and R.~Lysecky.
\newblock Hardware observability framework for minimally intrusive online
  monitoring of embedded systems.
\newblock In {\em Proc. 18th Int. Conf. on Engineering of Computer Based
  Systems (ECBS'11)}, pages 52--60, Las Vegas, USA, Apr. 2011. IEEE Computer
  Sociery.

\bibitem{Lee:15a}
J.~C. Lee and R.~Lysecky.
\newblock System-level observation framework for non-intrusive runtime
  monitoring of embedded systems.
\newblock {\em ACM Transactions on Design Automation of Electronic Systems},
  20(42):42:1--42:27, 2015.

\bibitem{lessig}
L.~Lessig.
\newblock {\em Code and other laws of cyberspace}.
\newblock Basic Books, 1999.

\bibitem{leucker07regular}
M.~Leucker and C.~S\'{a}nchez.
\newblock Regular linear temporal logic.
\newblock In {\em Proc. of The 4th Int'l Colloquium on Theoretical Aspects of
  Computing (ICTAC'07)}, volume 4711 of {\em LNCS}, pages 291--305. Springer,
  2007.

\bibitem{leucker09brief}
M.~Leucker and C.~Schallhart.
\newblock A brief account of runtime verification.
\newblock {\em Journal of Logic Algebraic Programming}, 78(5):293--303, 2009.

\bibitem{liskov:1991:sosp}
B.~Liskov, S.~Ghemawat, R.~Gruber, P.~Johnson, L.~Shrira, and M.~Williams.
\newblock Replication in the harp file system.
\newblock In {\em Proc. of the 13th ACM Symp. on Operating Systems Principles
  (SOSP'91)}, pages 226--238. ACM, 1991.

\bibitem{lomet:1977:sigsoft}
D.~B. Lomet.
\newblock Process structuring, synchronization, and recovery using atomic
  actions.
\newblock {\em SIGSOFT Software Engineering Notes}, 2(2):128--137, Mar. 1977.

\bibitem{LS03dis}
A.~Lomuscio and M.~J. Sergot.
\newblock Deontic interpreted systems.
\newblock {\em Studia Logica}, 75(1):63--92, 2003.

\bibitem{loreti2017:distributed-com}
D.~Loreti, F.~Chesani, A.~Ciampolini, and P.~Mello.
\newblock Distributed compliance monitoring of business processes over
  mapreduce architectures.
\newblock In {\em Proc. of the 8th ACM/SPEC Int'l Conf. on Performance
  Engineering Companion (ICPE'17)}, pages 79--84. ACM, 2017.

\bibitem{lourenco:2009:padtad}
J.~M. Lourenço, R.~J. Dias, J.~Luís, M.~Rebelo, and V.~Pessanha.
\newblock Understanding the behavior of transactional memory applications.
\newblock In {\em Proc. of the 7th Workshop on Parallel and Distributed
  Systems: Testing, Analysis, and Debugging (PADTAD'09)}, pages 31--39,
  Chicago, Illinois, 2009. ACM.

\bibitem{luckham_power_2005}
D.~C. Luckham.
\newblock {\em The Power of Events - An Introduction to Complex Event
  Processing in Distributed Enterprise Systems}.
\newblock ACM, 2005.

\bibitem{Lundqvist:99a}
T.~Lundqvist and P.~Stenstrom.
\newblock Timing anomalies in dynamically scheduled microprocessors.
\newblock In {\em Proc. of the 20th IEEE Real-Time Systems Symp. (RTSS'99)}.
  IEEE Computer Society, 1999.

\bibitem{luo13enforceMOP}
Q.~Luo and G.~Ro\c{s}u.
\newblock {EnforceMOP}: A runtime property enforcement system for multithreaded
  programs.
\newblock In {\em Proc. of the 2013 Int'l Symp. on Software Testing and
  Analysis (ISSTA'13)}, pages 156--166. ACM, 2013.

\bibitem{LCC+17pvr}
B.~Luteberget, J.~J. Camilleri, C.~Johansen, and G.~Schneider.
\newblock Participatory verification of railway infrastructure by representing
  regulations in railcnl.
\newblock In {\em Proc. of the 15th Int'l Conf. on Software Engineering and
  Formal Methods, (SEFM'17)}, volume 10469 of {\em LNCS}, pages 87--103.
  Springer, 2017.

\bibitem{maler2004monitoring}
O.~Maler and D.~Nickovic.
\newblock Monitoring temporal properties of continuous signals.
\newblock In {\em Proc. of the Joint Int'l Confs. on Formal Modelling and
  Analysis of Timed Systems (FORMATS'04) and Formal Techniques in Real-Time and
  Fault-Tolerant Systems (FTRTFT'04)}, volume 3253 of {\em LNCS}, pages
  152--166. Springer, 2004.

\bibitem{maler13monitoring}
O.~Maler and D.~Nickovic.
\newblock Monitoring properties of analog and mixed-signal circuits.
\newblock {\em STTT}, 15(3):247--268, 2013.

\bibitem{Manjhi:2005:FFI:1053724.1054115}
A.~Manjhi, V.~Shkapenyuk, K.~Dhamdhere, and C.~Olston.
\newblock Finding (recently) frequent items in distributed data streams.
\newblock In {\em Proc of the 21st Int'l Conf. on Data Engineering (ICDE'05)},
  pages 767--778, Washington, DC, USA, 2005. IEEE Computer Society.

\bibitem{margara_processing_2011}
A.~Margara and G.~Cugola.
\newblock Processing flows of information: From data stream to complex event
  processing.
\newblock In {\em Proc. of the 5th ACM Int'l Conf. on Distributed Event-Based
  Systems (DEBS'11)}, pages 359--360. ACM, 2011.

\bibitem{meyer:eiffel}
B.~Meyer.
\newblock {\em {E}iffel: The Language}.
\newblock Prentice Hall, 1992.

\bibitem{Misra:1982:FRE:867576}
J.~Misra and D.~Gries.
\newblock Finding repeated elements.
\newblock {\em Science of Computer Programming}, 2(2):143--152, 1982.

\bibitem{mittal07solving}
N.~Mittal, A.~Sen, and V.~K. Garg.
\newblock Solving computation slicing using predicate detection.
\newblock {\em IEEE Transactions on Parallel and Distributed Systems (TPDS)},
  18(12):1700--1713, 2007.

\bibitem{MS13ecc}
C.~Molina{-}Jim{\'{e}}nez and S.~K. Shrivastava.
\newblock Establishing conformance between contracts and choreographies.
\newblock In {\em Proc. of the IEEE 15th Conference on Business Informatics
  (CBI'13)}, pages 69--78. {IEEE} Computer Society, 2013.

\bibitem{Morris:1978:CLN:359619.359627}
R.~Morris.
\newblock Counting large numbers of events in small registers.
\newblock {\em Communications of the ACM}, 21(10):840--842, Oct. 1978.

\bibitem{Mozafari:2017:AQE:3035918.3056098}
B.~Mozafari.
\newblock Approximate query engines: Commercial challenges and research
  opportunities.
\newblock In {\em Proc. of the 2017 ACM Int'l Conf. on Management of Data
  (SIGMOD'17)}, pages 521--524. ACM, 2017.

\bibitem{mozafarisnappydata}
B.~Mozafari, J.~Ramnarayan, S.~Menon, Y.~Mahajan, S.~Chakraborty, H.~Bhanawat,
  and K.~Bachhav.
\newblock Snappydata: {A} unified cluster for streaming, transactions and
  interactice analytics.
\newblock In {\em Proc. of the 8th Biennial Conf. on Innovative Data Systems
  Research (CIDR'17)}, 2017.

\bibitem{mcsherry-naiad2013}
D.~G. Murray, F.~McSherry, R.~Isaacs, M.~Isard, P.~Barham, and M.~Abadi.
\newblock Naiad: a timely dataflow system.
\newblock In {\em {ACM} {SIGOPS} 24th Symposium on Operating Systems
  Principles, {SOSP} '13, Farmington, PA, USA, November 3-6, 2013}, pages
  439--455, 2013.

\bibitem{Nakamoto2008}
S.~Nakamoto.
\newblock Bitcoin: A peer-to-peer electronic cash system.
\newblock {\em Bitcoin.org}, 2008.

\bibitem{Nam:17a}
T.~H. Nam.
\newblock {\em Cache Memory Aware Priority Assignment and Scheduling Simulation
  of Real-Time Embedded Systems}.
\newblock PhD thesis, Université de Bretagne Occidentale, Brest, France, Jan.
  2017.

\bibitem{DBLP:journals/crossroads/Nelson12}
J.~Nelson.
\newblock Sketching and streaming algorithms for processing massive data.
\newblock {\em ACM Crossroads}, 19(1):14--19, 2012.

\bibitem{sstl}
L.~Nenzi, L.~Bortolussi, V.~Ciancia, M.~Loreti, and M.~Massink.
\newblock Qualitative and quantitative monitoring of spatio-temporal
  properties.
\newblock In {\em Proc. of the 6th Int'l Conf. on Runtime Verification
  (RV'15)}, volume 9333 of {\em LNCS}, pages 21--37. Springer, 2015.

\bibitem{Nghiem+Others/2010/Monte}
T.~Nghiem, S.~Sankaranarayanan, G.~E. Fainekos, F.~Ivancic, A.~Gupta, and G.~J.
  Pappas.
\newblock Monte-carlo techniques for falsification of temporal properties of
  non-linear hybrid systems.
\newblock In {\em Proc. of the 13th ACM Int'l Conf. on Hybrid Systems:
  Computation and Control (HSCC'10)}, pages 211--220. ACM, 2010.

\bibitem{harmonia}
T.~Nguyen, E.~Bartocci, D.~Ni\v{c}kovi\'{c}, R.~Grosu, S.~Jaksic, and
  K.~Selyunin.
\newblock The {HARMONIA} project: Hardware monitoring for automotive
  systems-of-systems.
\newblock In {\em Proc. of 7th Int'l Symp. On Leveraging Applications of Formal
  Methods, Verification and Validation (ISoLA'16)}, volume 9953 of {\em LNCS},
  pages 371--379. Springer, 2016.

\bibitem{Nickovic2018}
D.~Nickovic, O.~Lebeltel, O.~Maler, and T.~Ferrere.
\newblock {AMT} 2.0: Qualitative and quantitative trace analysis with extended
  signal temporal logic.
\newblock In {\em Proc. of the Int'l Conf. on Tools and Algorithms for the
  Construction and Analysis of Systems (TACAS'18)}, volume 10806 of {\em LNCS},
  pages 303--319. Springer, 2018.

\bibitem{Orme:08a}
W.~Orme.
\newblock Debug and trace for multicore {SoCs}: How to build an efficient and
  effective debug and trace system for complex, multicore {SoCs}.
\newblock ARM White paper, Sept. 2008.

\bibitem{orosa:2016:euromicro}
L.~Orosa and J.~M. Lourenço.
\newblock A hardware approach for detecting, exposing and tolerating high level
  atomicity violations.
\newblock In {\em Proc. of the 24th Euromicro Int'l Conf. on Parallel,
  Distributed, and Network-Based Processing (PDP'16)}, pages 159--167. IEEE
  Computer Society, 2016.

\bibitem{pace09challenges}
G.~J. Pace and G.~Schneider.
\newblock Challenges in the specification of full contracts.
\newblock In {\em Proc. of the 7th Int'l Conf. on Integrated Formal Methods
  (iFM'09)}, volume 5423 of {\em LNCS}, pages 292--306. Springer, 2009.

\bibitem{pant2017smooth}
Y.~V. Pant, H.~Abbas, and R.~Mangharam.
\newblock Smooth operator: Control using the smooth robustness of temporal
  logic.
\newblock In {\em Proc. of the 2017 IEEE Conference on Control Technology and
  Applications (CCTA'17)}, pages 1235--1240. IEEE, 2017.

\bibitem{PCP+16aba}
R.~Pardo, C.~Colombo, G.~Pace, and G.~Schneider.
\newblock An automata-based approach to evolving privacy policies for social
  networks.
\newblock In {\em Proc. of the 16th Int'l Conf. on Runtime Verification
  (RV'16)}, volume 10012 of {\em LNCS}, pages 285--301. Springer, 2016.

\bibitem{PKS+16sep}
R.~Pardo, I.~Kelly\'erov\'a, C.~S\'anchez, and G.~Schneider.
\newblock Specification of evolving privacy policies for online social
  networks.
\newblock In {\em Proc. of the 23rd International Symposium on Temporal
  Representation and Reasoning (TIME'16)}, pages 70--79. IEEE Computer Society,
  2016.

\bibitem{PSS18tek}
R.~Pardo, C.~S\'anchez, and G.~Schneider.
\newblock Timed epistemic knowledge bases for social networks.
\newblock In {\em Proc. of the 22nd Int'l Symposium on Formal Methods (FM'18)},
  volume 10951 of {\em LNCS}, pages 185--202. Springer, 2018.

\bibitem{DBLP:conf/cade/PassmoreI17}
G.~O. Passmore and D.~Ignatovich.
\newblock Formal verification of financial algorithms.
\newblock In {\em Proc. of the 26th Int'l Conf. on Automated Deduction
  (CADE'17)}, volume 10395 of {\em LNCS}, pages 26--41. Springer, 2017.

\bibitem{patel:2013:sigcomm}
P.~Patel, D.~Bansal, L.~Yuan, A.~Murthy, A.~Greenberg, D.~A. Maltz, R.~Kern,
  H.~Kumar, M.~Zikos, H.~Wu, C.~Kim, and N.~Karri.
\newblock Ananta: Cloud scale load balancing.
\newblock In {\em Proc. of the ACM SIGCOMM 2013 Conf. (SIGCOMM '13)}, SIGCOMM
  '13, pages 207--218. ACM, 2013.

\bibitem{Pellizzoni:08a}
R.~Pellizzoni, P.~Meredith, M.~Caccamo, and G.~Rosu.
\newblock Hardware runtime monitoring for dependable {COTS}-based real-time
  embedded systems.
\newblock In {\em Proc. of the IEEE Real-Time Systems Symp. (RTSS'08)}, pages
  481--491. IEEE Computer Society, 2008.

\bibitem{Phan:11a}
L.~T.~X. Phan, I.~Lee, and O.~Sokolsky.
\newblock A semantic framework for mode change protocols.
\newblock In {\em Proc. of the 17th IEEE Real-Time and Embedded Technology and
  Applications Symposium (RTAS'11)}, pages 91--100. IEEE Computer Society,
  2011.

\bibitem{PSt18ayp}
P.~Picazo-S\'anchez, G.~Schneider, and J.~Tapiador.
\newblock After you, please: Browser extensions order attacks and
  countermeasures.
\newblock Under submission.

\bibitem{Pike:11a}
L.~Pike, S.~Niller, and N.~Wegmann.
\newblock Runtime verification for ultra-critical systems.
\newblock In {\em Proc. of the 2nd Int'l Conf. on Runtime Verification
  (RV'11)}, volume 7186 of {\em LNCS}, pages 310--324. Springer, 2011.

\bibitem{PAS+17mdm}
S.~Pinisetty, T.~Antignac, D.~Sands, and G.~Schneider.
\newblock {Monitoring Data Minimisation}.
\newblock Technical report, Chalmers and University of Gothenburg, 2017.

\bibitem{PSS18rvh}
S.~Pinisetty, D.~Sands, and G.~Schneider.
\newblock Runtime verification of hyperproperties for deterministic programs.
\newblock In {\em Proc. of the 6th Int'l Conf. on Formal Methods in Software
  Engineering (FormaliSE'18)}. ACM, 2018.

\bibitem{RCPinto2014rv}
R.~C. Pinto and J.~Rufino.
\newblock Towards non-invasive run-time verification of real-time systems.
\newblock In {\em Proc. of the 26th Euromicro Conf. on Real-Time Systems - WIP
  Session}, pages 25--28. Euromicro, July 2014.

\bibitem{pnueli06psl}
A.~Pnueli and A.~Zaks.
\newblock {PSL} model checking and run-time verification via testers.
\newblock In {\em Proc. of the 14th Int'l Symp. on Formal Methods (FM'06)},
  volume 4085 of {\em LNCS}, pages 573--586. Springer, 2006.

\bibitem{PS13fan}
H.~Prakken and G.~Sartor.
\newblock Formalising arguments about norms.
\newblock In {\em Proc. of the 26th Annual Conf. on Legal Knowledge and
  Information Systems (JURIX'13)}, volume 259 of {\em Frontiers in Artificial
  Intelligence and Applications}, pages 121--130. {IOS} Press, 2013.

\bibitem{PS12ddl}
C.~Prisacariu and G.~Schneider.
\newblock A dynamic deontic logic for complex contracts.
\newblock {\em Journal of Logic and Algebraic Programming}, 81(4):458--490, May
  2012.

\bibitem{DBLP:journals/corr/Prybila0HW17}
C.~Prybila, S.~Schulte, C.~Hochreiner, and I.~Weber.
\newblock Runtime verification for business processes utilizing the bitcoin
  blockchain.
\newblock {\em CoRR}, abs/1706.04404, 2017.

\bibitem{Rahmatian:12a}
M.~Rahmatian, H.~Kooti, I.~G. Harris, and E.~Bozorgzadeh.
\newblock Adaptable intrusion detection using partial runtime reconfiguration.
\newblock In {\em Proc. of the IEEE 30th Int'l Conf. on Computer Design
  (ICCD'12)}, pages 147--152. IEEE Computer Society, 2012.

\bibitem{raman2014model}
V.~Raman, A.~Donz{\'e}, M.~Maasoumy, R.~M. M, A.~Sangiovanni-Vincentelli, and
  S.~A. Seshia.
\newblock Model predictive control with signal temporal logic specifications.
\newblock In {\em Proc. of the 53rd Annual Conf. on Decision and Control
  (CDC'14)}, pages 81--87. IEEE, 2014.

\bibitem{stlce}
V.~Raman, A.~Donz{\'{e}}, D.~Sadigh, R.~M.~Murray, and S.~A. Seshia.
\newblock Reactive synthesis from signal temporal logic specifications.
\newblock In {\em Proc. HSCC'15: the 18th International Conference on Hybrid
  Systems: Computation and Control}, pages 239--248. ACM, 2015.

\bibitem{ravichandran:2014:pact}
K.~Ravichandran, A.~Gavrilovska, and S.~Pande.
\newblock {DeSTM:} harnessing determinism in {STMs} for application
  development.
\newblock In {\em Proc. of the Int'l Conf. on Parallel Architectures and
  Compilation (PACT)}, pages 213--224. ACM, 2014.

\bibitem{RayGDBBG15}
R.~Ray, A.~Gurung, B.~Das, E.~Bartocci, S.~Bogomolov, and R.~Grosu.
\newblock Xspeed: Accelerating reachability analysis on multi-core processors.
\newblock In {\em Proc. of the 11th Int'l Haifa Verification Conf. on Hardware
  and Software: Verification and Testing (HVC'15)}, volume 9434 of {\em LNCS},
  pages 3--18. Springer, 2015.

\bibitem{RegerR-RV2015}
G.~Reger and D.~Rydeheard.
\newblock From first-order temporal logic to parametric trace slicing.
\newblock In {\em Proc. of the 6th Int'l Conf. on Runtime Verification
  (RV'15)}, LNCS, pages 216--232. Springer, 2015.

\bibitem{Reinbacher:14b}
T.~Reinbacher, M.~Fugger, and J.~Brauer.
\newblock Runtime verification of embedded real-time systems.
\newblock {\em Formal Methods in System Design}, 24(3):203--239, June 2014.

\bibitem{Reinbacher:14a}
T.~Reinbacher, K.~Y. Rozier, and J.~Schumann.
\newblock Temporal-logic based runtime observer pairs for system health
  management of real-time systems.
\newblock In {\em Proc. 20th Int. Conf. on Tools and Algorithms for the
  Construction and Analysis of Systems (TACAS'14)}, volume 8413 of {\em LNCS},
  pages 357--372. Springer, 2014.

\bibitem{Intel:16-PT}
J.~Reinders.
\newblock {\em Intel Processor Tracing}.
\newblock Intel Corporation, Sept. 2013.

\bibitem{rizk}
A.~Rizk, G.~Batt, F.~Fages, and S.~Soliman.
\newblock On a continuous degree of satisfaction of temporal logic formulae
  with applications to systems biology.
\newblock In {\em Proc. of the 6th Int'l Conf. on Computational Methods in
  Systems Biology (CMSB'08)}, volume 5307 of {\em LNCS}, pages 251--268.
  Springer, 2008.

\bibitem{filtering}
A.~Rodionova, E.~Bartocci, D.~Ni\v{c}kovi\'{c}, and R.~Grosu.
\newblock Temporal logic as filtering.
\newblock In {\em Proc. of HSCC 2016: the 19th International Conference on
  Hybrid Systems: Computation and Control}, pages 11--20. {ACM}, 2016.

\bibitem{Rufino:16a}
J.~Rufino.
\newblock Towards integration of adaptability and non-intrusive runtime
  verification in avionic systems.
\newblock {\em ACM SIGBED Review}, 13(1):60--65, Jan. 2016.
\newblock (Special Issue on 5th Embedded Operating Systems Workshop).

\bibitem{Rufino:16b-OSPERT}
J.~Rufino and I.~Gouveia.
\newblock Timeliness runtime verification and adaptation in avionic systems.
\newblock In {\em Proc. of the 12th Workshop on Operating Systems Platforms for
  Embedded Real-Time applications (OSPERT'16)}, pages 14--20, Toulouse, France,
  July 2016. Euromicro.

\bibitem{ruleml}
{RuleML Conferences}.
\newblock \url{http://2018.ruleml-rr.org}.

\bibitem{russo10cfs}
A.~Russo and A.~Sabelfeld.
\newblock Dynamic vs. static flow-sensitive security analysis.
\newblock In {\em Proc. of the 23rd IEEE Computer Security Foundations
  Symposium (CSF'10)}, pages 186--199. IEEE Computer Society, 2010.

\bibitem{sabelfeld03jsac}
A.~Sabelfeld and A.~C. Myers.
\newblock Language-based information-flow security.
\newblock {\em Journal on Selected Areas in Communications}, 21(1):5--19, 2003.

\bibitem{SainiGLC16}
A.~Saini, M.~S. Gaur, V.~Laxmi, and M.~Conti.
\newblock Colluding browser extension attack on user privacy and its
  implication for web browsers.
\newblock {\em Computers {\&} Security}, 63:14--28, 2016.

\bibitem{BHSS08sia}
M.~B. Salem, S.~Hershkop, and S.~J. Stolfo.
\newblock A survey of insider attack detection research.
\newblock In {\em Insider Attack and Cyber Security - Beyond the Hacker}, pages
  69--90. Springer, 2008.

\bibitem{sanchez10regular}
C.~S\'anchez and M.~Leucker.
\newblock Regular linear temporal logic with past.
\newblock In {\em Proc. of the 11th Int'l Conf. on Verification, Model
  Checking, and Abstract Interpretation, (VMCAI'10)}, volume 5944 of {\em
  LNCS}, pages 295--311. Springer, 2010.

\bibitem{insulin}
S.~Sankaranarayanan, C.~Miller, R.~Raghunathan, H.~Ravanbakhsh, and G.~E.
  Fainekos.
\newblock A model-based approach to synthesizing insulin infusion pump usage
  parameters for diabetic patients.
\newblock In {\em Proc. of the 50th Annual Allerton Conference on
  Communication, Control, and Computing}, pages 1610--1617. IEEE, 2012.

\bibitem{fragososantos15tgc}
J.~F. Santos, T.~Jensen, T.~Rezk, and A.~Schmitt.
\newblock {Hybrid Typing of Secure Information Flow in a JavaScript-like
  Language}.
\newblock In {\em Proc. of the 10th Int'l Symp. on on Trustworthy Global
  Computing (TGC'15)}, volume 9533 of {\em LNCS}. Springer, 2015.

\bibitem{schneider:1990:cacm}
F.~B. Schneider.
\newblock Implementing fault-tolerant services using the state machine
  approach: A tutorial.
\newblock {\em ACM Computing Surveys}, 22(4):299--319, Dec. 1990.

\bibitem{Schultz-Moller:2009:DCE:1619258.1619264}
N.~P. Schultz-M{\o}ller, M.~Migliavacca, and P.~Pietzuch.
\newblock Distributed complex event processing with query rewriting.
\newblock In {\em Proc. of the 3rd ACM Int'l Conf. on Distributed Event-Based
  Systems (DEBS'09)}, pages 4:1--4:12. ACM, 2009.

\bibitem{SelyuninJNRHBNG17}
K.~Selyunin, S.~Jaksic, T.~Nguyen, C.~Reidl, U.~Hafner, E.~Bartocci,
  D.~Nickovic, and R.~Grosu.
\newblock Runtime monitoring with recovery of the {SENT} communication
  protocol.
\newblock In {\em Proc. of the the 29th Int'l Conf. on Computer Aided
  Verification (CAV'17)}, volume 10426 of {\em LNCS}, pages 336--355. Springer,
  2017.

\bibitem{SelyuninNBNG16}
K.~Selyunin, T.~Nguyen, E.~Bartocci, D.~Ni\v{c}kovi\'{c}, and R.~Grosu.
\newblock Monitoring of {MTL} specifications with ibm's spiking-neuron model.
\newblock In {\em Proc. of the 2016 Design, Automation {\&} Test in Europe
  Conference (DATE'16)}, pages 924--929. {IEEE}, 2016.

\bibitem{sen04efficient}
K.~Sen, A.~Vardhan, G.~Agha, and G.~Rosu.
\newblock Efficient decentralized monitoring of safety in distributed systems.
\newblock In {\em Proc. of 26th Int'l Conf. on Software Engineering (ICSE
  2004)}, pages 418--427. IEEE CS Press, 2004.

\bibitem{SER12sdl}
A.~Shabtai, Y.~Elovici, and L.~Rokach.
\newblock {\em A Survey of Data Leakage Detection and Prevention Solutions}.
\newblock Springer Briefs in Computer Science. Springer, 2012.

\bibitem{rao:2011:vldb}
J.~R. E.~J. Shekita and S.~Tata.
\newblock Using paxos to build a scalable, consistent, and highly available
  datastore.
\newblock {\em Proc. of the VLDB Endowment}, 4(4):243--254, Jan. 2011.

\bibitem{Shobaki:01a}
M.~E. Shobaki and L.~Lindh.
\newblock A hardware and software monitor for high-level system-on-chip
  verification.
\newblock In {\em Proc. of the 2nd IEEE Int'l Symp. on Quality Electronic
  Design (ISQED 2001)}, pages 56--61, 2001.

\bibitem{spoth:2017:cidr:adaptive}
W.~Spoth, B.~S. Arab, E.~S. Chan, G.~Dieter, A.~Ghoneimy, B.~Glavic,
  B.~Hammerschmidt, O.~Kennedy, S.~Lee, Z.~H. Liu, X.~Niu, and Y.~Yang.
\newblock Adaptive schema databases.
\newblock In {\em Proc. of the 8th Biennial Int'l on Innovative Data Systems
  Research (CIDR'17)}. CIDRDB, 2017.

\bibitem{Stoller2012}
S.~D. Stoller, E.~Bartocci, J.~Seyster, R.~Grosu, K.~Havelund, S.~A. Smolka,
  and E.~Zadok.
\newblock Runtime verification with state estimation.
\newblock In {\em Proc. of the 2nd Int'l Conf. on Runtime Verification
  (RV'11)}, volume 7186 of {\em LNCS}, pages 193--207. Springer, 2011.

\bibitem{szabo}
N.~Szabo.
\newblock Smart contracts: Building blocks for digital markets.
\newblock {\em Extropy}, 16, 1996.

\bibitem{Todman:15a}
T.~Todman, S.~Stilkerich, and W.~Luk.
\newblock In-circuit temporal monitors for runtime verification of
  reconfigurable designs.
\newblock In {\em Proc. of the 52nd Annual Design Automation Conference
  (DAC'15)}, pages 50:1--50:6. ACM, 2015.

\bibitem{TsankovMDB14}
P.~Tsankov, S.~Marinovic, M.~T. Dashti, and D.~A. Basin.
\newblock Decentralized composite access control.
\newblock In M.~Abadi and S.~Kremer, editors, {\em Proc. of the 3rd Int'l Conf.
  Principles of Security and Trust (POST'14)}, volume 8414 of {\em LNCS}, pages
  245--264. Springer, 2014.

\bibitem{ulus14timed}
D.~Ulus, T.~Ferr\`{e}re, E.~Asarin, and O.~Maler.
\newblock Timed pattern matching.
\newblock In {\em Proc. of the 12th Int'l Conf. on Formal Modeling and Analysis
  of Timed Systems (FORMATS'14)}, volume 8711 of {\em LNCS}, pages 222--236.
  Springer, 2014.

\bibitem{ulus16online}
D.~Ulus, T.~Ferrère, E.~Asarin, and O.~Maler.
\newblock Online timed pattern matching using derivatives.
\newblock In {\em Proc. of TACAS'16}, volume 9636 of {\em LNCS}, pages
  736--751, Berlin, Germany, 2016. Springer.

\bibitem{vachharajani04micro}
N.~Vachharajani, M.~J. Bridges, J.~Chang, R.~Rangan, G.~Ottoni, J.~A. Blome,
  G.~A. Reis, M.~Vachharajani, and D.~I. August.
\newblock Rifle: An architectural framework for user-centric information-flow
  security.
\newblock In {\em Proc. of the 37th Int'l Symp. on Microarchitecture
  (MICRO'04)}, pages 243--254. IEEE Computer Society, 2004.

\bibitem{vale:2016:taaco}
T.~M. Vale, J.~A. Silva, R.~J. Dias, and J.~M. Lourenço.
\newblock Pot: Deterministic transactional execution.
\newblock {\em ACM Transactions on Architecture and Code Optimzation},
  13(4):52:1--52:24, Dec. 2016.

\bibitem{10.1007/978-3-642-28872-2_1}
W.~M.~P. van~der Aalst.
\newblock Distributed process discovery and conformance checking.
\newblock In {\em Proc. of the 15th Int'l Conf. on Fundamental Approaches to
  Software Engineering (FASE'12)}, volume 7212 of {\em LNCS}, pages 1--25,
  Berlin, Heidelberg, 2012. Springer.

\bibitem{vardi_church_2008}
M.~Y. Vardi.
\newblock From {C}hurch and {P}rior to {PSL}.
\newblock In {\em 25 Years of Model Checking - History, Achievements,
  Perspectives}, volume 5000 of {\em LNCS}, pages 150--171. Springer, 2008.

\bibitem{viswanathan00foundations}
M.~Viswanathan.
\newblock {\em Foundations for the run-time analysis of software systems}.
\newblock PhD thesis, University of Pennsylvania, 2000.

\bibitem{volpano96jcs}
D.~Volpano, C.~Irvine, and G.~Smith.
\newblock A sound type system for secure flow analysis.
\newblock {\em Journal of Computer Security}, 4(2-3), Jan. 1996.

\bibitem{Watterson:07a}
C.~Watterson and D.~Heffernan.
\newblock Runtime verification and monitoring of embedded systems.
\newblock {\em IET software}, 1(5):172--179, Oct. 2007.

\bibitem{DBLP:conf/bpm/WeberXRGPM16}
I.~Weber, X.~Xu, R.~Riveret, G.~Governatori, A.~Ponomarev, and J.~Mendling.
\newblock Untrusted business process monitoring and execution using blockchain.
\newblock In {\em Proc. of the 14th Int'l Conf. on Business Process Management
  (BPM'16)}, volume 9850 of {\em LNCS}, pages 329--347. Springer, 2016.

\bibitem{weil:2006:osdi}
S.~A. Weil, S.~A.Brandt, E.~L. Miller., D.~D.~E. Long, and C.~Maltzahn.
\newblock {Ceph:} a scalable, high-performance distributed file system.
\newblock In {\em Proc. of the 7th Symp. on Operating Systems Design and
  Implementation (OSDI'06)}, OSDI '06, pages 307--320. USENIX Association,
  2006.

\bibitem{White:2007:NEP:1265530.1265567}
W.~White, M.~Riedewald, J.~Gehrke, and A.~Demers.
\newblock What is ``next'' in event processing?
\newblock In {\em Proc. of the 26th ACM SIGMOD-SIGACT-SIGART Symp. on
  Principles of Database Systems (PODS'07)}, pages 263--272, New York, NY, USA,
  2007. ACM.

\bibitem{Wilhelm:08a}
R.~Wilhelm, J.~Engblom, A.~Ermedahl, N.~Holsti, S.~Thesing, D.~Whalley,
  G.~Bernat, C.~Ferdinand, R.~Heckmann, T.~Mitra, F.~Mueller, I.~Puaut,
  P.~Puschner, J.~Staschulat, and P.~Stenstrom.
\newblock The worst-case execution time problem - overview of methods and
  survey of tools.
\newblock {\em ACM Transactions on Embedded Computing Systems}, 7(3), Apr.
  2008.

\bibitem{WongpiromsarnTM12}
T.~Wongpiromsarn, U.~Topcu, and R.~M. Murray.
\newblock Receding horizon temporal logic planning.
\newblock {\em {IEEE} Transactions on Automatation and Control},
  57(11):2817--2830, 2012.

\bibitem{ethereum}
G.~Wood.
\newblock Ethereum: A secure decentralised generalised transaction ledger.
\newblock {\em Ethereum Project Yellow Paper}, 151:1--32, 2014.

\bibitem{Woodruff:2012:TBD:2213977.2214063}
D.~P. Woodruff and Q.~Zhang.
\newblock Tight bounds for distributed functional monitoring.
\newblock In {\em Proc. of the 44th Annual ACM Symposium on Theory of Computing
  (STOC'12)}, pages 941--960. ACM, 2012.

\bibitem{VW51dl}
G.~H.~V. Wright.
\newblock Deontic logic.
\newblock {\em Mind}, 60:1--15, 1951.

\bibitem{wsla}
{WSLA}.
\newblock \url{www.research.ibm.com/wsla/}.

\bibitem{Wyner15}
A.~Z. Wyner.
\newblock From the language of legislation to executable logic programs.
\newblock In {\em Logic in the Theory and Practice of Lawmaking}, volume~2 of
  {\em Legisprudence Library}, pages 409--434. Springer, 2015.

\bibitem{mining}
J.~Xiaoqing, A.~Donz{\'e}, J.~V. Deshmukh, and S.~A. Seshia.
\newblock Mining {R}equirements from {C}losed-loop {C}ontrol {M}odels.
\newblock In {\em Proc. of the {ACM} International Conference on Hybrid
  Systems: Computation and Control (HSCC'13)}, pages 43--52. ACM, 2013.

\bibitem{xie:2014:sosp}
C.~Xie, C.~Su, M.~Kapritsos, Y.~Wang, N.~Yaghmazadeh, L.~Alvisi, and
  P.~Mahajan.
\newblock {Salt:} combining {ACID} and {BASE} in a distributed database.
\newblock In {\em Proc. of the 11th USENIX Conf. on Operating Systems Design
  and Implementation (OSDI'14)}, pages 495--509. USENIX Association, 2014.

\bibitem{xie:2015:sosp}
C.~Xie, C.~Su, C.~Littley, L.~Alvisi, M.~Kapritsos, and Y.~Wang.
\newblock High-performance {ACID} via modular concurrency control.
\newblock In {\em Proc. of the 25th Symp. on Operating Systems Principles
  (SOSP'15)}, pages 279--294. ACM, 2015.

\bibitem{YaghoubiF2017acc}
S.~Yaghoubi and G.~Fainekos.
\newblock Hybrid approximate gradient and stochastic descent for falsification
  of nonlinear systems.
\newblock In {\em Proc. the 2017 American Control Conference (ACC'17)}, pages
  529--534. IEEE, 2017.

\bibitem{Yang2012}
H.~Yang, B.~Hoxha, and G.~Fainekos.
\newblock Querying parametric temporal logic properties on embedded systems.
\newblock In {\em Proc. of the 24th IFIP WG 6.1 Int'l Conf. on Testing Software
  and Systems (ICTSS'12)}, volume 7641 of {\em LNCS}, pages 136--151. Springer,
  2012.

\bibitem{yang16pldi}
J.~Yang, T.~Hance, T.~H. Austin, A.~Solar-Lezama, C.~Flanagan, and S.~Chong.
\newblock Precise, dynamic information flow for database-backed applications.
\newblock In {\em Proc. of the 37th ACM SIGPLAN Conf. on Programming Language
  Design and Implementation (PLDI'16)}, pages 631--647. ACM, June 2016.

\bibitem{YLA+17smd}
Y.~Ye, T.~Li, D.~Adjeroh, and S.~S. Iyengar.
\newblock A survey on malware detection using data mining techniques.
\newblock {\em ACM Computing Surveys}, 50(3):41:1--41:40, June 2017.

\bibitem{Yi:2013:OTD:3118739.3118856}
K.~Yi and Q.~Zhang.
\newblock Optimal tracking of distributed heavy hitters and quantiles.
\newblock {\em Algorithmica}, 65(1):206--223, Jan. 2013.

\bibitem{yu2017:verifying-tempo}
B.~Yu, Z.~Duan, C.~Tian, and N.~Zhang.
\newblock Verifying temporal properties of programs: A parallel approach.
\newblock {\em Journal of Parallel and Distributed Computing}, 2017.

\bibitem{zaharia_spark:_2010}
M.~Zaharia, M.~Chowdhury, M.~J. Franklin, S.~Shenker, and I.~Stoica.
\newblock Spark: Cluster computing with working sets.
\newblock In {\em Proc. of the 2nd USENIX Workshop on Hot Topics in Cloud
  Computing (HotCloud'10)}. USENIX Association, 2010.

\end{thebibliography}

\end{document}